\documentclass[journal]{IEEEtran}
\usepackage{amsthm}
\usepackage{amsmath}
\usepackage{amssymb}
\usepackage{algorithm}
\usepackage{algorithmic}
\usepackage{enumerate}
\usepackage{graphicx}
\usepackage{color}
\usepackage{subfigure}
\usepackage{booktabs}
\usepackage{multirow}
\usepackage{multicol}
\usepackage{cite}
\usepackage{picins}
\usepackage{url}
\usepackage[justification=centering]{caption}
\usepackage[utf8]{inputenc}
\graphicspath{{figures/}}
\ifCLASSINFOpdf
\else
\fi
\begin{document}
%
\title{Efficient reversible data hiding via two layers of double-peak embedding}
%
%
%


\author{Fuhu Wu,
	Jian Sun,
	Shun Zhang$^\ast$, Zhili Chen, Hong Zhong
\IEEEcompsocitemizethanks{\IEEEcompsocthanksitem F. Wu, J. Sun, S. Zhang and H Zhong are with School of Computer Science and Technology, Anhui University, Hefei 230601, China. \protect
E-mail: wufuhu@ahu.edu.cn; e20201050@stu.ahu.edu.cn; szhang@ahu.edu.cn; zhongh@ahu.edu.cn;
\IEEEcompsocthanksitem Z. Chen is with Software Engineering Institute, East China Normal University, Shanghai 200062, China. \protect
E-mail: zhlchen@sei.ecnu.edu.cn.
\IEEEcompsocthanksitem $^\ast$ Corresponding author (Shun Zhang)}
}

\markboth{Journal of \LaTeX\ Class Files,~Vol.~0, No.~0, February~2022}%
{Shell \MakeLowercase{\textit{et al.}}: Efficient reversible data hiding via two-layer embedding strategy}

%



\maketitle

\begin{abstract}
Reversible data hiding continues to attract significant attention in recent years. In particular, an increasing number of authors focus on the higher significant bit (HSB) plane of an image which can yield more redundant space. On the other hand, the lower significant bit planes are often ignored for embedding in existing schemes due to their harm to the embedding rate. This paper proposes an efficient reversible data hiding scheme via a double-peak two-layer embedding (DTLE) strategy with prediction error expansion. The higher six-bit planes of the image are assigned as the HSB plane, and double prediction error peaks are applied in either embedding layer. This makes fuller use of the redundancy space of images compared with the one error peak strategy. Moreover, we carry out the median-edge detector pre-processing for complex images to reduce the size of the auxiliary information. A series of experimental results show that our DTLE approach achieves up to $83\%$ higher embedding rate on real-world datasets while guaranteeing better image quality.
\end{abstract}

\begin{IEEEkeywords}
Reversible data hiding, higher significant bit, prediction error expansion, median-edge detector, embedding rate
\end{IEEEkeywords}

%
\IEEEpeerreviewmaketitle

\section{Introduction}
\IEEEPARstart{W}{ith} the maturity of the $5$G mobile communication technology and dramatic proliferation of mobile devices, the information industry has developed explosively. This leads to the protection issue on users' data privacy and security, which has attracts special attention from the government and industry. Data hiding has become one of the important technologies of protecting data \cite{CC04,MPJ12}. It embeds secret data into the images that can be transmitted by public channels. However, data hiding may cause permanent damage to the original image after the extraction of secret data, then it has limited applications in the areas with high-security requirements, such as medical diagnosis, military image processing, etc. Afterwards, reversible data hiding (RDH) is studied by some authors for enabling the complete reconstruction of original images after the extraction \cite{SNZ04,KYK18,WZC21}.

There appear various methods of reversible data hiding. The commonly used methods include lossless compression, integer transform, difference expansion (DE), histogram shifting (HS) and prediction error expansion (PEE) \cite{ ZHL13,GS15,Tian03,NSA06}. Specifically, the lossless compression method obtains a redundant space for embedding secret data while compressing some trivial contents with maintaining the image quality \cite{FK15,AXA19}. The integer transform approach modifies some specified frequency domain coefficients of the main signal to embed data \cite{XXS18,QQZ20,CC20}. The DE method allows further exploration of the image while freeing up more redundant space from the difference values between pixels for data embedding \cite{SCH15,MJ19,TWC20}. The HS method focuses on data embedding in the histogram peaks \cite{LLY13,SLV22,HCZ22}.

The PEE method improves the embedding capacity (EC) via the correlations among adjacent pixels in the image. Li et al. \cite{LLL13} combined PEE with pixel-value-ordering to embed secret data by the prediction error between the maximum and minimum values of each block. Peng et al. \cite{PLY14}  extended Li et al.'s work and presented an improved PVO-based RDH method which utilized a new histogram-modification strategy and computed new differences to embed data.
Li et al. \cite{LWL18} added a new location map marker bit to enable the embedding of multiple largest-valued (or smallest-valued) pixels of blocks. Tang et al. \cite{TZL20} presented a fluctuation based sorting strategy and embedded additional information by utilizing an improved rhombus predictor.

Recently some authors paid main attention to the higher significant bit (HSB) plane for more redundancy space \cite{YXZ19,PP18,PYQ18,YST21}. Liu et al. \cite{LCW16} divided the image into blocks and counted pixels on the bit planes selected from the third bit to the seventh bit, which solved the problem that histogram zeros may not exist. However, the scheme was limited by the size of blocks and gave up the embedding space of the highest bit plane, which resulted in a lower EC. Wang et al. \cite{WYW17} decomposed the pixels to get the HSB and lower significant bit (LSB) planes. The HSB plane pixels were paired to generate difference and the secret data bits were embedded in the histogram. The approach failed to balance between EC and image quality, resulting in the final image with a low peak signal-to-noise ratio (PSNR). Recently, Kumar et al. \cite{KJ20} proposed a two-layer embedding (TLE) strategy based on the HSB plane which embeded by two prediction error values. The scheme ignored the zero prediction-error case in embedding and made insufficient use of the image redundant space while losing the size control of auxiliary information. 

As mentioned above, the existing solutions for RDH still have two limitations.

(1) \textbf{Ignoring the adoption of the sixth-bit plane.}
The lower bit planes have the disadvantage of providing smaller embedding space due to poor correlations while reducing image distortion \cite{WYW17,KJ20}. This presents us a chance of utilizing the sixth-bit plane appropriately.

(2) \textbf{Lacking the size control of auxiliary information.}
Some complex images have large location maps for marking possible overflow pixels, which leads to an amount of auxiliary information and the drop of EC \cite{KJ20}. This inspires us to reduce the size of auxiliary information by pre-processing.

To address these issues, we propose a reversible data hiding scheme based on two-layer embedding and PEE, which maximizes the utilization of the redundant space while reducing the distortion of the image by exploiting lower bit planes. We also deploy the median-edge detector (MED) approach \cite{WSS00} for pre-processing complex images which used to be applied in encrypted images.

The main contributions of our work are as follows:

\begin{itemize}
    \item We design a new approach DTLE, an efficient reversible data hiding scheme via the two-layer embedding, which implements the embedding at two prediction error peaks in either layer.
    \item The design of DTLE takes into account the sixth-bit plane and the eight surrounding pixels in the embedding process for the HSB plane. The value changes involved in the sixth-bit plane result in smaller variations of pixel values and improves image quality.
    \item Our proposal DTLE utilizes MED pre-processing before embedding, which significantly reduces the size of auxiliary information for complex images. Experimental results demonstrate that our designed scheme outperforms the state-of-the-art schemes in terms of EC and PSNR.
\end{itemize}

The remaining parts of this paper are organized as follows. We first introduce the related schemes in Section \ref{sec:relatedwork}.
Section \ref{sec:PS} describes the details of the proposed scheme. Experimental results with comparisons are presented in Section \ref{sec:result}. Finally, we conclude in Section \ref{sec:con}.

\section{Related Work}\label{sec:relatedwork}
In this section, as a review, we first introduce Liu et al.'s BPTI approach \cite{LCW16}. Then, we present Wang et al.'s SBDE embedding \cite{WYW17}. Moreover, Kumar et al.'s TLE embedding is also presented. Actually, our proposed method is motivated by SBDE and TLE embeddings to improve EC and PSNR.


\subsection{Liu et al.'s BPTI approach \cite{LCW16}}
Liu et al. embedded the secret data at the peak point in the $18\times 18$ blocks by HS. This scheme extracted $n$-bit planes from an $8$-bit plane for each pixel to generate the bit plane truncation image (BPTI), in which the bit planes were selected from the third bit to the seventh bit.
The peak point $p$ and zero point $z$ from the histogram of the current block were recorded. Secret bits were embedded in the point $p$ while values between the points of $z$ and $p$ were shifted. The current block was skipped if no zero point existed inside.

\begin{figure}[tb]
	\centering
    \includegraphics[width=2.5cm]{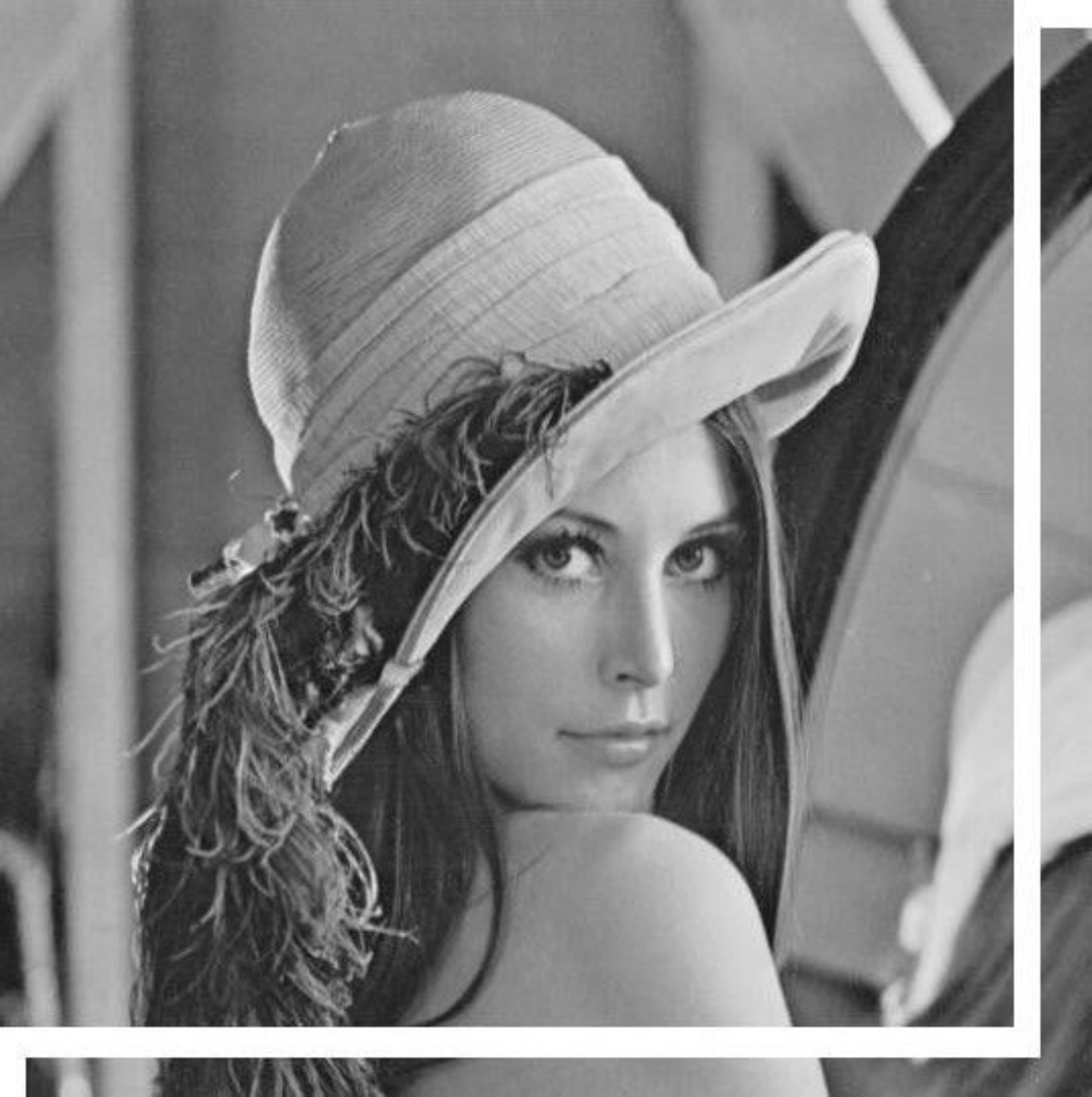}\\
	\caption{Region division of original image.}
	\label{fig:tr}
\end{figure}

The original image was divided into two areas according to Fig.~\ref{fig:tr}. The pixels in the top left area denoted by $M$ were used for embedding secret data, while the bottom right area $N$ played the role of storing auxiliary information such as location maps and block size, to ensure the reversibility.

\subsection{Wang et al.'s SBDE embedding \cite{WYW17}}
Wang et al. proposed a RDH scheme for embedding secret data in the HSB plane using significant-bit-difference expansion (SBDE). The value of each pixel ranging from $0$ to $255$ for a greyscale image $I_o$ of size $h \times w$ was represented by 8 bits. The pixel value ${P_{i,j}}$ of the $i$-th row and $j$-th column in the original image could be represented by two parts as

\begin{equation}
\label{eq:11}
{P_{i,j}} = \sum\limits_{k = n}^7 {{c_k} \times {2^k} + \sum\limits_{k = 0}^{n - 1} {{c_k} \times {2^k}} },
\end{equation}
where
\begin{equation}
\label{eq:22}
{h_{i,j}} = \sum\limits_{k = n}^7 {{c_k} \times {2^k}},
\end{equation}
\begin{equation}
\label{eq:33}
{l_{i,j}} = \sum\limits_{k = 0}^{n - 1} {{c_k} \times {2^k}}.
\end{equation}
The ${h_{i,j}}$ denotes the pixel value in HSB plane while ${l_{i,j}}$ the pixel value in LSB, ${c_k} \in \left\{ {0,1} \right\}$ represented the value in the $k$-th bit plane of the pixel and $n$ ranges from $0$ to $7$. Then, the pixel value in HSB can be updated by a new decimal number $h_{i,j}$ ranging from $0$ to $255$,

\begin{equation}
\label{eq:new}
h_{i,j} = (\sum\limits_{k = n}^7 {{c_k}}  \times {2^k})/{2^n}
\end{equation}

Following this, all pixels in the image can be represented by two values $h_{i,j}$ and $l_{i,j}$ in HSB and LSB, respectively, due to the bit plane decomposition.

Take an example by setting $P_{i,j}=215$ whose binary sequence is $\{11010111\}$ and $n=2$. Then ${h_{i,j}} = (\sum\limits_{k = 2}^7 {{c_k} \times {2^k}})/2^n  = 53$ and  ${l_{i,j}} = \sum\limits_{k = 0}^1 {{c_k} \times {2^k}}  = 3$.
To begin the embedding,
first, the HSB plane $I_{H\!S\!B}=\{h_{i,j}\}$ was derived from the decomposition of the original image $I_o$ by Eq. \ref{eq:22} with $h_{i,j}$ values given by Eq. \ref{eq:new}. Then, the scheme paired the pixel values to obtain the difference

\begin{equation}
\label{eq:44}
{d_{i,j}} = {I_{H\!S\!B}}(i,j) - {I_{H\!S\!B}}(i,j - 1),
\end{equation}

where $i \in \left\{ {1,\ldots,h} \right\}$ and $j \in \left\{ {2,\ldots,w} \right\}$. Next, the remaining values in the first column are paired to obtain the difference

\begin{equation}
\label{eq:55}
{d_{i,1}} = {I_{H\!S\!B}}(i,1) - {I_{H\!S\!B}}(i - 1,1),
\end{equation}

where $i \in \left\{ {2,\ldots,h} \right\}$ and $d_{1,1}=I_{H\!S\!B}(1,1)$. 
The peak $a$ and the sub-peak $b$ were obtained from the difference values histogram, and the data were embedded in the $I_{H\!S\!B}(i,j)$ when $d_{i,j}=a$ or $b$. For the case of $a > b$, if the difference value $d_{i,j}>a$, $I_{H\!S\!B}(i,j)$ was shift towards right by $1$ bit except for the boundary value, and similarly $I_{H\!S\!B}(i,j)$ is shifted towards left if $d_{i,j}<b$. Difference values between $a$ and $b$ remain unchanged. Since the HSB plane was highly correlated between adjacent pixels, then the two (sub-)peak values were usually $a=1$ and $b=0$.


\subsection{Kumar et al.'s TLE embedding \cite{KJ20}}
Kumar et al. decomposed the original image $I_o$ into $I_{H\!S\!B}$ and $I_{L\!S\!B}$ planes by Eqs. \ref{eq:33} and \ref{eq:new} with $n=3$, and designed a two-layer embedding (TLE) method on the ${I_{H\!S\!B}}$ plane. They employed the chessboard pattern to obtain two predicted values ${p_1}$ and ${p_2}$ according to the four adjacent pixels of the current ${x_{i,j}}$. In the first-layer of the secret data embedding, the prediction error is ${e_1} = {x_{i,j}} - {p_1}$. When ${e_1}=1$, the pixel value ${x_{i,j}}$ was increased by a secret data bit $s_1 \in \left\{ {0,1} \right\}$. When ${e_1}>1$, it was also increased by $1$ bit to ensure reversibility, otherwise for ${e_1}<1$, the pixel value remained unchanged. In the second-layer of the data embedding, the prediction error is ${e_2} = x_{i,j}' - {p_2}$, where $x_{i,j}'$ was the renewed pixel value after the first-layer of embedding. When ${e_2}=-1$, the pixel was then decreased by a secret data bit $s_2 \in \left\{ {0,1} \right\}$ while decreased by $1$ bit when ${e_2}<-1$, otherwise it remained.

TLE utilized a peak value to make extensions in either embedding layer. It remained the pixels unchanged if embedding two identical bits of data in two layers, which leads to the increase of EC and a reduction of the distortion of image.

\begin{figure}[tb]
	\centering
	\includegraphics[width=10.6cm]{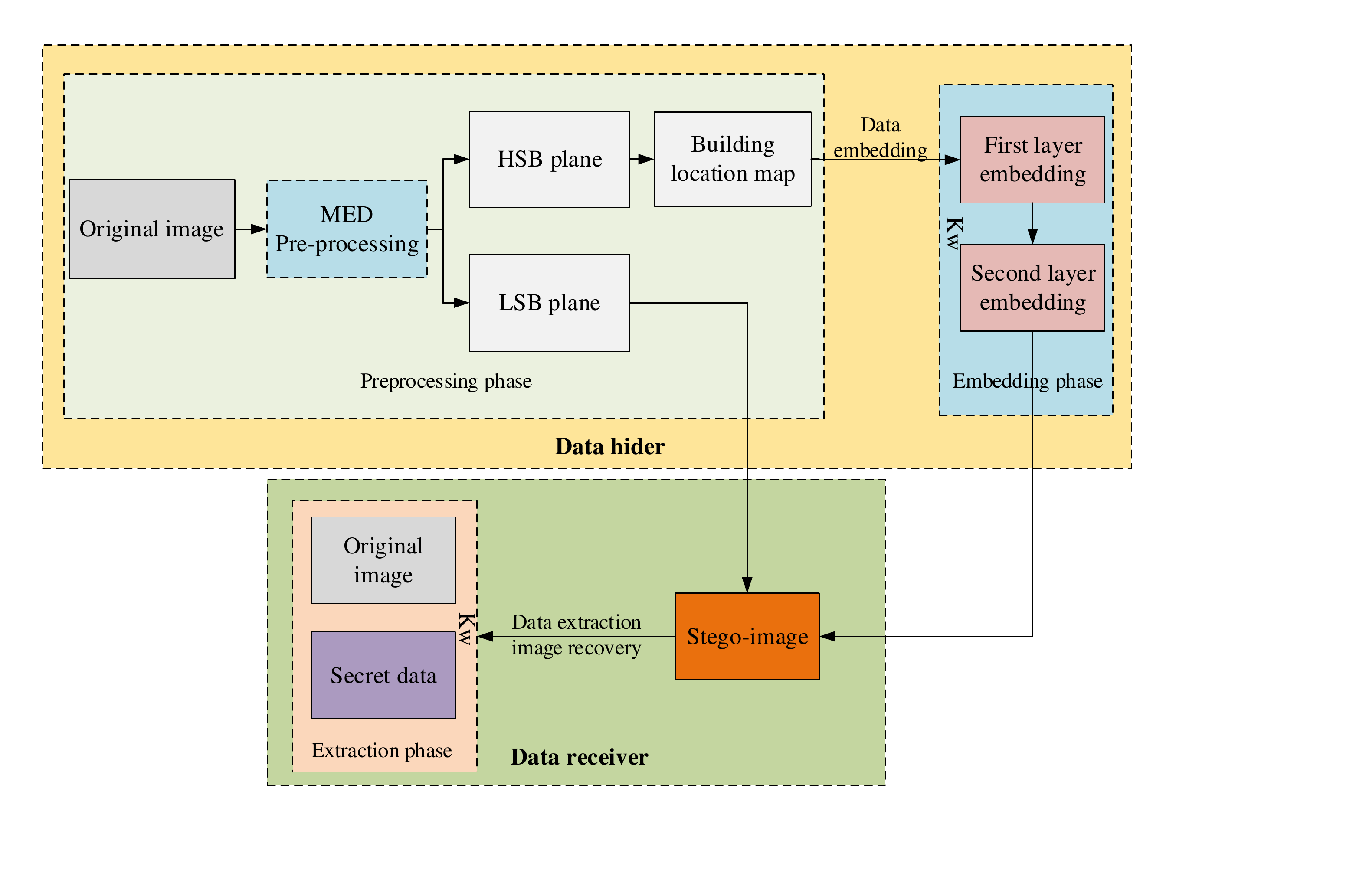}\\
	\caption{Framework of DTLE.}
	\label{fig:fo}
\end{figure}

\section{Proposed scheme}
\label{sec:PS}
In this section we describe DTLE, a reversible data hiding (RDH) approach in terms of double (sub-)peaks in each of two-layer embedding. We first present the DTLE three-phase scheme and then describe each phase in detail as shown in Fig.~\ref{fig:fo}. In the first phase, after being MED pre-processed, each original image of size $h\times w$ is decomposed into the HSB plane ${I_{H\!S\!B}}$ and LSB plane ${I_{L\!S\!B}}$, and the possible overflow pixels are processed to create a location map.
In the second phase, we embed the secret data into the ${I_{H\!S\!B}}$ plane via two-layer embedding and PEE together with the data hiding key ${k_w}$. In the last phase, the receiver extracts the secret data accurately and recovers the original image using ${k_w}$.

\subsection{Image pre-processing}
\label{subsec:IP}
\begin{figure}[tb]
	\centering
	\includegraphics[width=2.5cm]{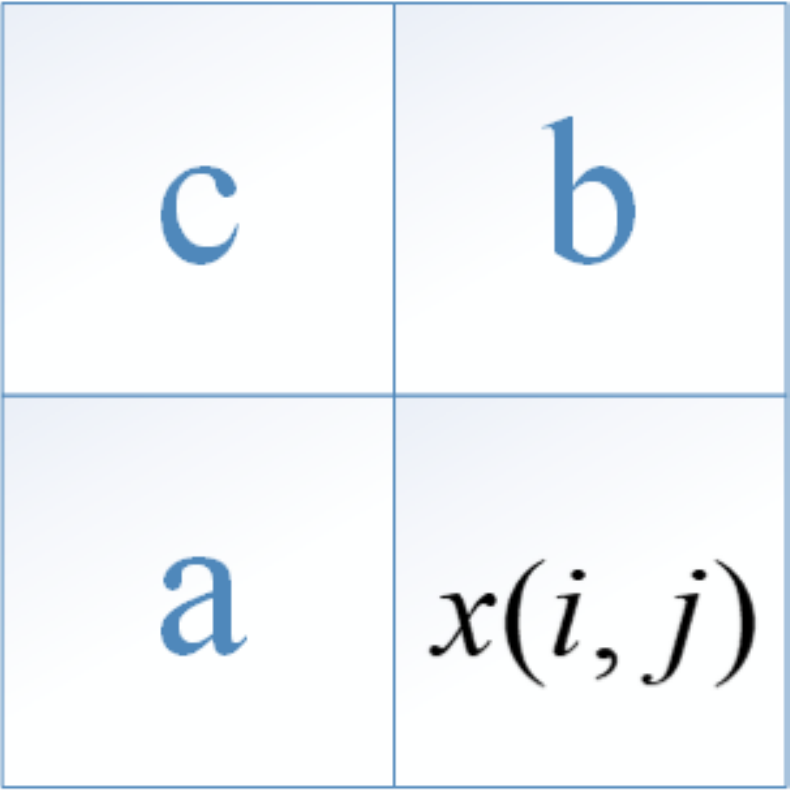}\\
	\caption{Context of the pixel $x(i,j)$ for MED predictor.}
	\label{fig:med}
\end{figure}
\subsubsection{MED processing}
For an original image $I_o$ of size $h \times w$, the pixel values can be predicted by the MED predictor \cite{WSS00} where the first row and column of pixels remain unchanged. Fig. \ref{fig:med} shows the context of a pixel $x{(i,j)}$ and three neighbouring pixels for MED, for which the predicted value ${p_x}(i,j)$ is given by

\begin{equation}
\label{eq:new11}
{p_x}(i,j) = \left\{ {\begin{array}{l}
{\max(a,b), \ \ \text{if} \quad c \le \min(a,b)},\\
{\min(a,b), \quad \text{if} \quad c \ge \max(a,b)},\\
{a + b - c, \quad \text{otherwise}}.
\end{array}} \right.
\end{equation}

\noindent
The difference value $D(i,j)$ between the current pixel $x{(i,j)}$ and the predicted value ${p_x}(i,j)$ is

\begin{equation}
\label{eq:new22}
D(i,j) = x(i,j) - {p_x}(i,j),\quad \quad 2 \le i \le h,\ 2 \le j \le w.
\end{equation}

Since there are usually positive and negative values in the difference image (matrix) $D_I$ from Eq. \ref{eq:new22}, we denote the smallest value of $D(i,j)$ by $e_{\min}$ and express the absolute value in $8$-bits. A histogram shift of $D_I$ gives a new difference image $D_I'$ whose pixel values are all nonnegative as

\begin{equation}
\label{eq:hsmed}
D'(i,j) = D(i,j) + |{e_{\min }}|,\quad \quad 2 \le i \le h,\ 2 \le j \le w.
\end{equation}

While the first row and column are critical to the recovery of the original image $I_o$ after MED pre-processing, they stay unchanged. We use an $1$-bit \emph{flag} to record whether there are any overflow pixels in the image $D_I'$. If the pixel is greater than or equal to $255$, we set the pixel value to $255$, record the excess with $8$ bits and take the \emph{flag} as $1$, otherwise, $0$. All the above auxiliary information is recorded as $\mathcal{A}$.

\subsubsection{Decomposition of bit planes}
The new difference image $D_I'$ is decomposed into the ${I_{H\!S\!B}}$ and ${I_{L\!S\!B}}$ planes by Eqs. \ref{eq:33} and \ref{eq:new}. In our scenario, since the maximum changes of ${I_{H\!S\!B}}$ for pixels is $\pm 2$ in data embedding, the image pixels need processing accordingly. We select the higher six-bit planes as the HSB plane which will be explained in Section \ref{sec:result}. We denote the pixel values of plane ${I_{H\!S\!B}}$ as $I_H$. The maximum value of ${I_{H\!S\!B}}$ is $\max  = {2^6} - 1 = 63$ while the minimum value $\min = 0$ can be calculated by setting $n = 2$ in Eq. \ref{eq:new}.

\subsubsection{Building location map} We modify pixel values and create a location map $LM$ to mark possible overflow pixels,

\begin{equation}
\label{eq:66}
{I_{H}}(i,j)=\left\{ \begin{array}{l}
{I_{H}}(i,j) - 2,\quad \text{if}\quad\;{I_{H}}(i,j) = 63,\\
{I_{H}}(i,j) + 2,\quad \text{if}\quad\;{I_{H}}(i,j) = 0,\\
{I_{H}}(i,j) - 1,\quad \text{if}\quad\;{I_{H}}(i,j) = 62,\\
{I_{H}}(i,j) + 1,\quad \text{if}\quad\;{I_{H}}(i,j) = 1,\\
{I_{H}}(i,j),\quad \quad \;\;\,\text{otherwise}.
\end{array} \right.
\end{equation}

\begin{itemize}
\item In the cases, ${I_{H}}(i,j) = 63$ or $0$, set $LM(i,j) = 1$;
\item In the cases, ${I_{H}}(i,j) = 62$ or $1$, set $LM(i,j) = 2$;
\item Otherwise, set $LM(i,j) = 0$.
\end{itemize}

Note that the location map $LM$ is defined in three cases whose values are represented in the decimal base rather than the binary system. $C_{L\!M}$ is obtained by lossless compression that uses arithmetic coding with a compressed length $L_{C\!L\!M}$.

\begin{figure}[tb]
    \begin{center}
        \subfigure[Histogram of original image $I_o$.]{\label{fig:NOMED}
        \includegraphics[width=0.465\linewidth]{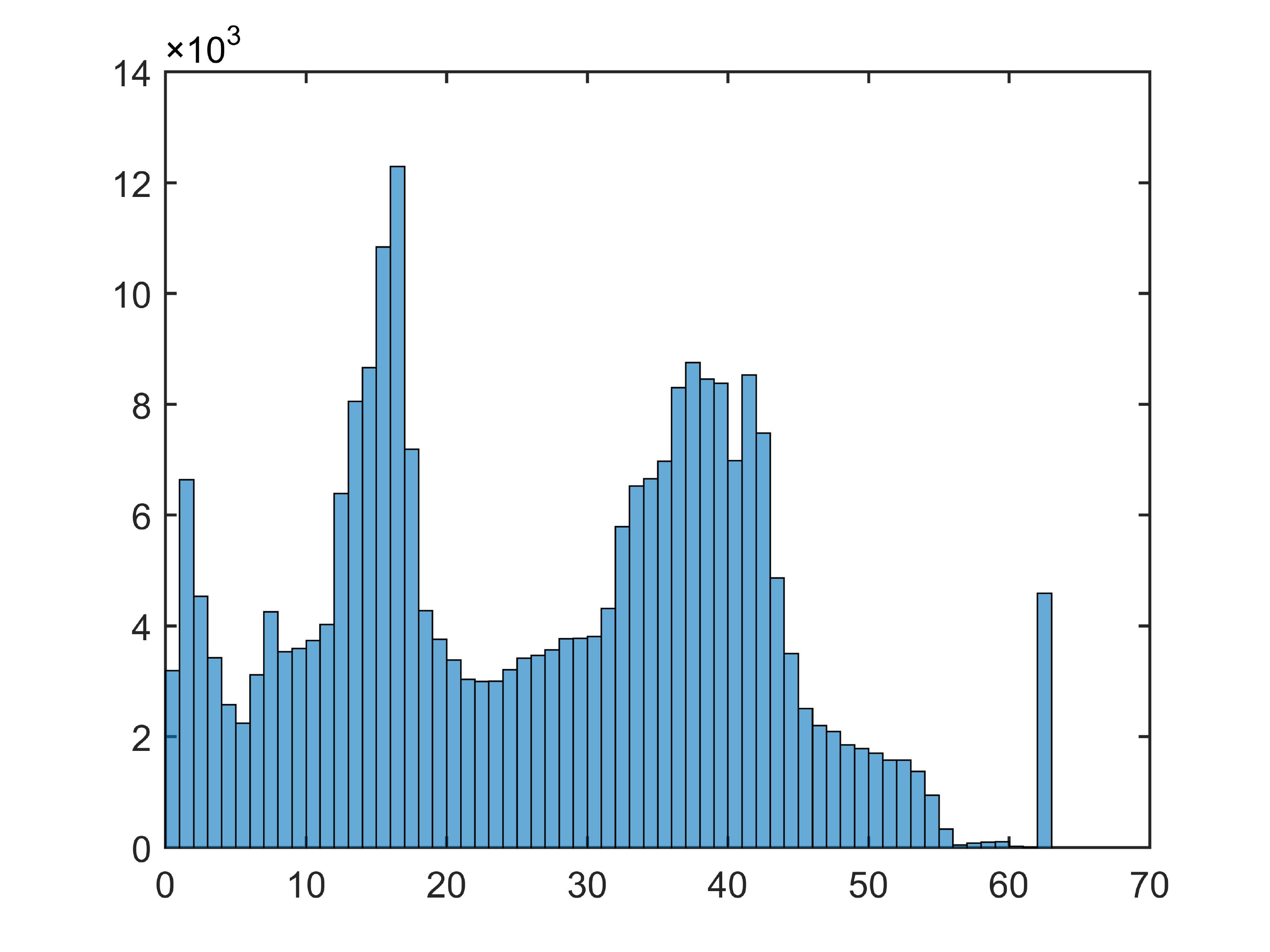}
        }
    \hfill
        \subfigure[Histogram of difference image $D_I'$.]{\label{fig:MED}
        \includegraphics[width=0.465\linewidth]{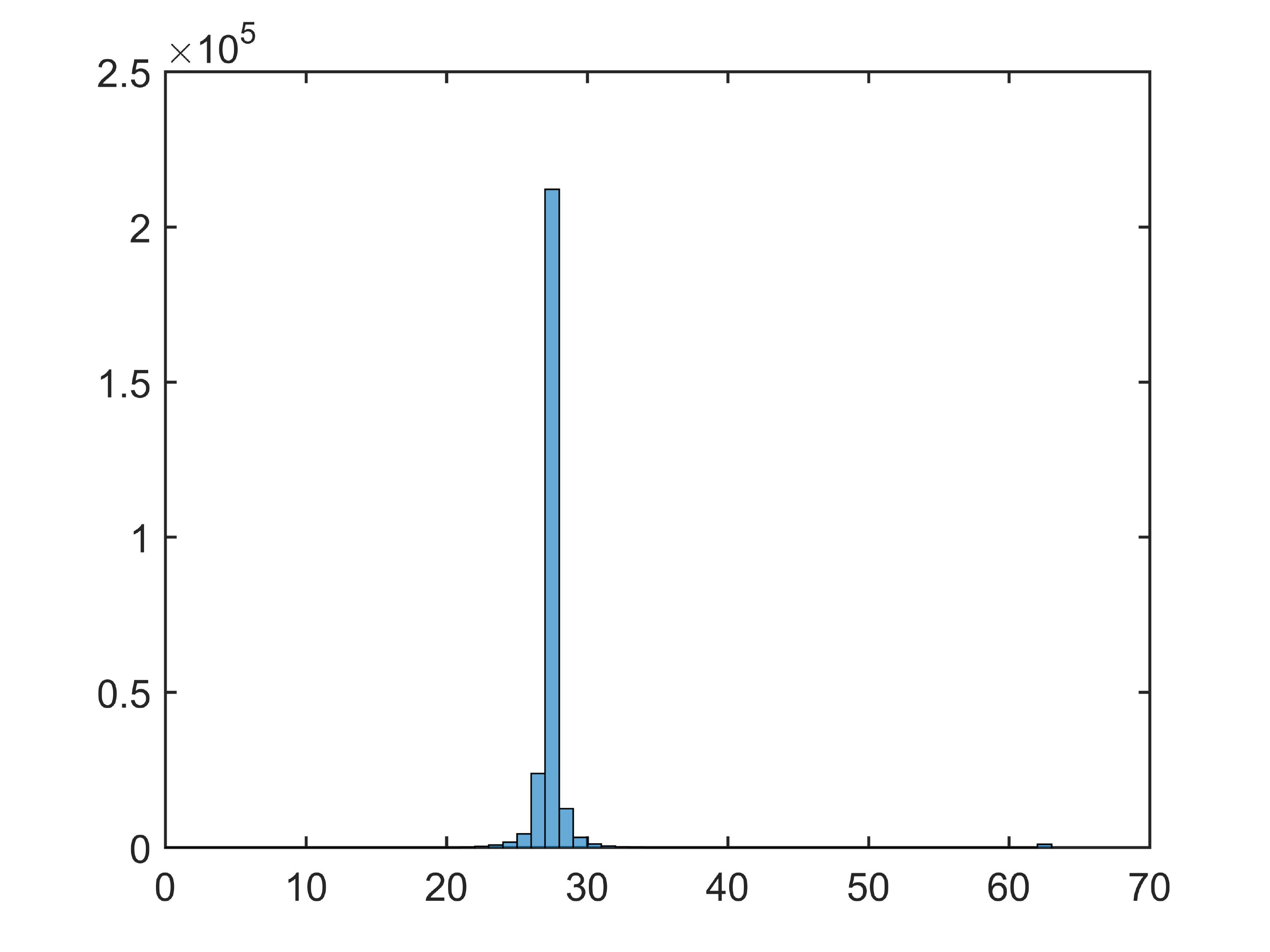}
        }
    \end{center}
    \caption{HSB plane histogram of image $Peppers$.}\label{fig:hsbp}
\end{figure}

As shown in Fig~\ref{fig:hsbp}, the HSB plane of original image has a large number of possible overflow pixels that need to be marked by the location map $LM$. In the image $D_I'$ produced by MED pre-processing, the pixels are more concentrated in the middle of the histogram and the number of possible overflow pixels is significantly reduced.

\begin{figure}[tb]
    \hspace{0.08cm}
	\begin{minipage}[t]{0.468\linewidth}
		\centering
	    \includegraphics[width=3.3cm]{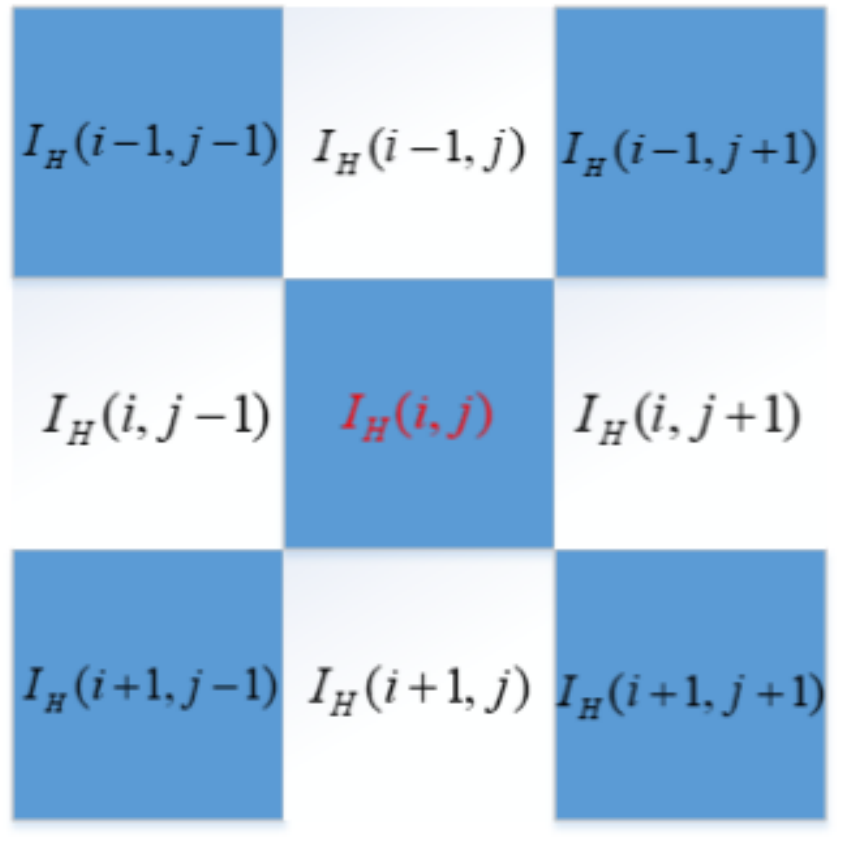}\\
	    \caption{Prediction pattern.}
	    \label{fig:pm}
	\end{minipage}\hspace{0.3cm}
    \begin{minipage}[t]{0.468\linewidth}
        \centering
	    \includegraphics[width=3.3cm]{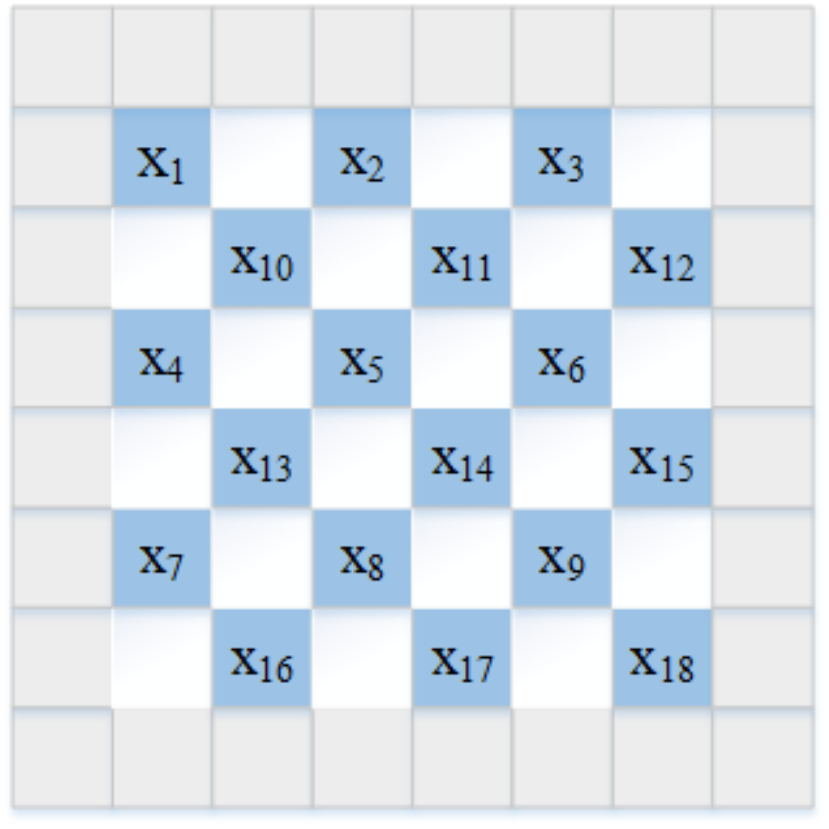}\\
	    \caption{Chessboard representation of image.}
	    \label{fig:cp}
    \end{minipage}
\end{figure}

\subsection{Data embedding}
\label{subsec:DE}
This section introduces the embedding process of the DTLE in detail and makes a comparative analysis with TLE \cite{KJ20}.
\subsubsection{Calculation of predicted values}
The predictor pair, $p_1$ and $p_2$, are obtained for embedding by eight neighboring pixel values on each pixel ${I_{H}(i,j)}$, as shown in Fig.~\ref{fig:pm}. 
The sequence in ascending order is obtained by sorting the neighbors and denoted by $\left( {z_1,z_2,\ldots,z_7,z_8} \right)$, then we define

\begin{equation}
\label{eq:77}
{p_1} = \left\lfloor {\frac{{{z_1} + {z_2} + {z_3} + {z_4} + {z_5} + {z_6}}}{6}} \right\rfloor,
\end{equation}

\begin{equation}
\label{eq:88}
{p_2} = \left\lfloor {\frac{{{z_3} + {z_4} + {z_5} + {z_6} + {z_7} + {z_8}}}{6}} \right\rfloor,
\end{equation}

\noindent
where $\left\lfloor \cdot \right\rfloor $ denotes the floor operator that gives the largest integer no more than the given value.

\subsubsection{Processing two-layer embeddings}
As shown in Fig.~\ref{fig:cp}, the HSB plane is scanned in the chessboard method that starts from left to right and top to bottom for secret data embedding. The blue-colored pixels are first used for embedding followed by white ones.
To be specific, the blue-colored pixels are divided into two groups that consisting of all the odd rows and even rows, respectively. The secret data are first embedded in odd rows (ordered $x_1,\ldots,x_9$) and secondly in even rows (ordered $x_{10},\ldots,x_{18}$). The  white-colored pixels are processed in the same manner, while the first and last row/columns remain unchanged to store auxiliary information like the location map.

\begin{figure}[tb]
    \begin{center}
        \subfigure[First-layer difference histogram.]{\label{fig:Fl}
        \includegraphics[width=0.46\linewidth]{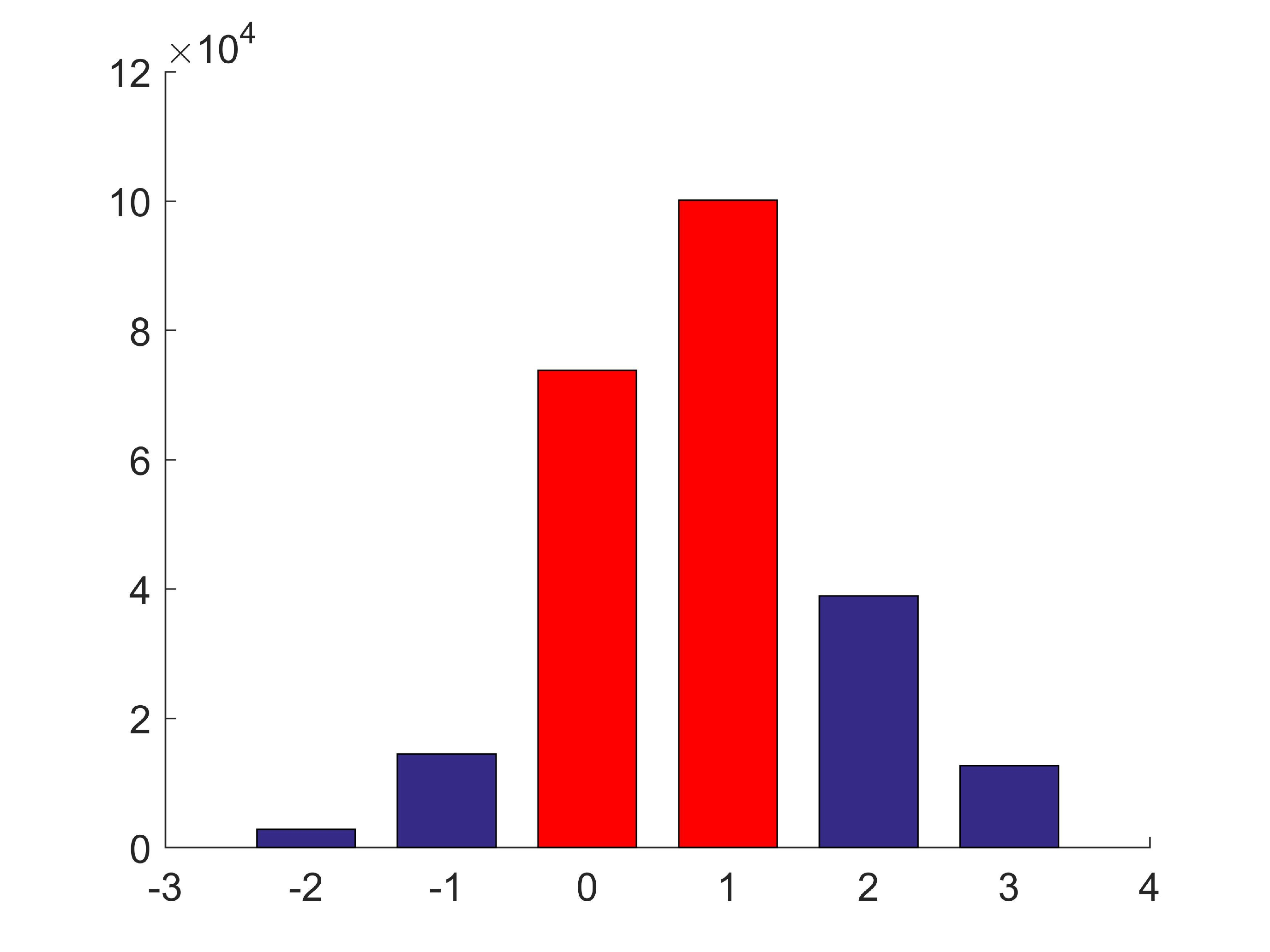}
        }
    \hfill
        \subfigure[Second-layer difference histogram.]{\label{fig:Sl}
        \includegraphics[width=0.47\linewidth]{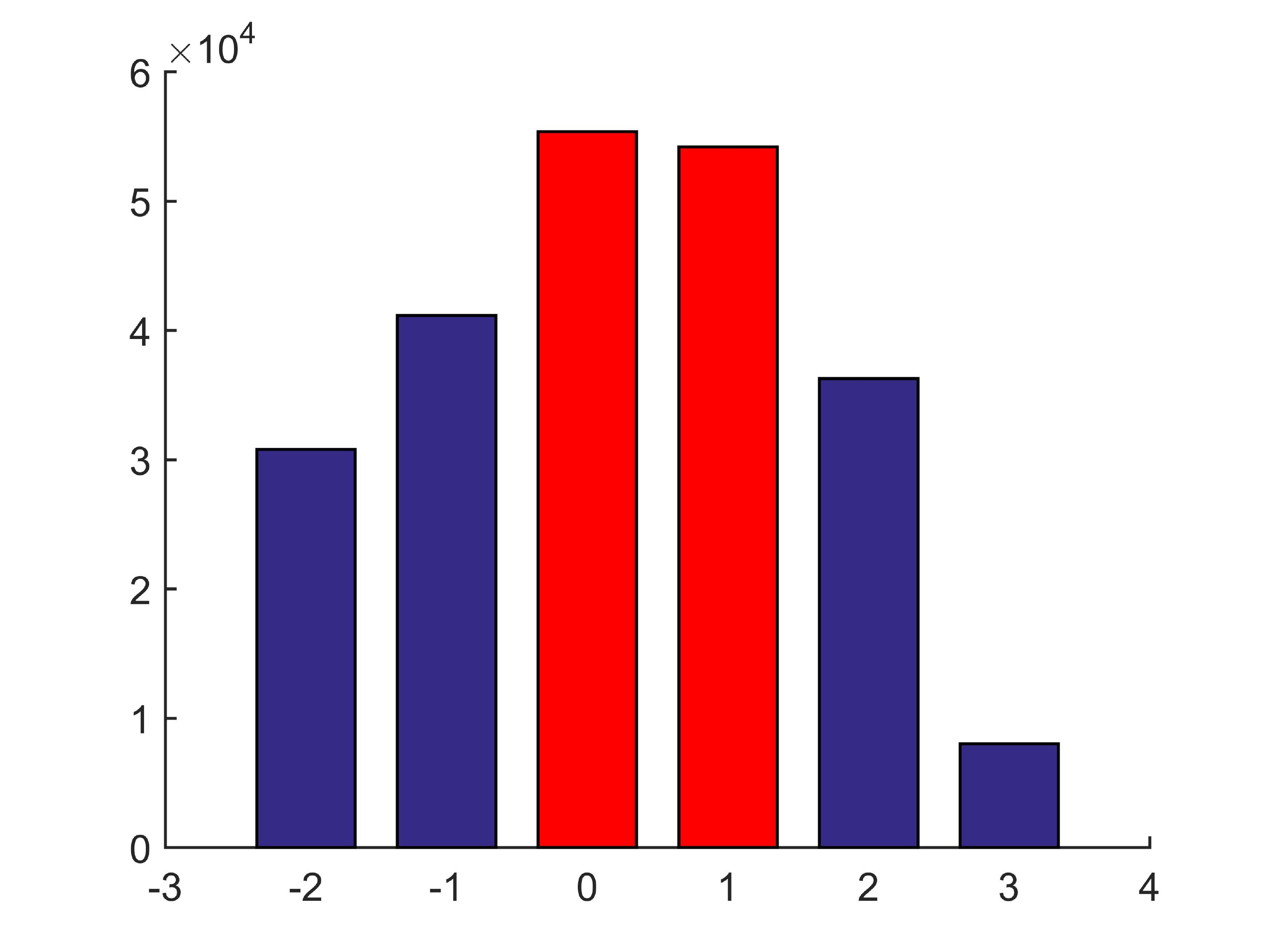}
        }
    \end{center}
    \caption{Prediction error histogram on test image Lena.}\label{fig:Peh}
\end{figure}

As shown in Fig.~\ref{fig:Peh}, the histograms of prediction error statistics emphasize the peak and sub-peak values, $1$ and $0$, in two layers. This inspires us to embed data also on the pixel values with the sub-peak error.
The two-layer embeddings are performed on each pixel by the pair ($p_1$, $p_2$). In the first layer, the prediction error is ${e_1} = {I_{H}}(i,j) - {p_1}$, 
and the secret data bit ${b_1} \in \left\{ {0,1} \right\}$ is embedded with renewing the pixel value,

\begin{equation}
\label{eq:99}
I{'_{H}}\left( {i,j} \right) = \left\{ \begin{array}{l}
{I_{H}}\left( {i,j} \right) + {b_1},\;\;\;\,\,\;\,\,\,\text{if}\;\;{e_1} = 1,\\
{I_{H}}\left( {i,j} \right) + 1,\;\,\quad \,\;\,\,{\kern 1pt} \text{if}\;\;{e_1} > 1,\\
{I_{H}}\left( {i,j} \right) - {b_1},\quad \quad \text{if}\;\;{e_1} = 0,\\
{I_{H}}\left( {i,j} \right) - 1,\quad \quad \,\,\text{if}\;\;{e_1} < 0.
\end{array} \right.
\end{equation}

Similarly in the second-layer embedding, the prediction error is ${e_2} = I{'_{H}}(i,j) - {p_2}$
while the secret data bit ${b_2} \in \left\{ {0,1} \right\}$ is embedded with renewing

\begin{equation}
\label{eq:1010}
I{_{H}''}\left( {i,j} \right) = \left\{ \begin{array}{l}
I{_{H}'}\left( {i,j} \right) + {b_2},\;\;\;\;\,\;\,\,\text{if}\,\,\;{\kern 1pt} {\kern 1pt} {e_2} = 1,\\
I{_{H}'}\left( {i,j} \right) + 1,\;\;\,\quad \,\;{\kern 1pt} \text{if}\;\;\;{e_2} > 1,\\
I{_{H}'}\left( {i,j} \right) - {b_2},\quad \;\,\;\,\text{if}\;\;\;{e_2} = 0,\\
I{_{H}'}\left( {i,j} \right) - 1,\quad \quad \,\,\text{if}\;\;\;{e_2} < 0.
\end{array} \right.
\end{equation}


\begin{figure*}[tb]
	\centering
	\includegraphics[width=18cm]{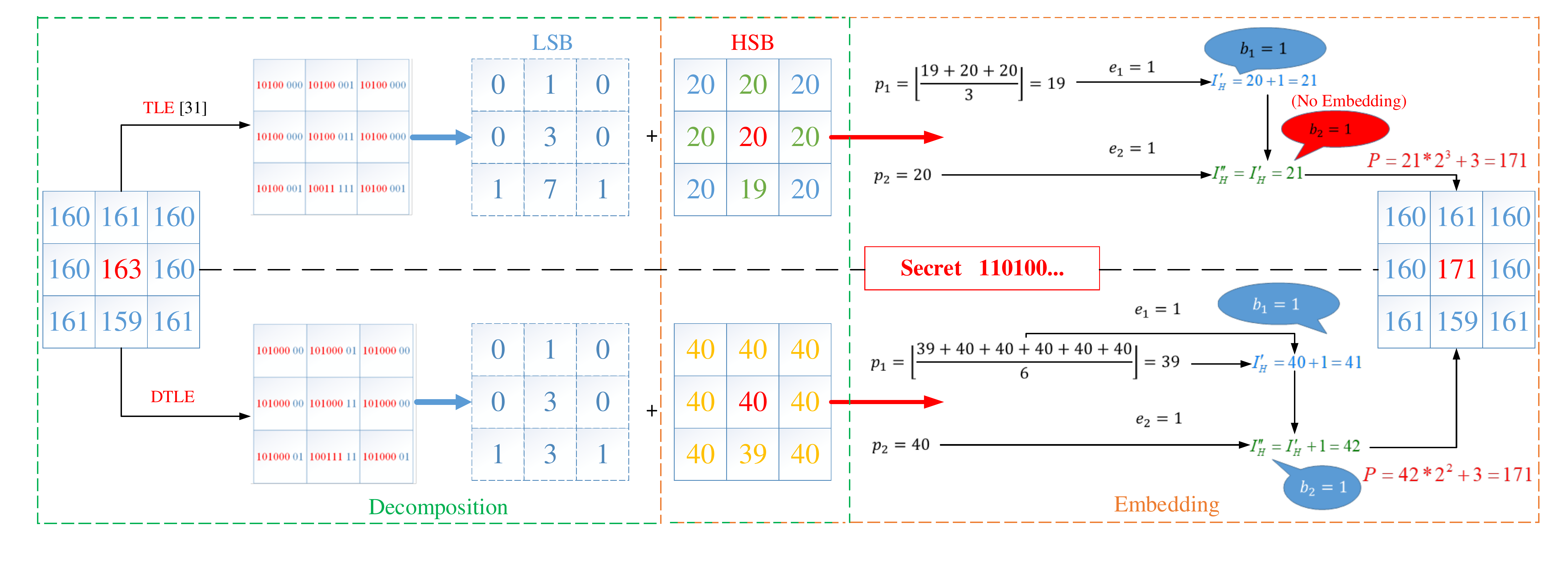}\\
	\caption{Comparative example of TLE and DTLE embeddings.}
	\label{fig:Eo}
\end{figure*}

The embedding steps can be presented as follows.

\begin{enumerate}[\textbf {Step} 1:]
    \item  Perform MED predictor and histogram shift on original image $I_o$ to obtain image $D_I'$ and auxiliary information $\mathcal{A}$;
    \item  Decompose the $D_I'$ into the planes ${H\!S\!B}$ and ${L\!S\!B}$ with $n = 2$, by Eqs. \ref{eq:33} and \ref{eq:new};
    \item  Process the ${I_{H\!S\!B}}$ plane using Eq. \ref{eq:66} to obtain the location map $LM$. Compress the $LM$ using arithmetic coding to obtain the compressed location map $C_{L\!M}$ with length $L_{C\!L\!M}$;
    \item  Extract the predefined sequence of $L\!S\!B\!s$ 
    and add it to the end of secret data $\mathcal{S}$ along with $C_{L\!M}$ and $\mathcal{A}$;
    \item  Embed secret data for blue-colored pixels firstly and white-colored pixels secondly, by Eqs. \ref{eq:99} and \ref{eq:1010};
    \item  Replace the $L\!S\!B\!s$ sequence in the first and last rows/columns by the position information $K_{\text{end}}$ of the last secret data-carrying pixel and the length of the compressed location map $L_{C\!L\!M}$ in order;
    \item  Build the stego-image $I_e$ by combining the plane ${I''_{H\!S\!B}}$ and the plane ${I_{L\!S\!B}}$.
\end{enumerate}

\subsubsection{Embedding example}
\label{subsubsec:Aa}
This part takes an embedding example to illustrate our DTLE scheme with comparisons to TLE by Kumar et al. \cite{KJ20}, as shown in Fig.~\ref{fig:Eo}.
We consider to embed the secret $(b_1b_2)=(11)$ on the central pixel with value $163$ by both schemes.


Firstly, the HSB and LSB planes are obtained by the decomposition of each pixel. The first six binary numbers in red stand for the HSB plane in DTLE while the first five binary numbers in TLE, and the later blue numbers represent the respective LSB plane. The TLE method takes the higher five-bit planes as the embedding plane, while DTLE selects the higher six-bit planes.

Secondly in the embedding process, TLE utilizes the neighbouring pixels from four directions in HSB and one peak value in either layer. Averaging the smallest three values of neighbouring pixels gives ${p_1}$ while the largest three values present ${p_2}$. In the first-layer embedding, ${e_1} = {I_{H}}(i,j) - {p_1} = 1$, and the value of ${I'_{H}}(i,j)$ is $21$ due to the embedding of ${b_1} = 1$. However, in the second-layer embedding, since ${e_2} = {I'_{H}}(i,j) - {p_2} = 1$ does not allow the embedding, the pixel remain unchanged and ${I''_{H}}(i,j)=21$. The HSB and LSB planes are combined to obtain a pixel value of $171$, and the pixel value is changed with difference $171 - 163 = 8$. As for our DTLE, $p_1$ and $p_2$ are calculated by Eqs. \ref{eq:77} and \ref{eq:88}, respectively, with $6$ neighbors. In the first-layer embedding, ${e_1}= 1$, and ${I'_{H}}(i,j)$ are calculated by Eq. \ref{eq:99} with embedding data ${b_1} = 1$. In the second-layer embedding,  ${e_2}= 1$ and ${I''_{H}}(i,j)$ are calculated by Eq. \ref{eq:1010} with allowing the embedding of ${b_2} = 1$. Afterwards, the pixel value is changed also with difference $171 - 163 = 8$.

Given the above example, the TLE method modifies the pixel with difference ${2^3} = 8$ per bit after embedding in the higher five-bit planes while DTLE embeds in the higher six-bit planes with pixel difference ${2^2} = 4$ indeed. This leads to less distortion and higher quality of stego-image for ETLE. Moreover, TLE allows embedding only one bit of secret data while our proposed DTLE embeds two bits. Similar examples take place often in embeddings.

On the contrary, there are a small minority of cases in complex images that
DTLE (without MED) can not directly perform embedding in some layer while TLE achieves embedding.
For instance, assume that a (central) pixel can be embedded by both schemes due to its normal distributed neighbors.
However, if its corner neighbors change greatly on pixel values, DTLE (without MED) would fail to perform embedding while TLE succeeds since the four direct neighbors remain unchanged. To solve such issues,
the MED pre-processing helps greatly avoid the curse, which will be illustrated in experimental Section \ref{sec:result}.


\subsection{Data extraction and image recovery}
\begin{figure}[tb]
	\centering
	\includegraphics[width=15.6cm]{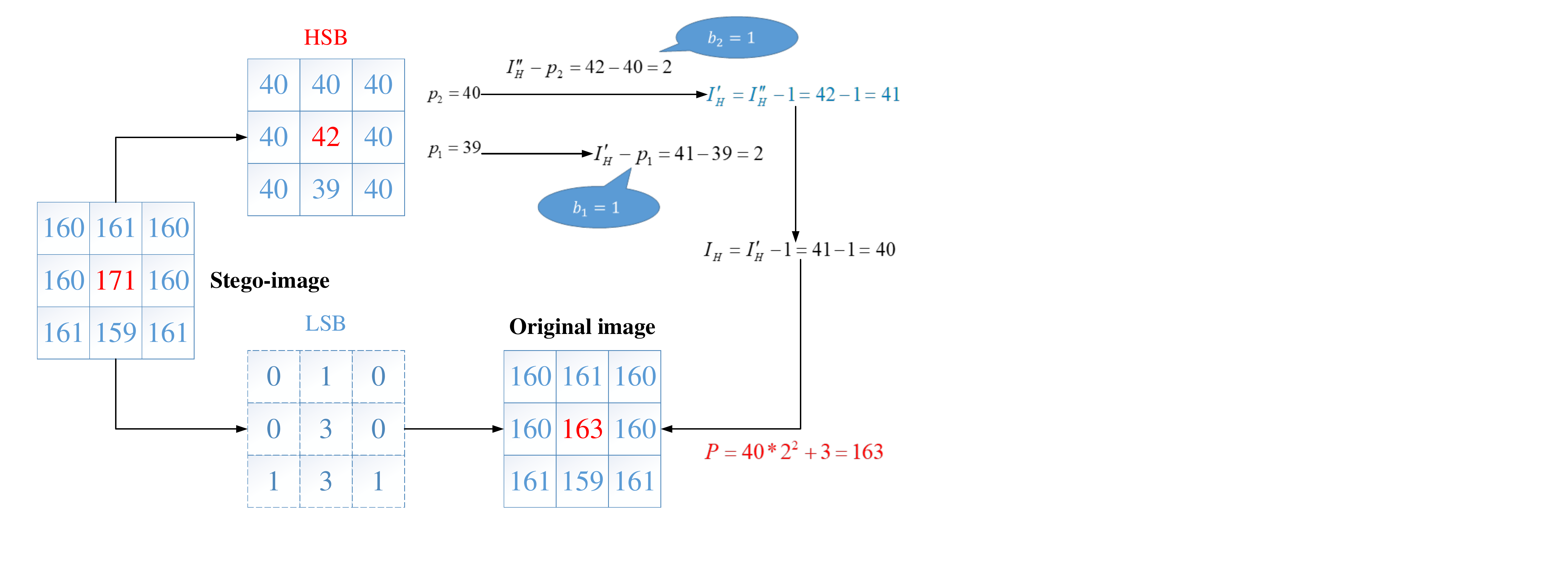}\\
	\caption{Example of extraction process for DTLE.}
	\label{fig:Txa}
\end{figure}

This section first describes the reverse process of DTLE embedding phase as shown in Fig.~\ref{fig:Txa} that is followed to extract the hidden secret data and recover the original image. Here, $I_{H}''$ is the current pixel at the receiver's side, $I_{H}'$ is the pixel after first-layer reverse, $I_{H}$ is the pixel after second-layer reverse, and $p_1$ and $p_2$ are the predictor pair of the $I''$ in the embedded image by Eqs. \ref{eq:77} and \ref{eq:88}, respectively.

\begin{itemize}
    \item First-layer reverse for obtaining $I_H'$:
    \begin{itemize}
        \item[-] if $e_2=I_{H}''-{p_2} \in \{-1,2\}$, we can extract the secret data $b_2=1$. The value $I_{H}''$ decreases by $1$ when $e_2=2$ and increases by 1 when $e_2=-1$;
        \item[-] if $e_2=I_{H}''-{p_2} \in \{0,1\}$, we can extract the secret data $b_2=0$ and keep the value of $I_{H}''$ unchanged;
        \item[-] if $e_2=I_{H}''-{p_2} <-1$ or $I_{H}''-{p_2}>2$, there is no secret data hidden and the value of $I_{H}''$ is increased by $1$ or decreased by $1$, correspondingly.
    \end{itemize}
    \item Second-layer reverse for obtaining $I_H$:
    \begin{itemize}
        \item[-] if $e_1=I_{H}' - {p_1} \in \{  - 1,2\}$, we can extract the secret data $b_1=1$. The value of $I_{H}'$ decreases by 1 when $e_1=2$ and increases by 1 when $e_1=-1$;
        \item[-] if $e_1=I_{H}' - {p_1} \in \{  0,1\}$, we can extract the secret data $b_1=0$ and keep the value of $I_{H}'$ unchanged;
        \item[-] if $e_1=I_{H}' - {p_1} <  - 1$ or $I_{H}' - {p_1} >  2$, there is no secret data hidden and the value of $I_{H}'$ is increased by 1 or decreased by 1.
    \end{itemize}
\end{itemize}

Along with the reverse process of DTLE embedding, the receiver extracts the secret data from the stego-image $I_e$ and recovers the original image $I_o$. The steps of extraction and recovery are detailed as follows.
\begin{enumerate}[\textbf {Step} 1:]
\item Decompose stego-image $I_e$ into ${I''_{H\!S\!B}}$ and ${I_{L\!S\!B}}$;
\item Extract the final secret data-carrying pixel $K_{\text{end}}$ from the ${I''_{H\!S\!B}}$ and the compressed location map length $L_{C\!L\!M}$;
\item Extract all secret data from the white-colored pixels in reverse order of embedding;
\item Extract all secret data from the blue-colored pixels in reverse order of embedding;
\item Recover the ${I_{H\!S\!B}}$ plane with the extracted $L\!S\!B\!s$ sequence and decompress the location map $C_{L\!M}$ to obtain the location map $LM$;
\item Recover the ${I_{H\!S\!B}}$ plane using the location map $LM$ and shifts by Eq. \ref{eq:1111};
\item Combine ${I_{H\!S\!B}}$ and ${I_{L\!S\!B}}$ planes to obtain the $D_I'$;
\item The values of \emph{flag}, overflow pixels and $e_{\min}$ are obtained from the auxiliary information $\mathcal{A}$. The original image $I_o$ is recovered by the reverse of MED predictor.
\end{enumerate}
\begin{equation}
\label{eq:1111}
{I_{H}}(i,j) = \left\{ \begin{array}{l}
{I_{H}}(i,j) + 2,\quad \text{if}\;{I_{H}}(i,j) = 61 \\ \quad\quad\quad\quad\quad\quad\quad  \& \;LM(h) = 1,\\
{I_{H}}(i,j) - 2,\quad \text{if}\;{I_{H}}(i,j) = 2 \\ \quad\quad\quad\quad\quad\quad\quad  \& \;LM(h) = 1,\\
{I_{H}}(i,j) + 1,\quad \text{if}\;{I_{H}}(i,j) = 61 \\ \quad\quad\quad\quad\quad\quad\quad  \& \;LM(h) = 2,\\
{I_{H}}(i,j) - 1,\quad \text{if}\;{I_{H}}(i,j) = 2 \\ \quad\quad\quad\quad\quad\quad\quad  \& \;LM(h) = 2,\\
{I_{H}}(i,j),\quad \quad \;\;\;\text{otherwise}.
\end{array} \right.
\end{equation}

Note that during the extraction, when the sequence of LSBs is extracted, the first row of pixels in the ${I''_{H\!S\!B}}$ plane should be recovered and restored to perform the subsequent data extraction.

The extraction process of secret data is illustrated in Fig.~\ref{fig:Txa} by taking an example. The stego-image is first decomposed into the HSB and LSB planes. Then, the predictor pair $(p_1, p_2)$ is obtained by Eqs. \ref{eq:77} and \ref{eq:88}. The first-layer reverse gives ${e_2} = I_{H}'' - {p_2} = 2$. The secret data ${b_2}=1$ is extracted and the pixel value $I_{H}''$ is subtracted by $1$ to get $I_{H}'=41$. Afterwards, the second-layer reverse provides ${e_1} = I_{H}' - {p_1} = 2$. The secret data $b_1=1$ is extracted and the pixel value $I_{H}'$  is subtracted by $1$ to obtain $I_{H}=40$. Finally, the HSB plane $I_{H\!S\!B}$ is combined with the LSB plane to recover the original image $I_o$ with the reverse MED
predictor, while the secret data $({b_1} {b_2})=(11)$ are obtained.

\begin{table}[tb]
	\scriptsize
	\renewcommand{\arraystretch}{2.0}
	\centering
	\caption{Performance evaluation on different HSB planes.}
	\begin{tabular}{|c|c|c|c|c|}
    \hline
	\multirow{2}*{Test Images} & \multicolumn{2}{c|}{HSB ($n=3$)} & \multicolumn{2}{c|}{HSB $(n=2)$} \\
    \cline{2-5}
            & EC & PSNR & EC & PSNR \\
    \hline
	   $Lena$ & 340858 & 26.9105 & 264722 & 32.6351 \\
	\hline
       $Airplane$ & 380623 & 26.9893 & 332846 & 32.7203 \\
    \hline
       $Baboon$ & 191798 & 26.0179 & 121807 & 31.8554 \\
    \hline
       $Barbara$ & 380362 & 27.2762 & 337482 & 33.4667 \\
    \hline
       $Jetplane$ & 354665 & 27.0466 & 296815 & 32.8690 \\
    \hline
       $Boat$ & 313821 & 26.7775 & 242686 & 32.5074 \\
    \hline
       $Peppers$ & 268988 & 26.6848 & 278654 & 33.2095 \\
    \hline
       $Lake$ & 284372 & 26.6523 & 203341 & 32.3660 \\
    \hline
	\end{tabular}
	\label{tab:Pe}
\end{table}

\begin{table*}[htbp]
	\scriptsize
	\renewcommand{\arraystretch}{2.0}
	\centering
	\caption{Comparisons among different schemes performed on the test images.}
	\begin{tabular}{|c|c|c|c|c|c|c|}
    \hline
	\multirow{2}*{Images} & \multicolumn{3}{c|}{EC} & \multicolumn{3}{c|}{PSNR} \\
    \cline{2-7}
            & TLE \cite{KJ20} & DTLE-NoMED & DTLE & TLE \cite{KJ20} & DTLE-NoMED & DTLE \\
    \hline
	   $Lena$ & 213929 & 264722 & 223149 & 32.2635 & 32.6351 & 32.0282 \\
	\hline
       $Airplane$ & 230103 & 332846 & 291862 & 32.6108 & 32.7203 & 32.5418 \\
    \hline
       $Baboon$ & 116984 & 121807 & 102567 & 31.0517 & 31.8554 & 31.6002 \\
    \hline
       $Barbara$ & 245114 & 337482 & 379875 & 32.7483 & 33.4667 & 33.0256 \\
    \hline
       $Jetplane$ & 221322 & 296815 & 275308 & 32.4029 & 32.8690 & 32.3903 \\
    \hline
       $Boat$ & 202976 & 242686 & 234400 & 32.0961 & 32.5074 & 32.1335 \\
    \hline
       $Peppers$ & 208225 & 278654 & 407640 & 32.7738 & 33.2095 & 33.1824 \\
    \hline
       $Lake$ & 164698 & 203341 & 152301 & 31.6238 & 32.3660 & 31.7026 \\
    \hline
	\end{tabular}
	\label{tab:Coe}
\end{table*}

\begin{figure*}[tb]
    \begin{minipage}[t]{0.09\textwidth}
        \centering
        \subfigure[][$Lena$]{\includegraphics[width=2.2cm]{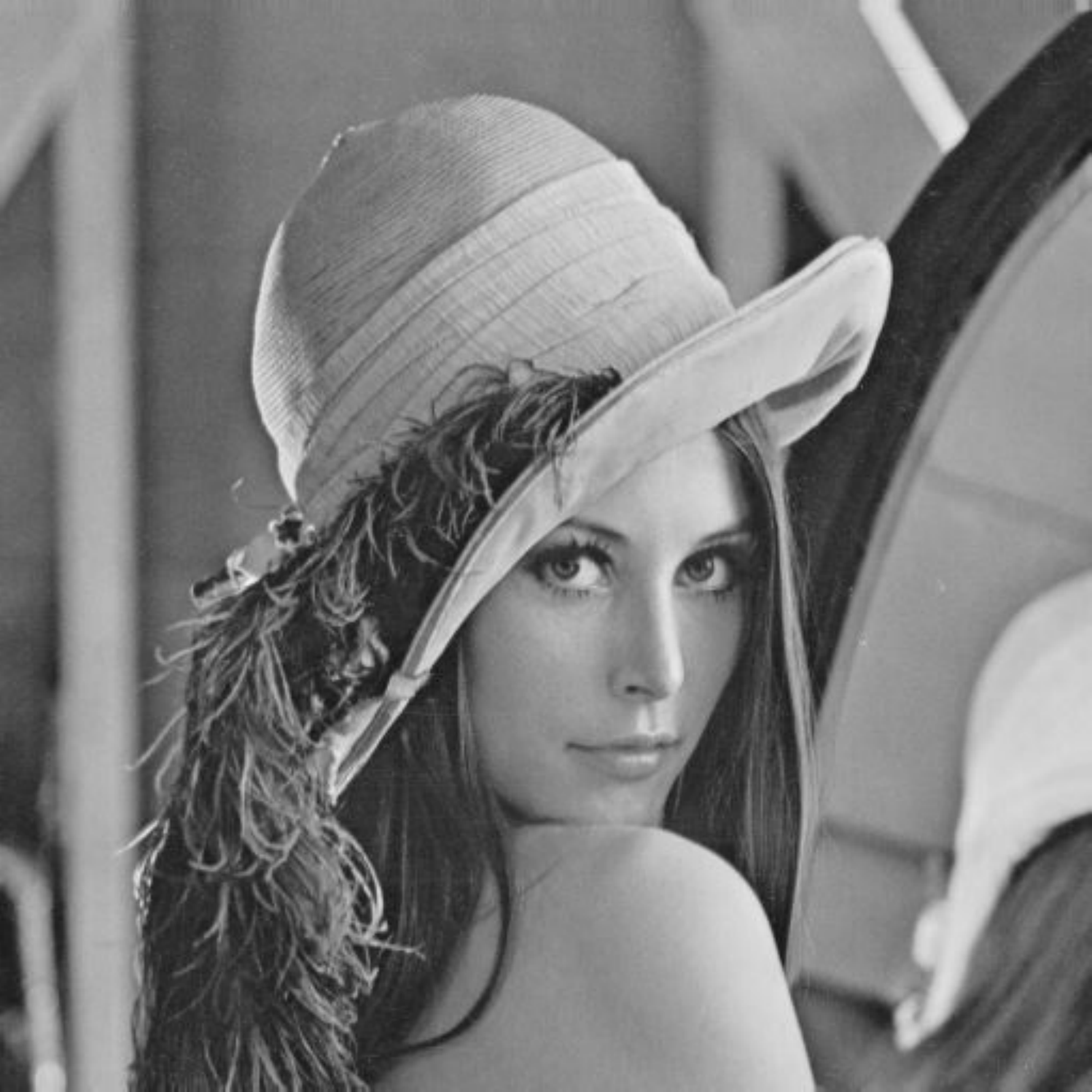}}
    \end{minipage}\hspace{0.5cm}
    \begin{minipage}[t]{0.09\textwidth}
        \centering
        \subfigure[][$Airplane$]{\includegraphics[width=2.2cm]{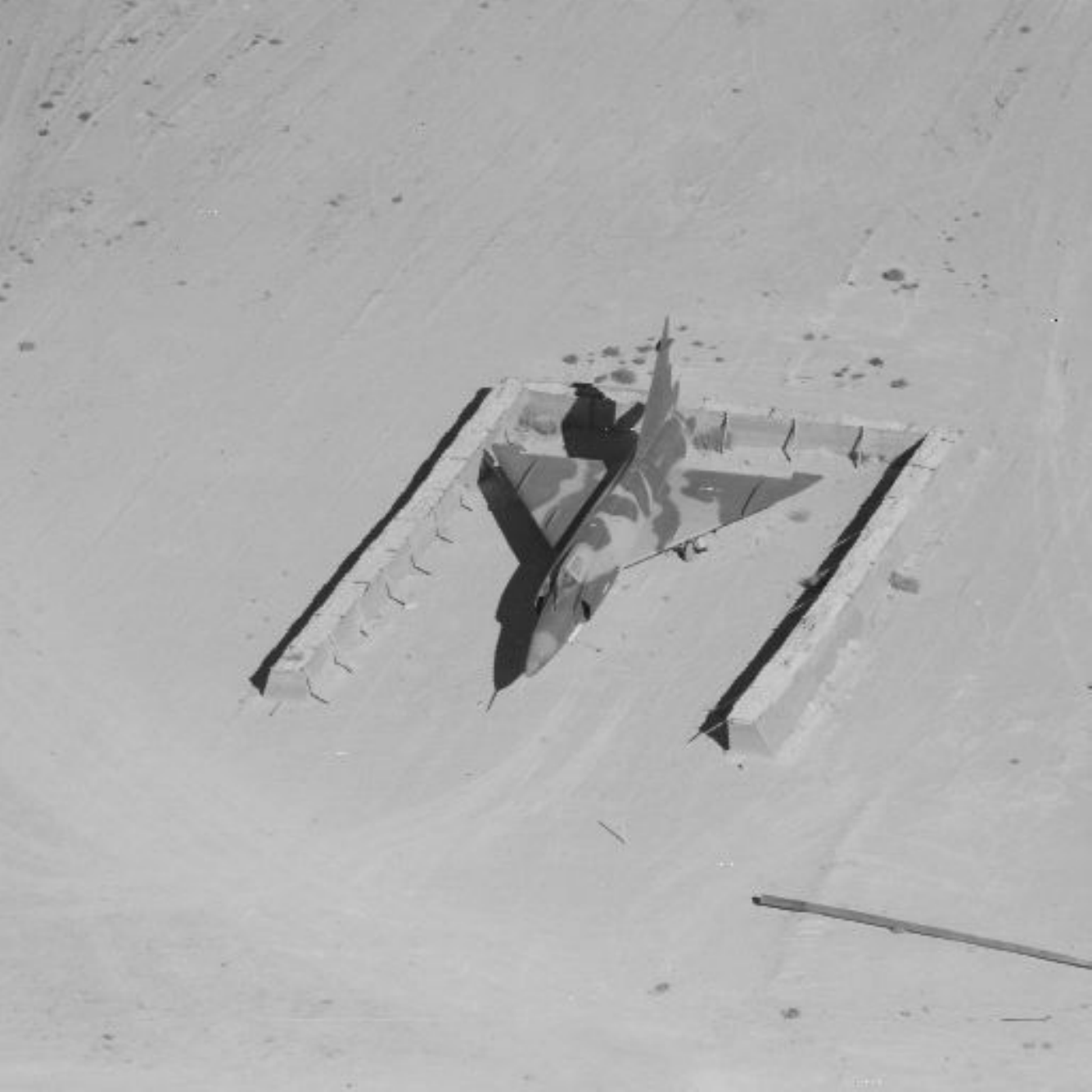}}
    \end{minipage}\hspace{0.5cm}
    \begin{minipage}[t]{0.09\textwidth}
        \centering
        \subfigure[][$Baboon$]{\includegraphics[width=2.2cm]{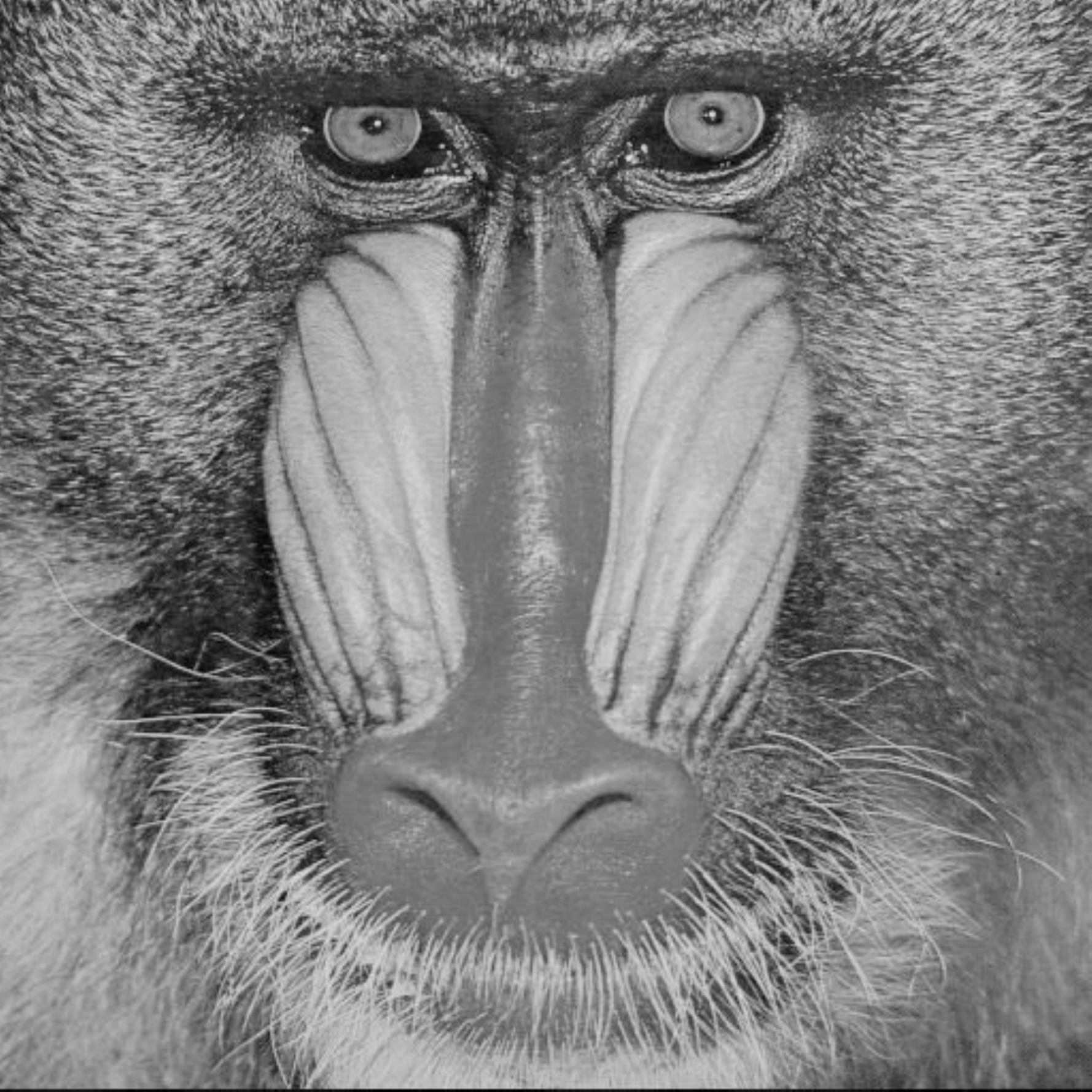}}
    \end{minipage}\hspace{0.5cm}
    \begin{minipage}[t]{0.09\textwidth}
        \centering
        \subfigure[][$Barbara$]{\includegraphics[width=2.2cm]{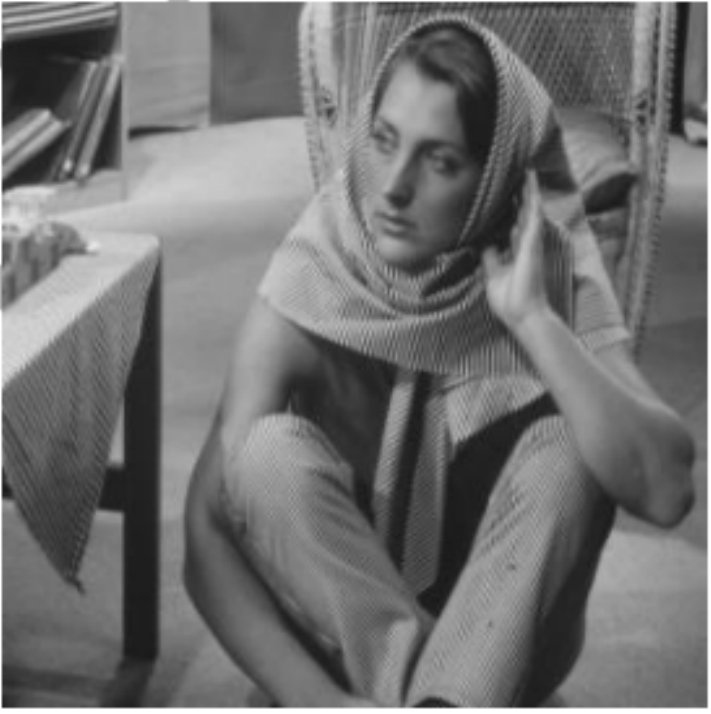}}
    \end{minipage}\hspace{0.5cm}
    \begin{minipage}[t]{0.09\textwidth}
        \centering
        \subfigure[][$Jetplane$]{\includegraphics[width=2.2cm]{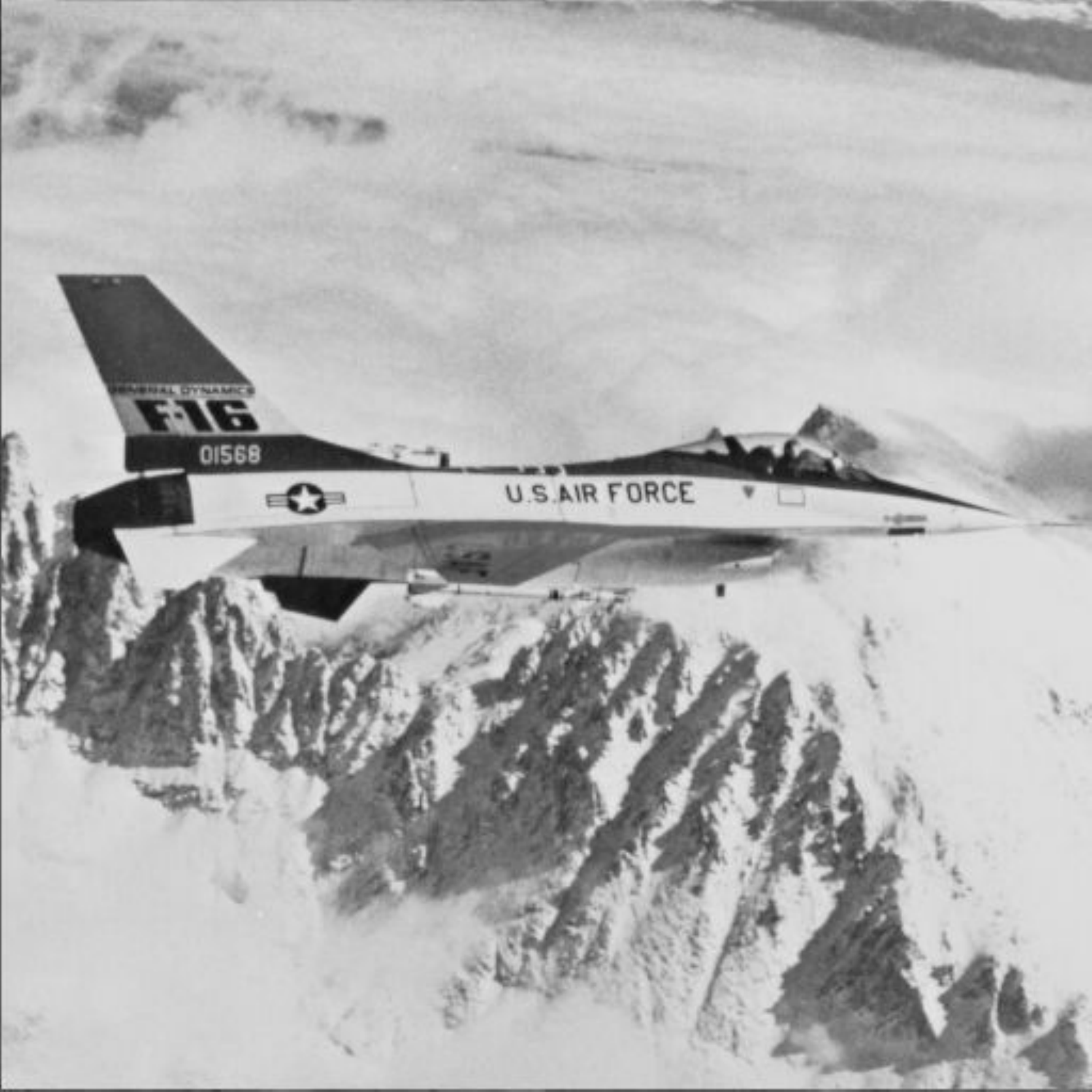}}
    \end{minipage}\hspace{0.5cm}
    \begin{minipage}[t]{0.09\textwidth}
        \centering
        \subfigure[][$Boat$]{\includegraphics[width=2.2cm]{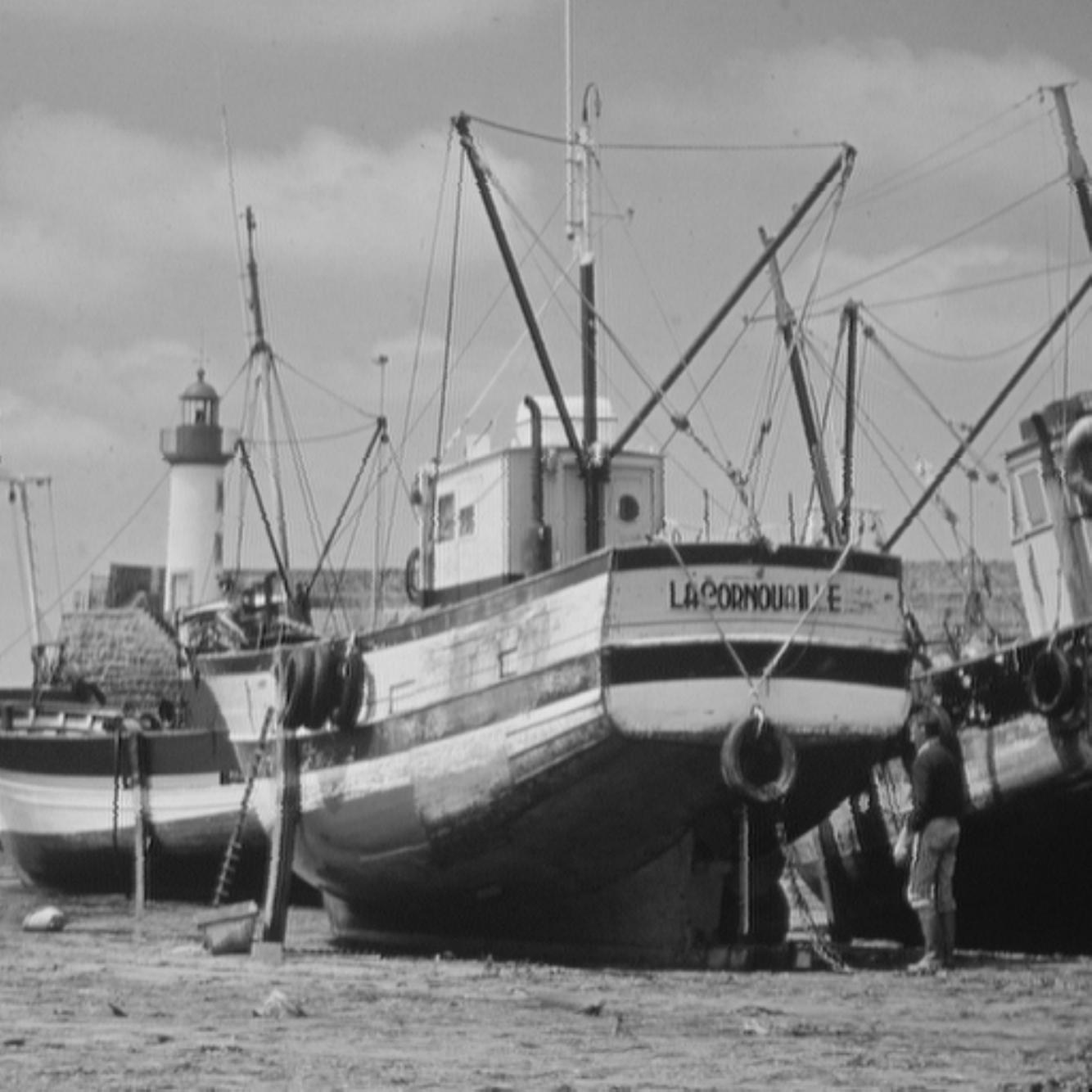}}
    \end{minipage}\hspace{0.5cm}
    \begin{minipage}[t]{0.09\textwidth}
        \centering
        \subfigure[][$Peppers$]{\includegraphics[width=2.2cm]{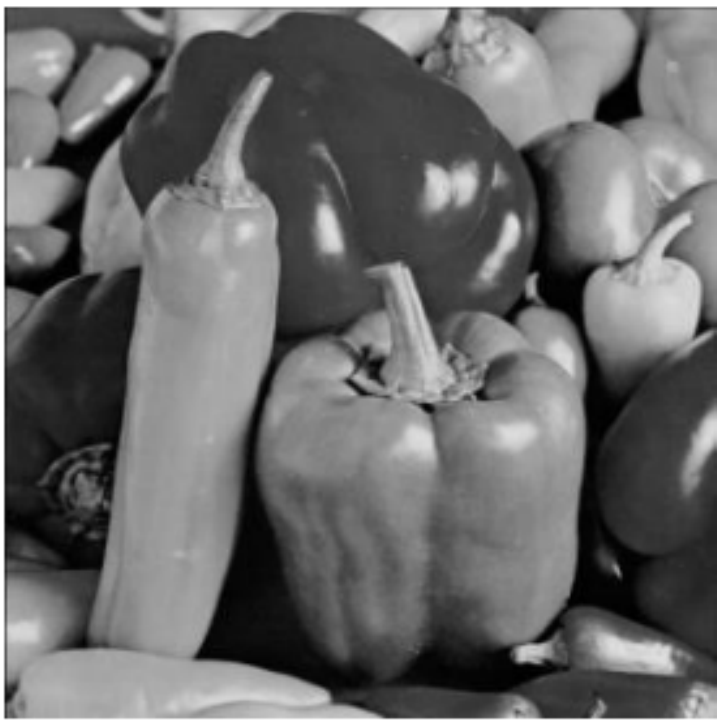}}
    \end{minipage}\hspace{0.5cm}
    \begin{minipage}[t]{0.09\textwidth}
        \centering
        \subfigure[][$Lake$]{\includegraphics[width=2.2cm]{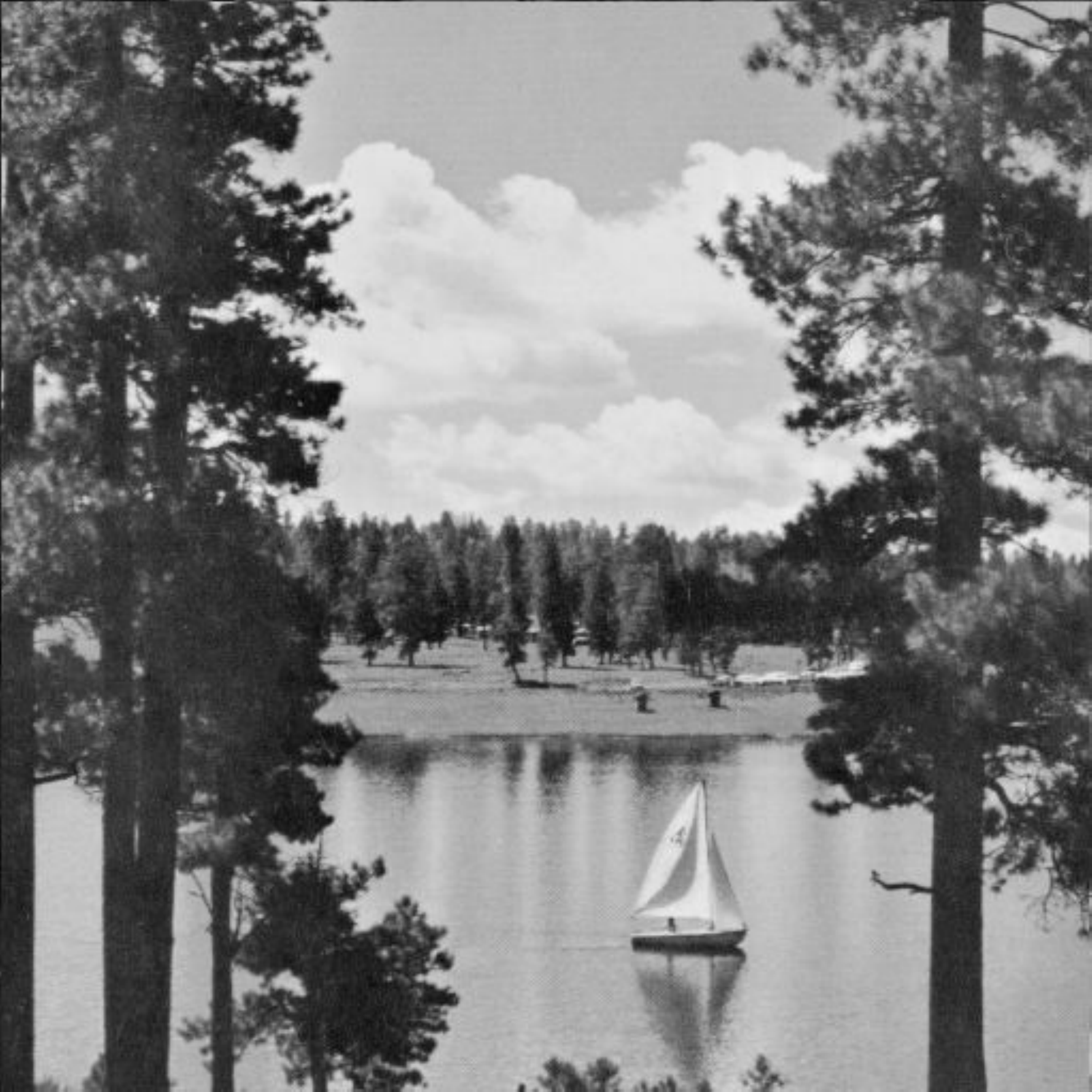}}
    \end{minipage}\hspace{0.5cm}
    \caption{Eight test images.}
    \label{fig:Et}
\end{figure*}

\begin{figure*}[tb]
    \begin{minipage}[t]{0.09\textwidth}
        \centering
        \subfigure[][$Lena$]{\includegraphics[width=2.2cm]{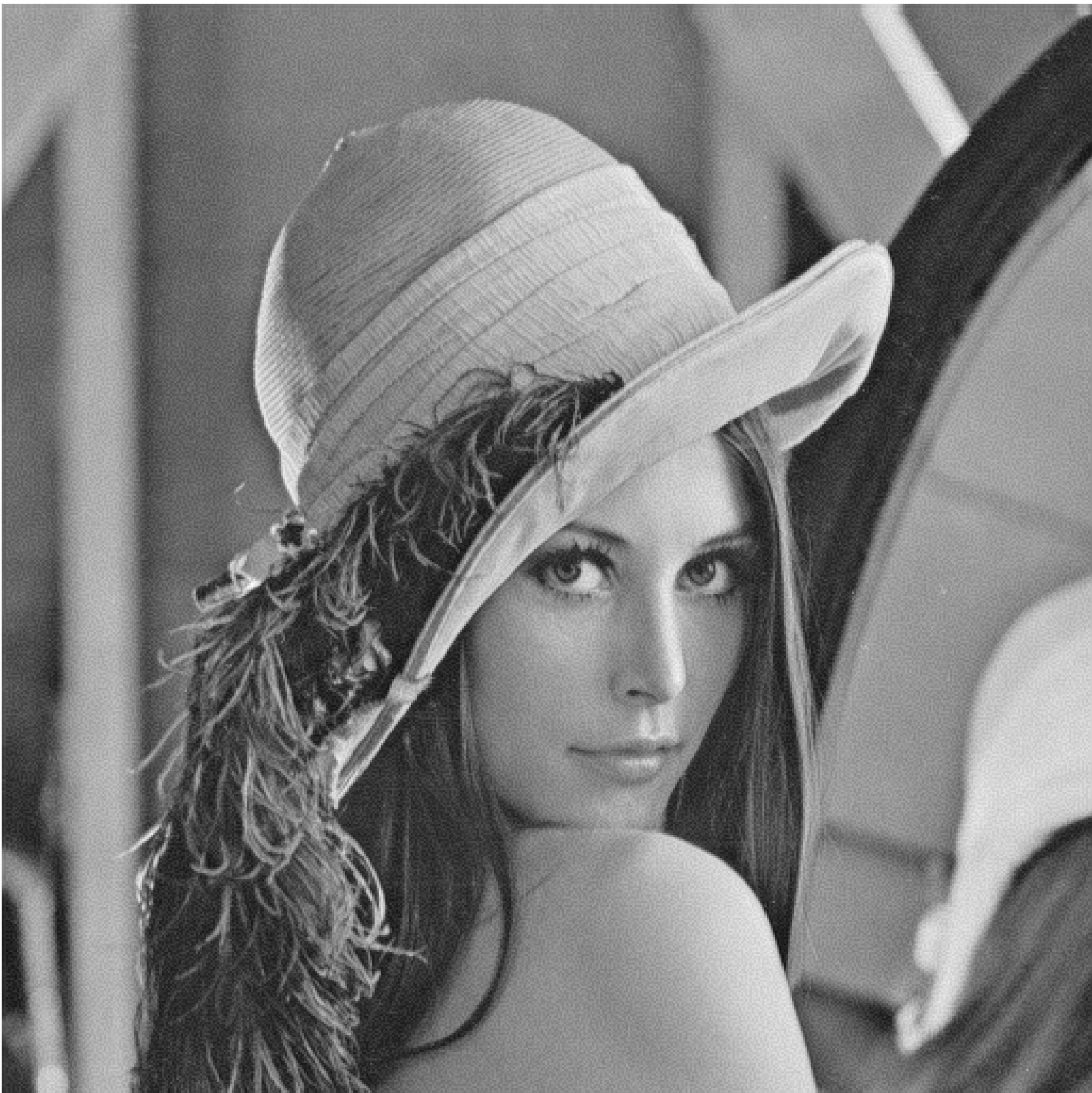}}
    \end{minipage}\hspace{0.5cm}
    \begin{minipage}[t]{0.09\textwidth}
        \centering
        \subfigure[][$Airplane$]{\includegraphics[width=2.2cm]{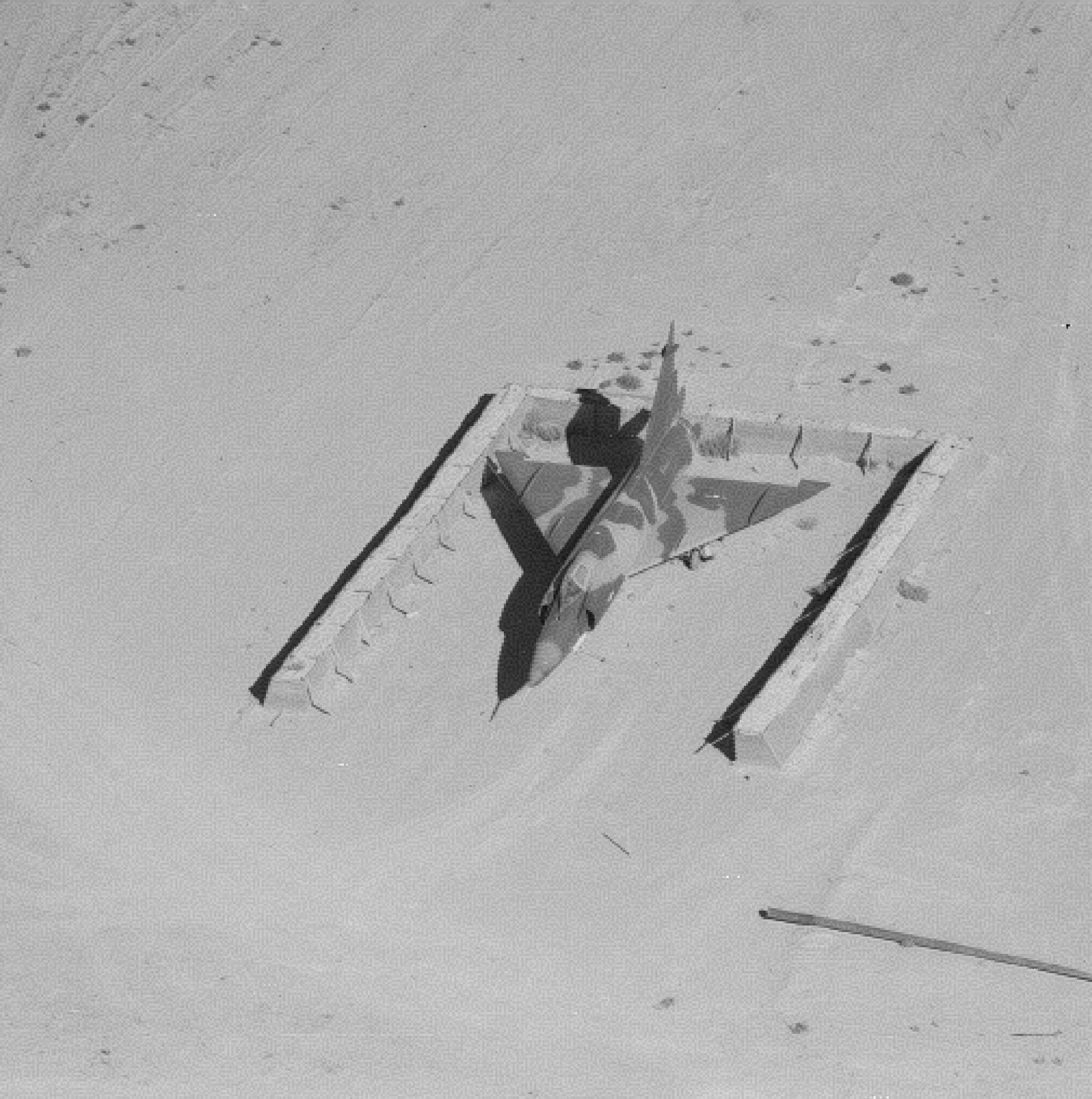}}
    \end{minipage}\hspace{0.5cm}
    \begin{minipage}[t]{0.09\textwidth}
        \centering
        \subfigure[][$Baboon$]{\includegraphics[width=2.2cm]{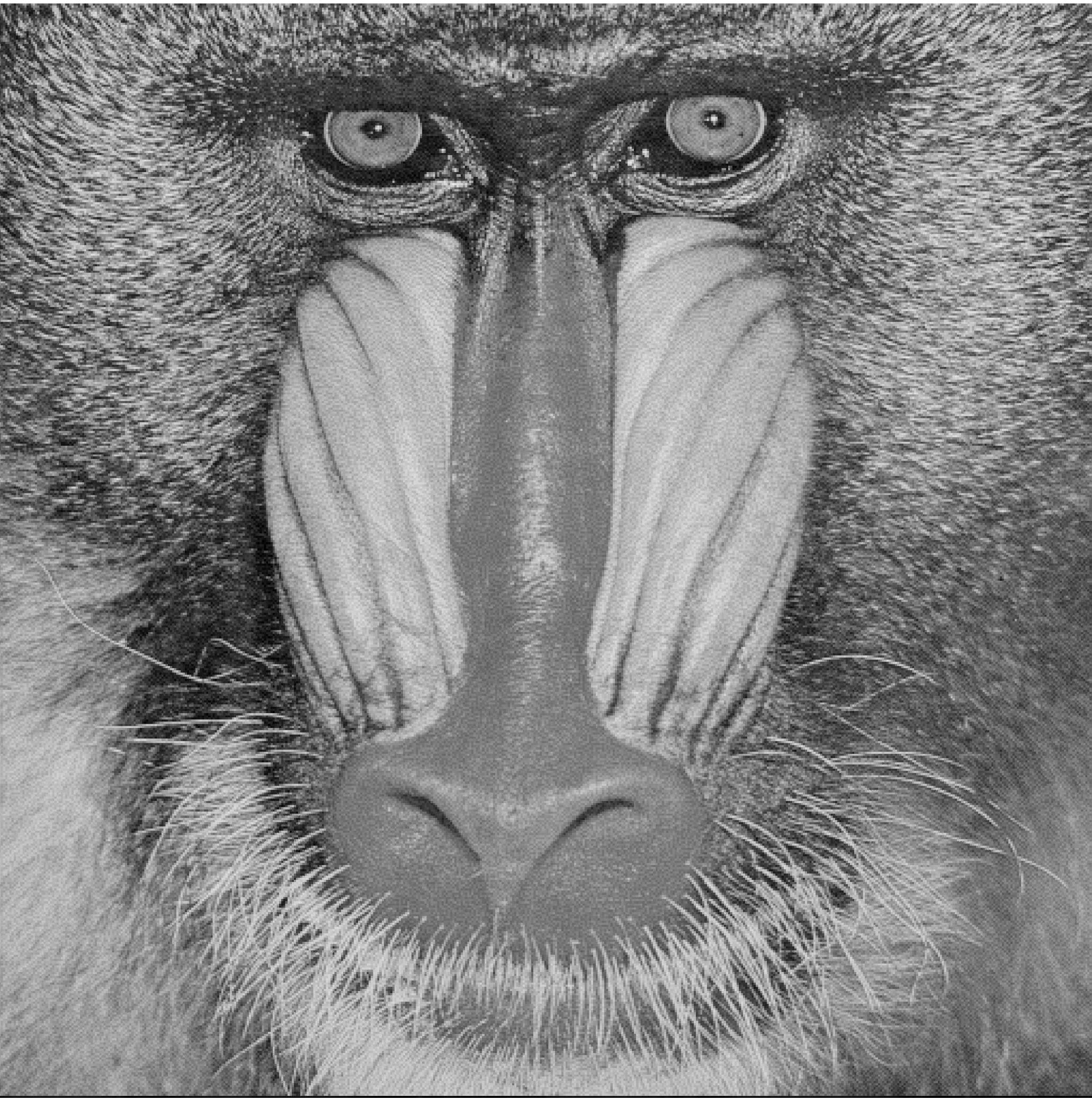}}
    \end{minipage}\hspace{0.5cm}
    \begin{minipage}[t]{0.09\textwidth}
        \centering
        \subfigure[][$Barbara$]{\includegraphics[width=2.2cm]{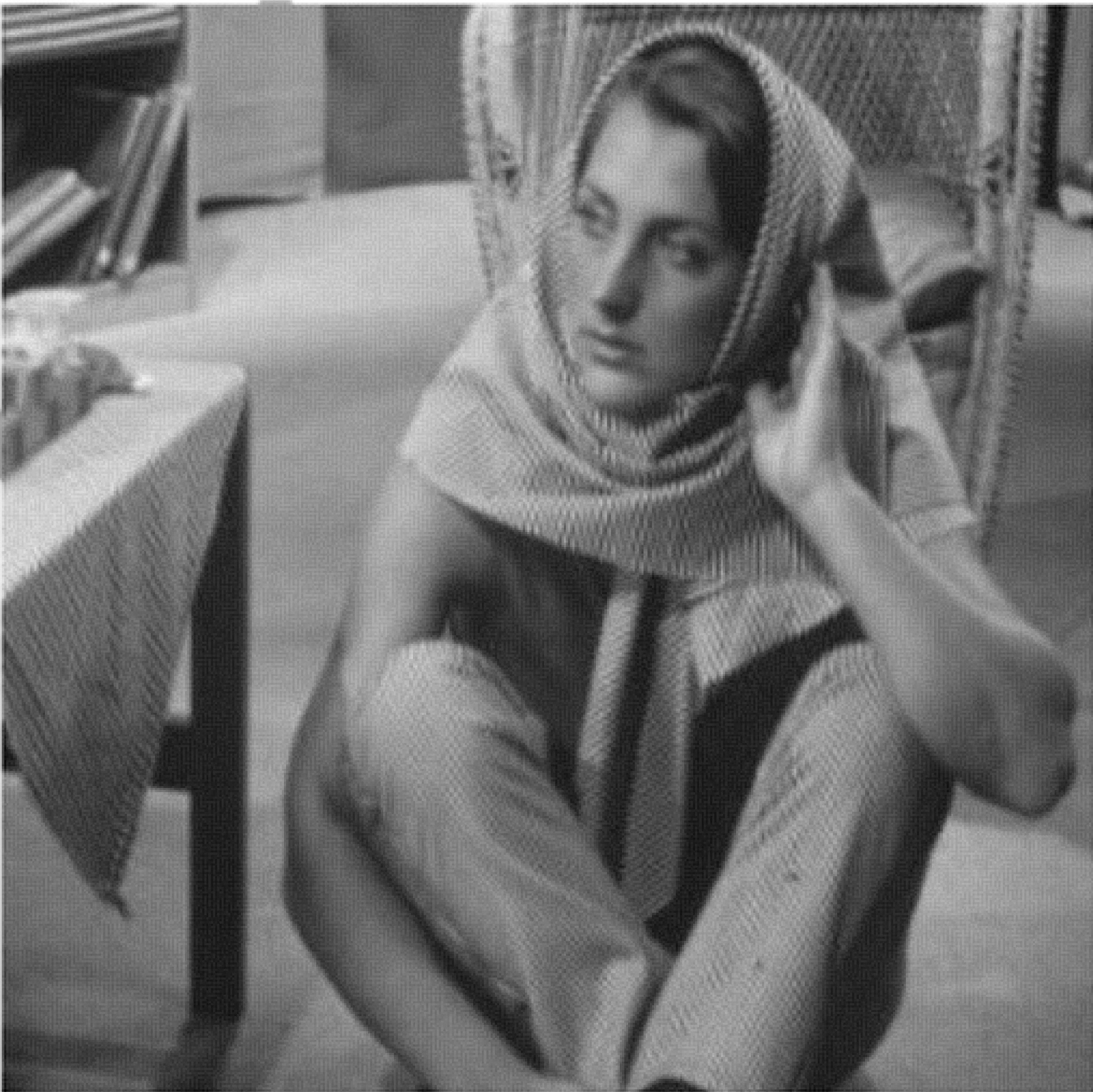}}
    \end{minipage}\hspace{0.5cm}
    \begin{minipage}[t]{0.09\textwidth}
        \centering
        \subfigure[][$Jetplane$]{\includegraphics[width=2.2cm]{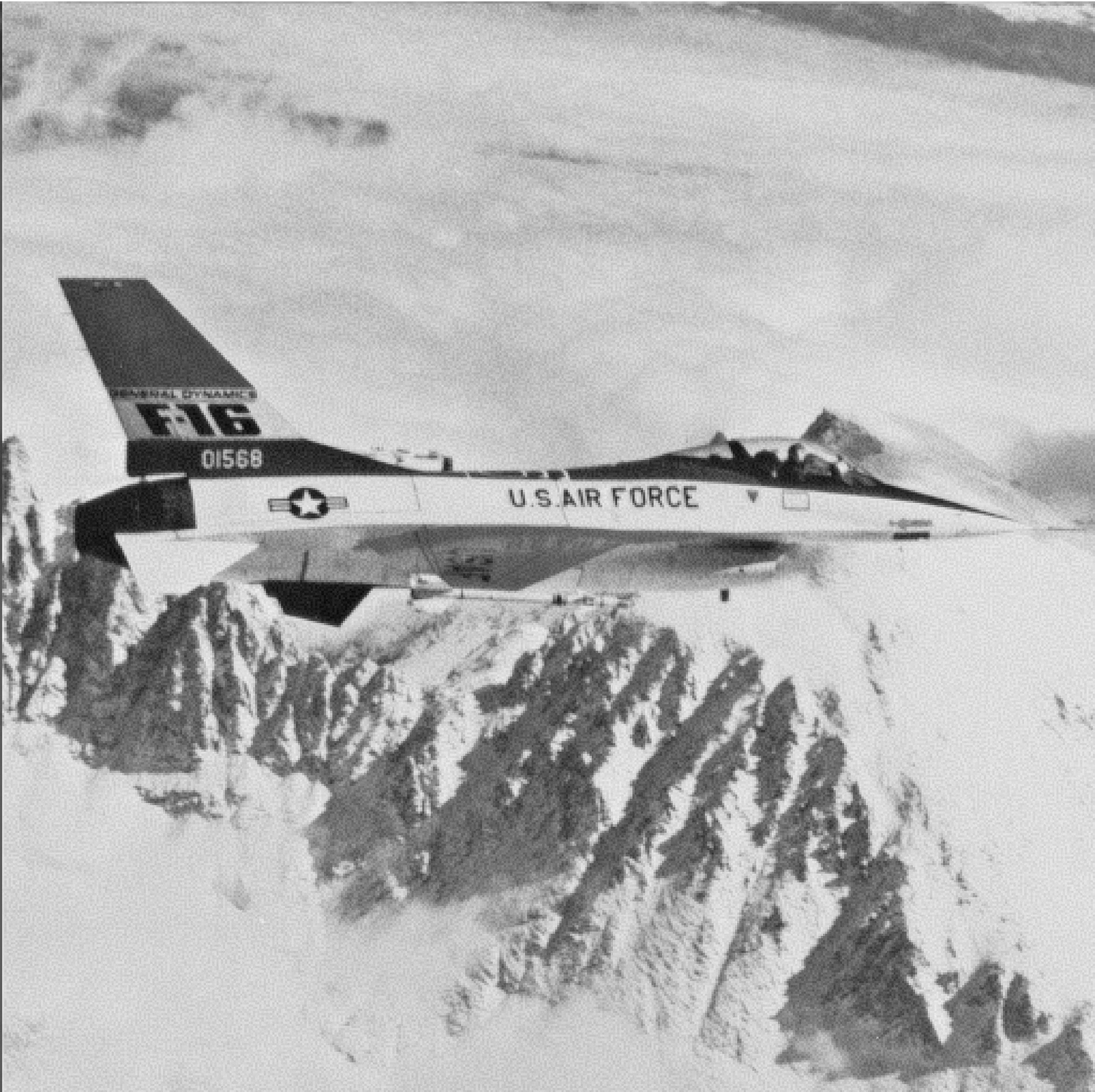}}
    \end{minipage}\hspace{0.5cm}
    \begin{minipage}[t]{0.09\textwidth}
        \centering
        \subfigure[][$Boat$]{\includegraphics[width=2.2cm]{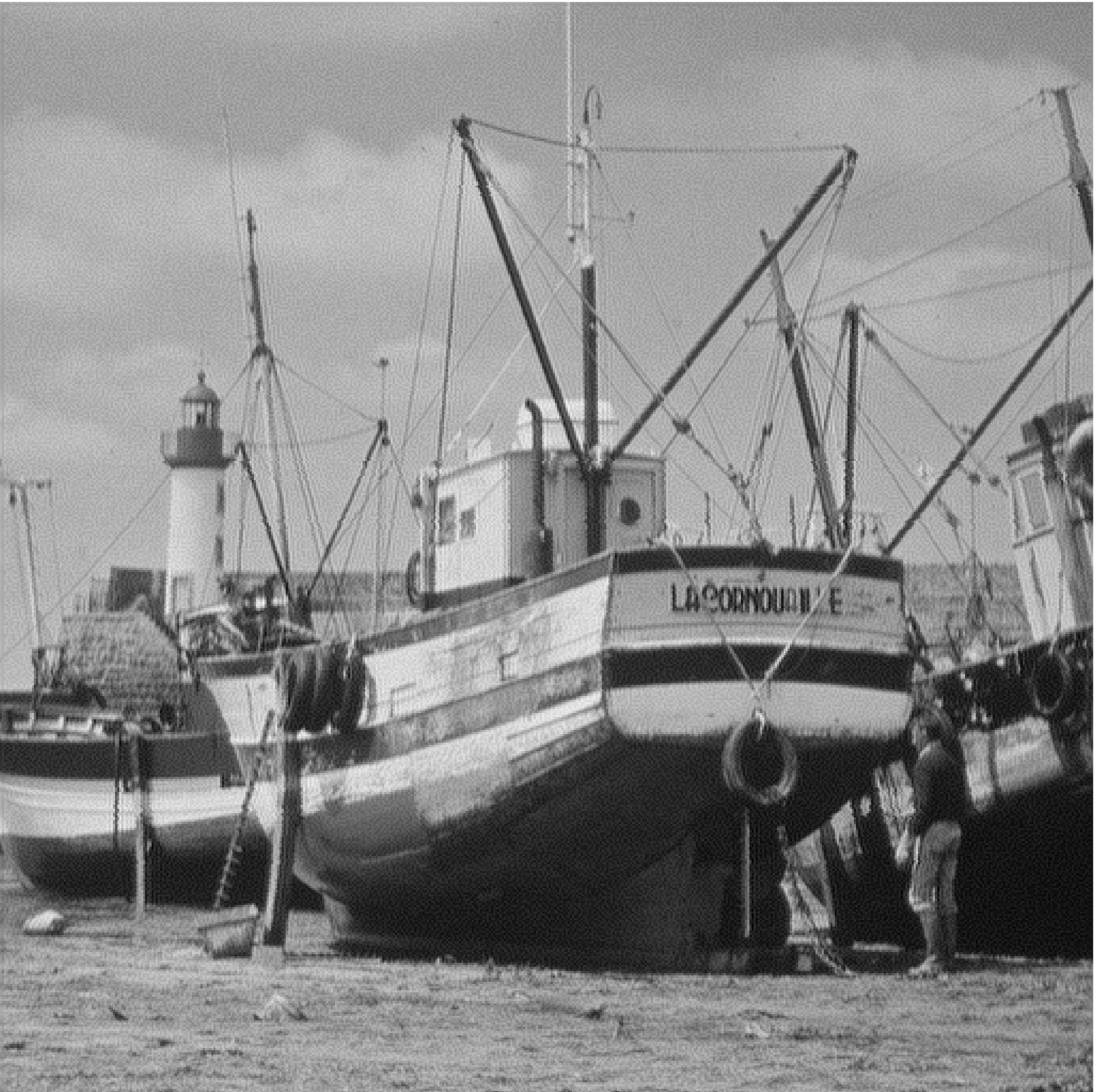}}
    \end{minipage}\hspace{0.5cm}
    \begin{minipage}[t]{0.09\textwidth}
        \centering
        \subfigure[][$Peppers$]{\includegraphics[width=2.2cm]{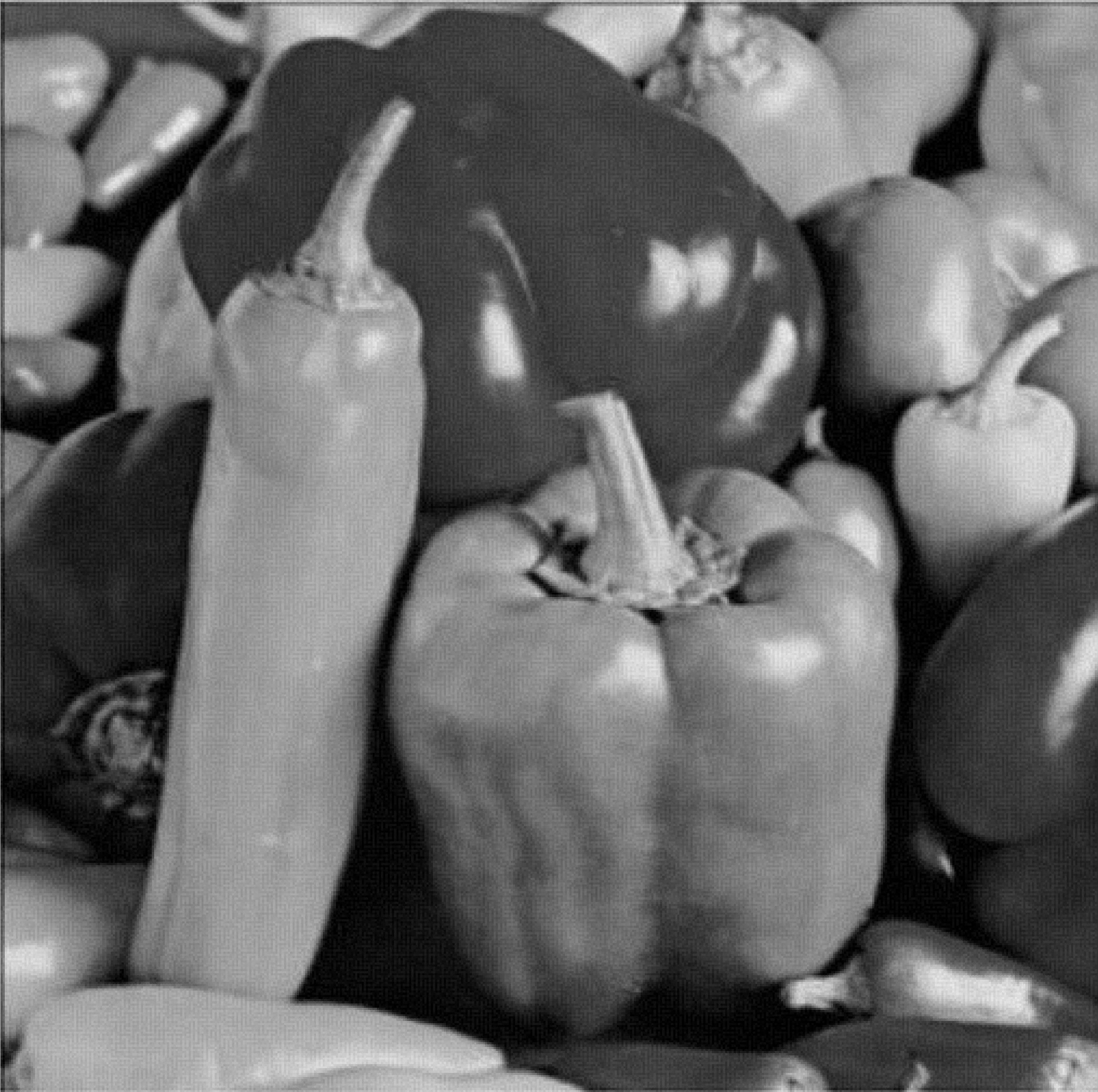}}
    \end{minipage}\hspace{0.5cm}
    \begin{minipage}[t]{0.09\textwidth}
        \centering
        \subfigure[][$Lake$]{\includegraphics[width=2.2cm]{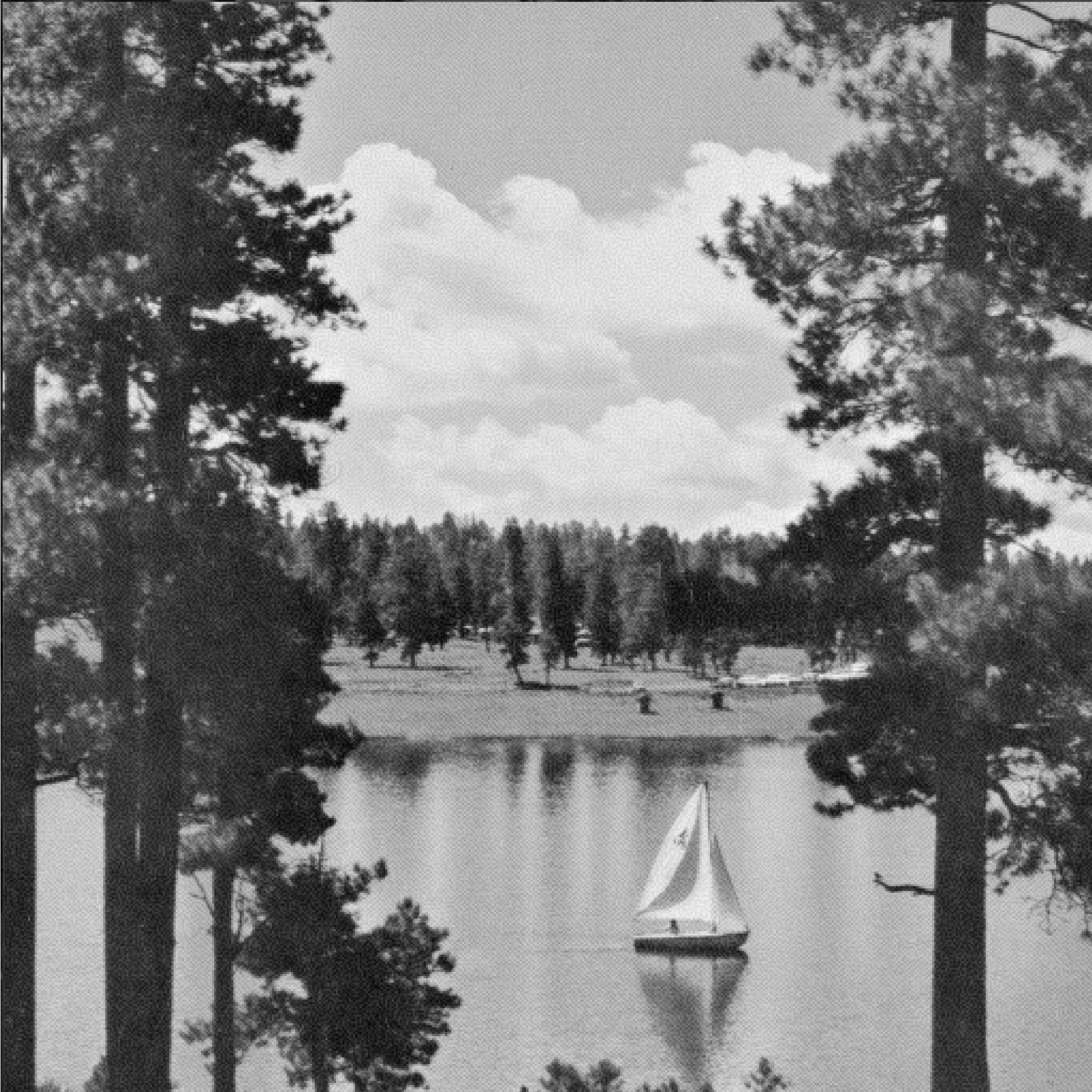}}
    \end{minipage}\hspace{0.5cm}
    \caption{Stego-images embedded with the maximum secret data by DTLE.}
    \label{fig:Sm}
\end{figure*}

\begin{figure*}[tb]
    \begin{minipage}[t]{0.21\textwidth}
        \centering
        \subfigure[][$Lena$]{\includegraphics[width=4.5cm]{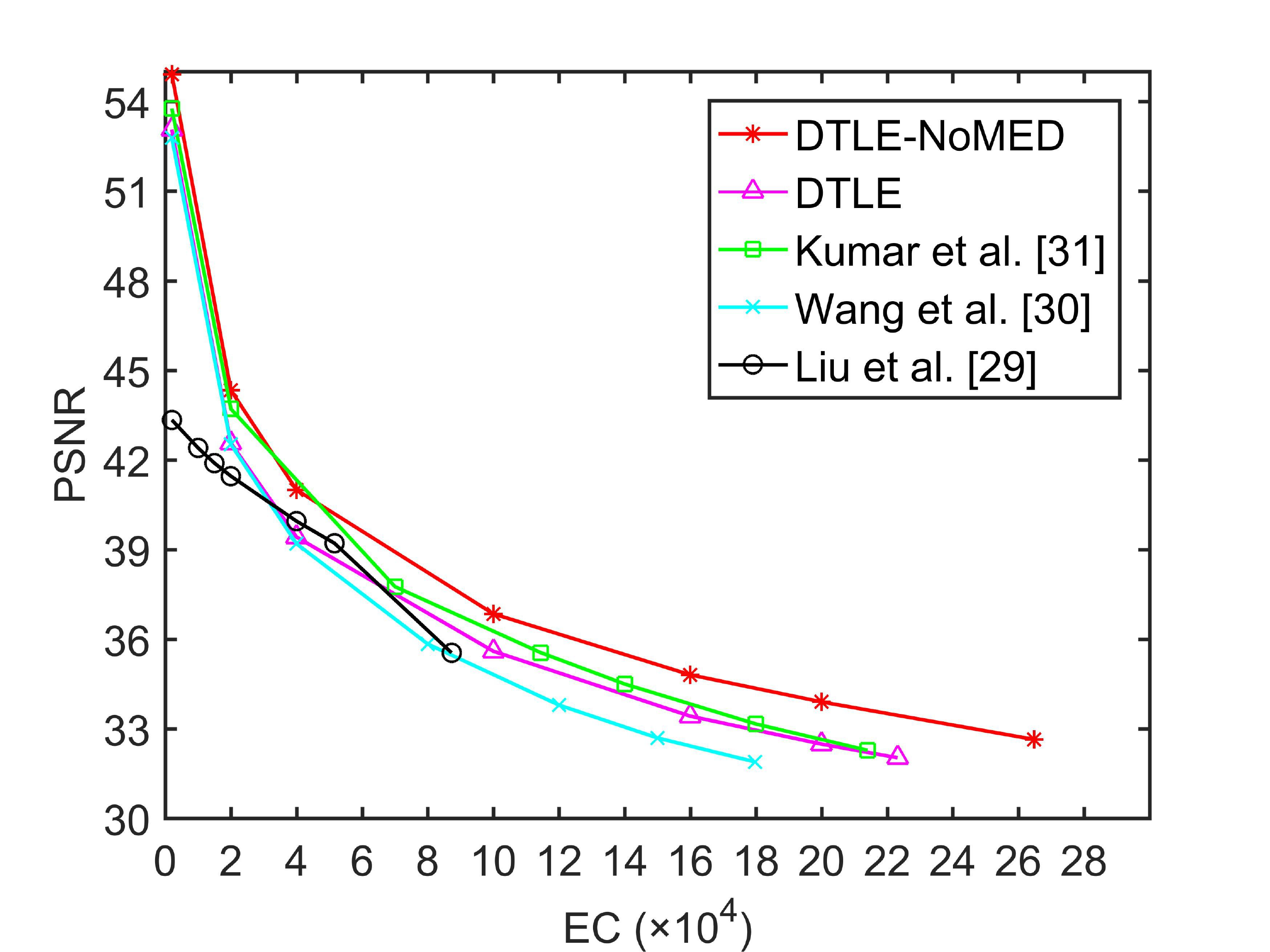}}
    \end{minipage}\hspace{0.6cm}
    \begin{minipage}[t]{0.21\textwidth}
        \centering
        \subfigure[][$Airplane$]{\includegraphics[width=4.5cm]{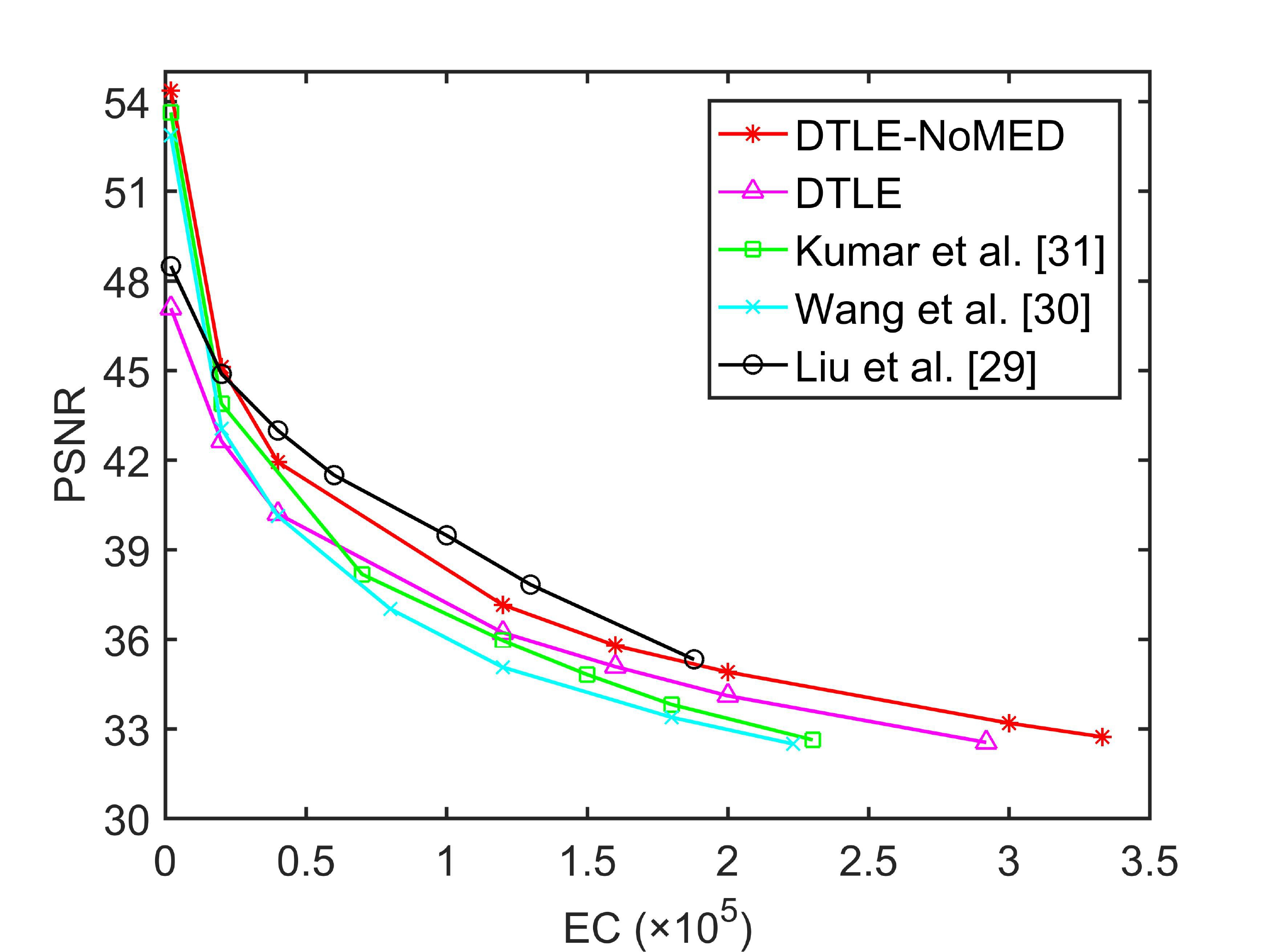}}
    \end{minipage}\hspace{0.6cm}
    \begin{minipage}[t]{0.21\textwidth}
        \centering
        \subfigure[][$Baboon$]{\includegraphics[width=4.5cm]{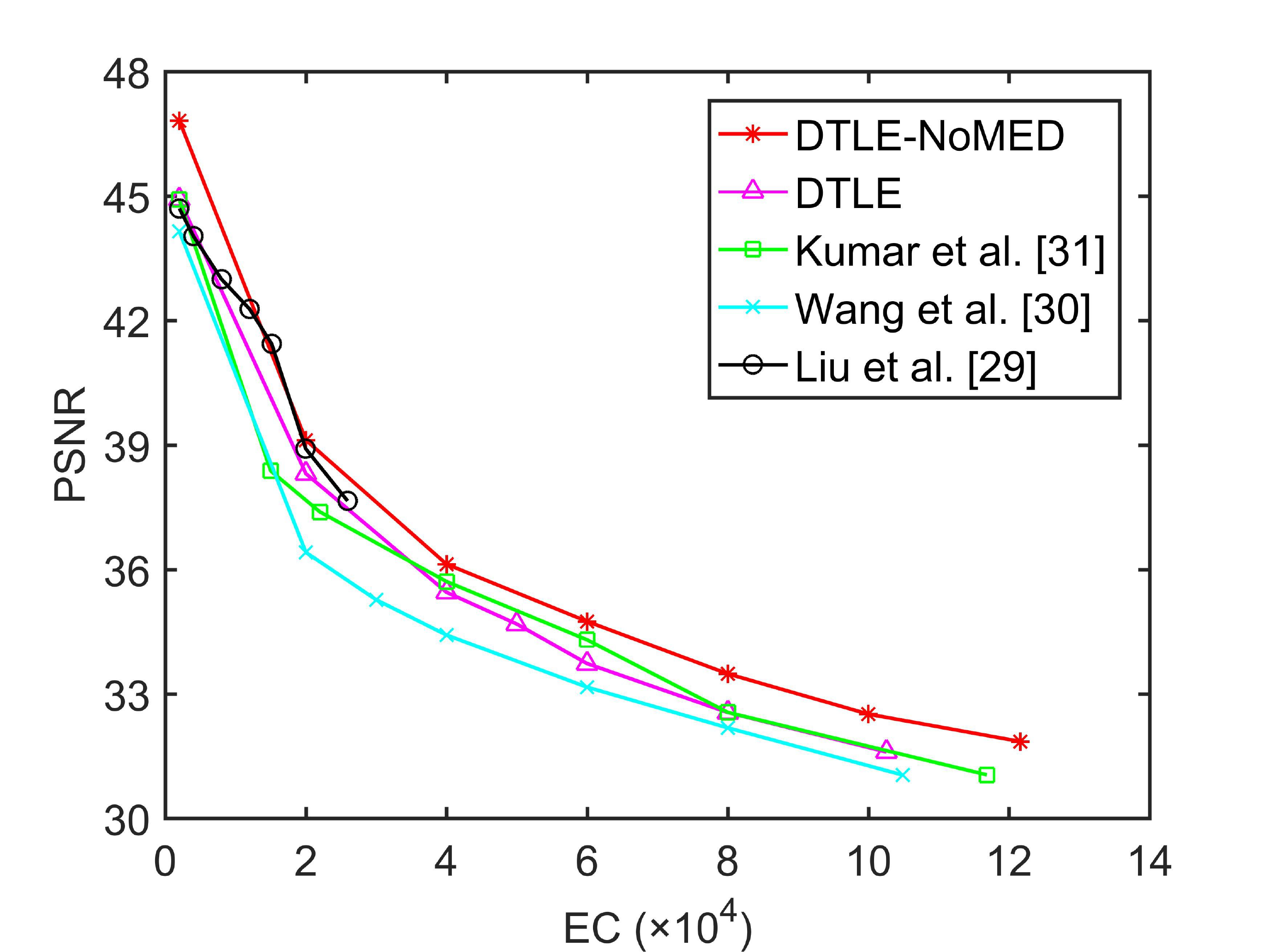}}
    \end{minipage}\hspace{0.6cm}
    \begin{minipage}[t]{0.21\textwidth}
        \centering
        \subfigure[][$Barbara$]{\includegraphics[width=4.5cm]{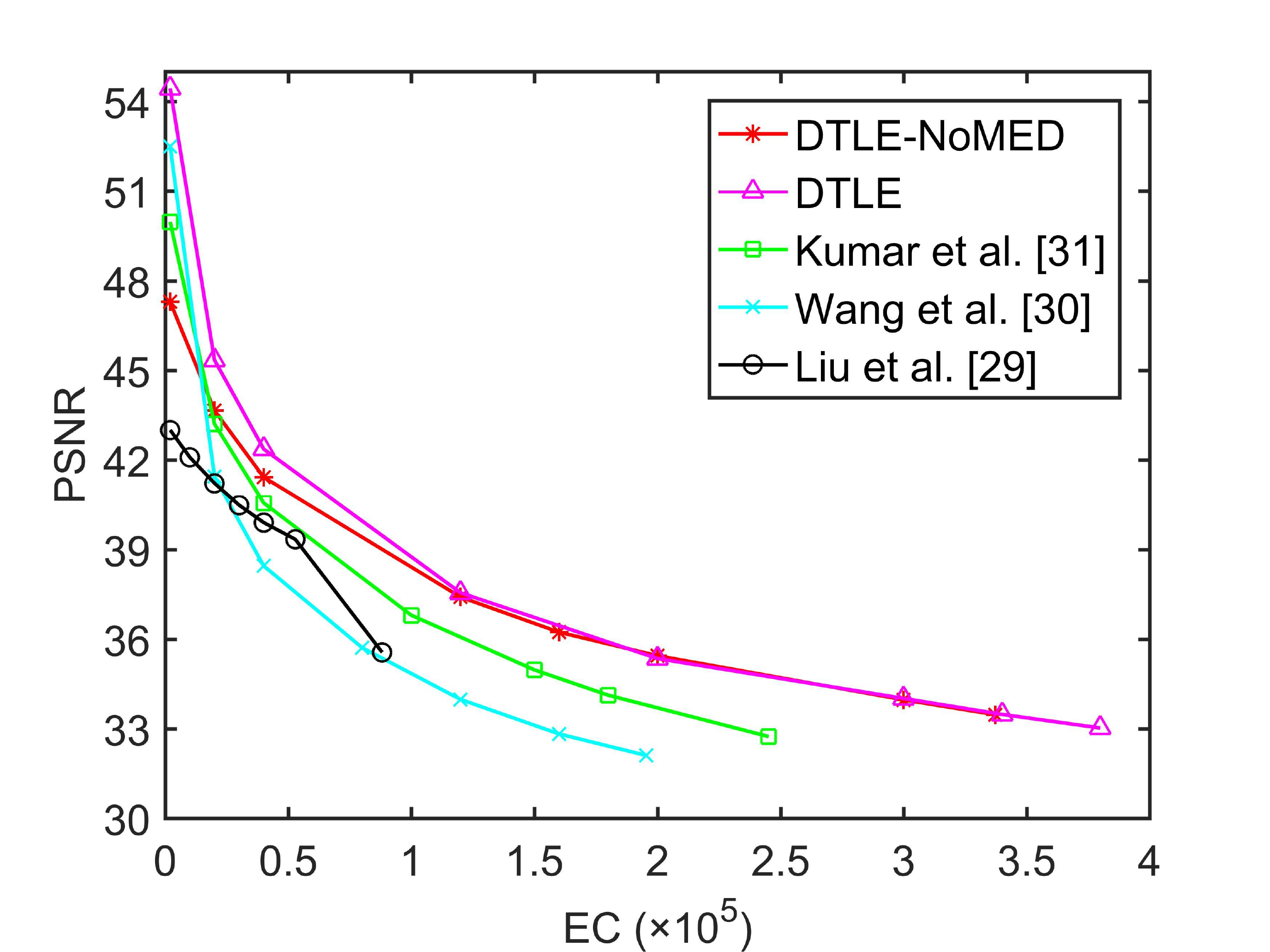}}
    \end{minipage}\hspace{0.6cm}

    \begin{minipage}[t]{0.21\textwidth}
        \centering
        \subfigure[][$Jetplane$]{\includegraphics[width=4.5cm]{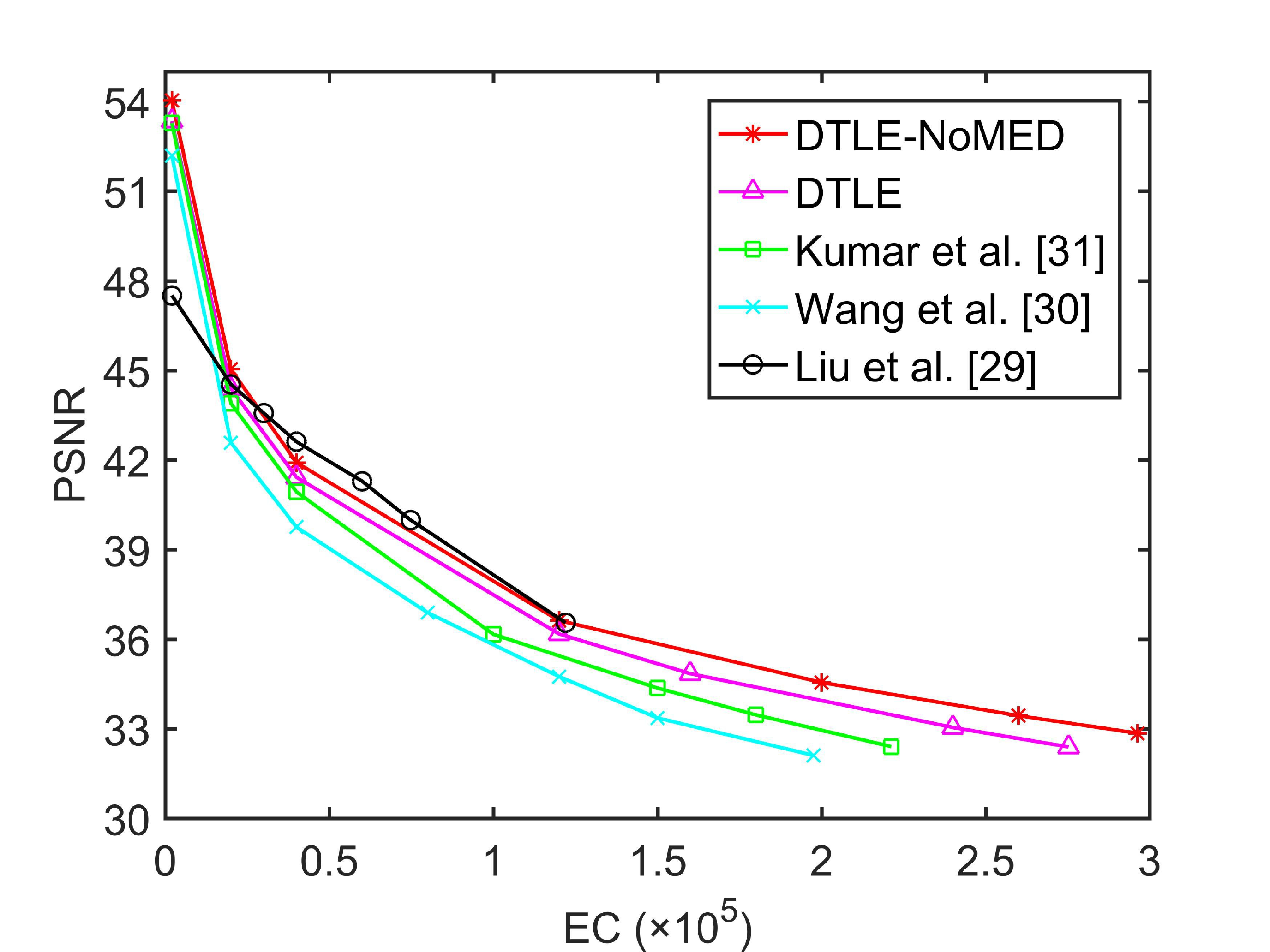}}
    \end{minipage}\hspace{0.6cm}
    \begin{minipage}[t]{0.21\textwidth}
        \centering
        \subfigure[][$Boat$]{\includegraphics[width=4.5cm]{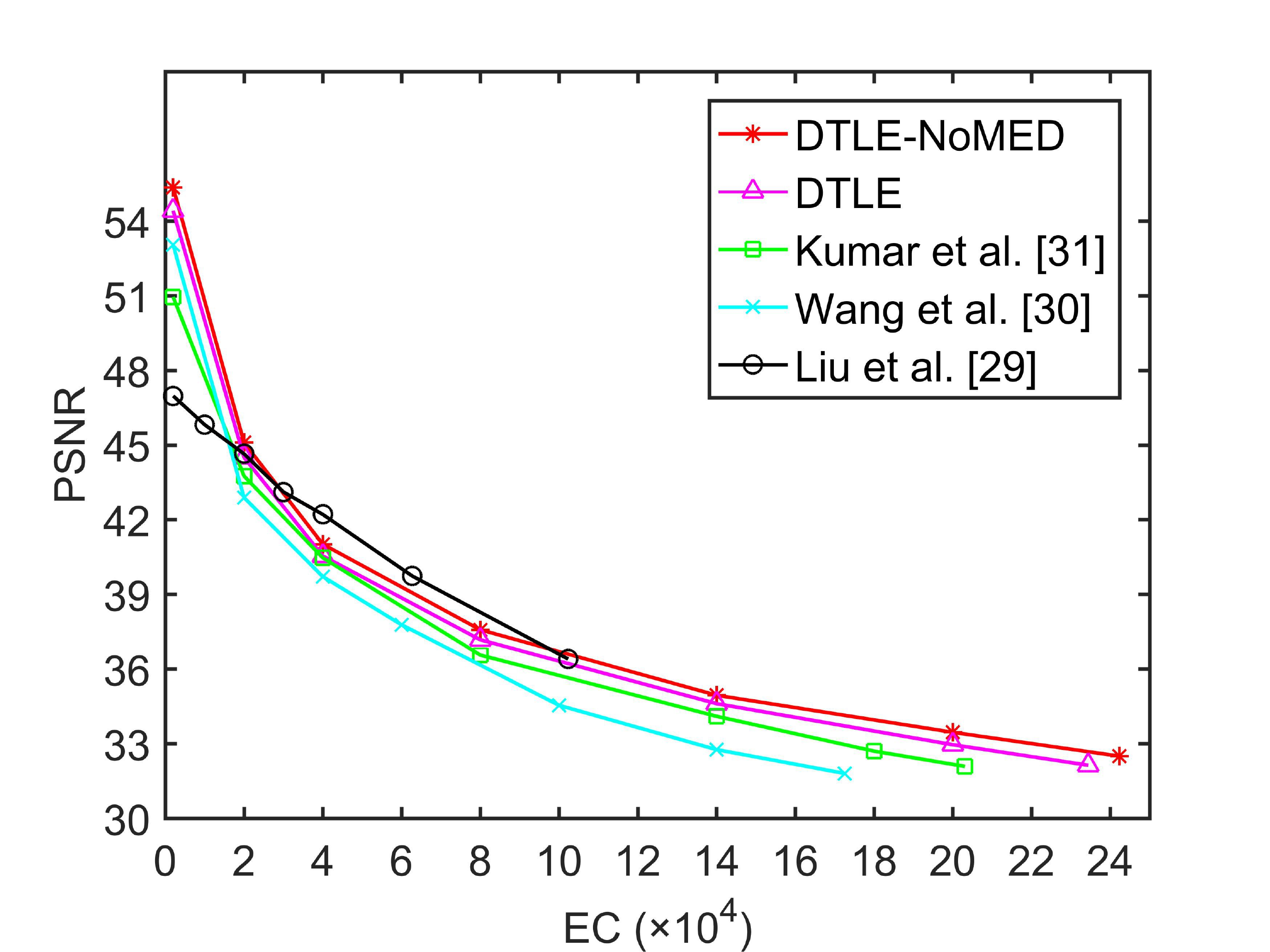}}
    \end{minipage}\hspace{0.6cm}
    \begin{minipage}[t]{0.21\textwidth}
        \centering
        \subfigure[][$Peppers$]{\includegraphics[width=4.5cm]{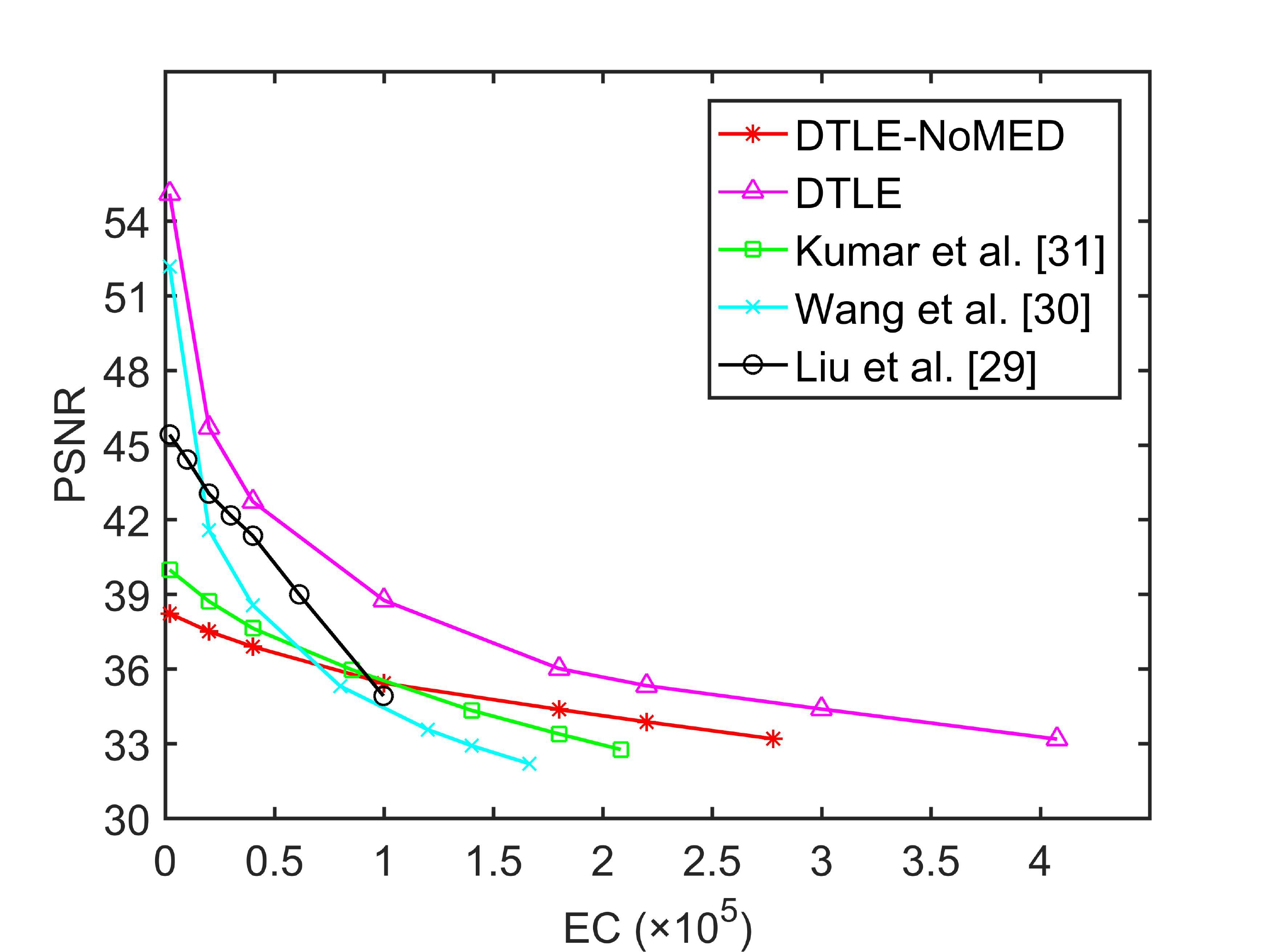}}
    \end{minipage}\hspace{0.6cm}
    \begin{minipage}[t]{0.21\textwidth}
        \centering
        \subfigure[][$Lake$]{\includegraphics[width=4.5cm]{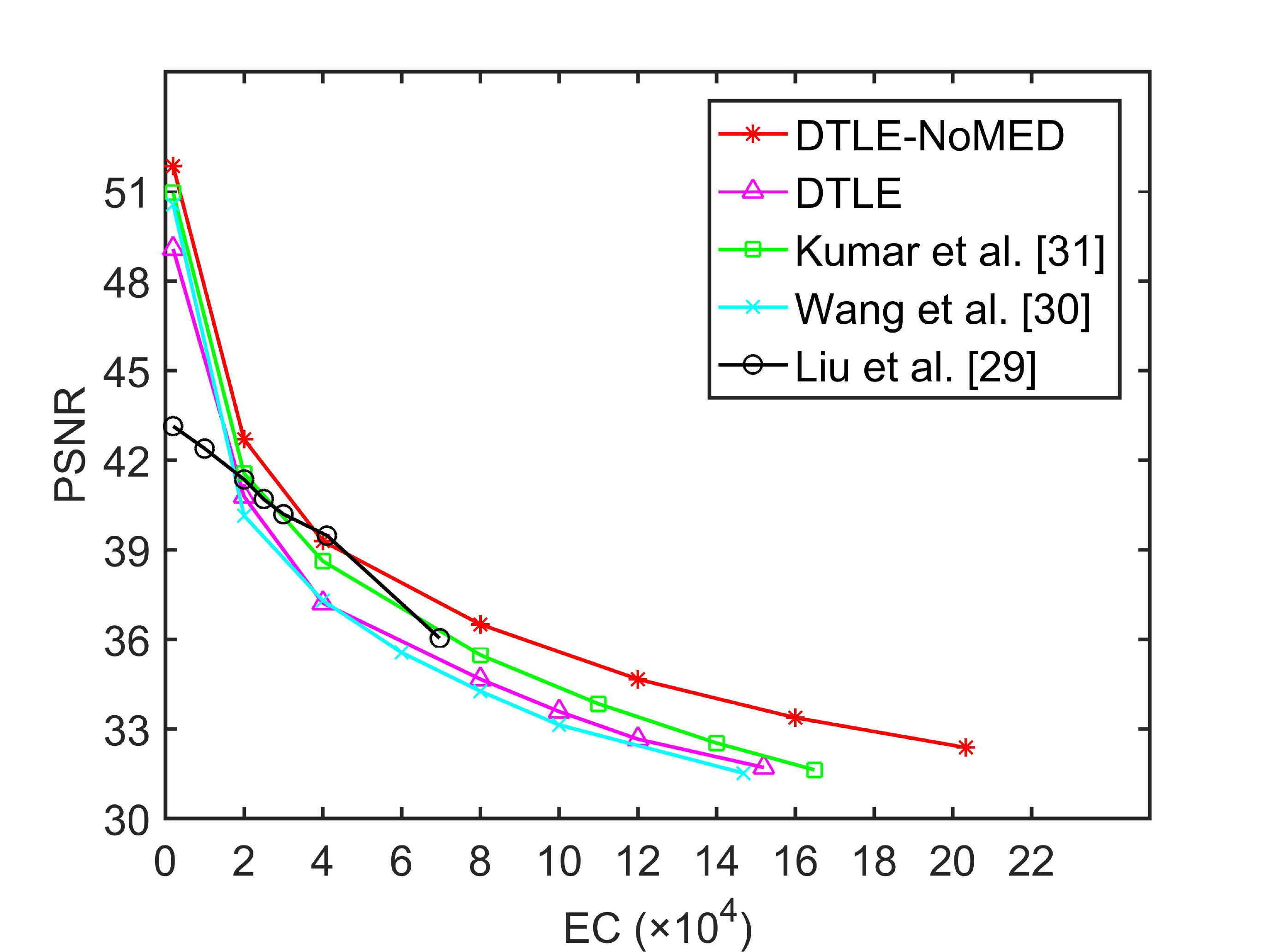}}
    \end{minipage}\hspace{0.6cm}
    \caption{PSNR and EC comparisons among different schemes performed on test images.}
    \label{fig:PE}
\end{figure*}

\section{Experimental results and analysis}
\label{sec:result}
In this section, our designed DTLE is compared with three state-of-the-art schemes \cite{LCW16,WYW17,KJ20} as well as the DTLE-NoMED scheme (i.e., DTLE without MED pre-processing). The metrics mainly include EC, embedding rate (ER) and PSNR, where EC denotes the embedded bits in a single image, ER stands for average bits per pixel (bpp) in a dataset and PSNR is a peak signal-to-noise ratio to measure image quality by decibels.

The experimental environment is based on a PC with a $4.00$ GHz AMD A$6$-$6420$B processor, $8.00$ GB of RAM, Windows $7$ operating system and Matlab R$2016$a. We first perform a series of experiments on eight common images: $Lena$, $Airplane$, $Baboon$, $Babara$, $Jetplane$, $Boat$, $Peppers$, $Lake$. Then we evaluate their performances on the three datasets: BOSSbase \cite{BFP11}, BOWS-2 \cite{BF17} and UCID \cite{SS03}. Among them, Both BOSSbase and BOWS-$2$ include $10,000$ greyscale images of $512\times 512$ size, while UCID involves $1338$ greyscale images of $512\times 384$ or $384\times 512$ size.

\subsection{Experimental results on test images}
The test images are presented in Fig.~\ref{fig:Et}, which are all $8$-bit greyscale images of size $512\times512$. The secret data consists of a randomly generated bit stream, and the stego-images embedded with the maximum secret data bits by DTLE is given in Fig.~\ref{fig:Sm}.

We evaluate the performances by EC and PSNR. We experiment with varying the HSB plane selections, i.e., the values of $n$, cf. Eq. \ref{eq:22}, whose results are shown in Table \ref{tab:Pe}. The EC with $n=3$ is higher that with $n=2$ while the PSNR values for $n=3$ are less than $30$. When the PSNR is generally below $30$, the distortion of the image can be detected by human eyes and the information transmission process is prone to data leakage. Although the EC decreases at $n=2$, the PSNR value increases significantly. This means that the HSB plane at $n=2$ is the best selection for the higher six-bit planes.

In Table \ref{tab:Coe}, the comparisons among DTLE, DTLE-NoMED and TLE \cite{KJ20} shows that DTLE-NoMED has the highest EC on all eight tested images:
at least $5\times 10^3$ bits on $Baboon$ and at most $10^5$ bits on $Airplane$ compared with TLE. Compared to the DTLE-NoMED, the DTLE has reduced EC on the six images while achieving great improvement on complex images $Peppers$ and $Barbara$. This shows that DTLE-NoMED makes full use of redundant space in images while some complex images producing huge auxiliary information leads to a reduction of EC. Usually we regard a type of image as a complex image in which the number of possible overflow pixels exceeds $5\%$ of the total pixels. The images pre-processed by MED in DTLE have more concentrated pixel values and smaller location maps and allow higher EC.


Further, we perform experimental comparisons among five schemes in terms of the tradeoff between EC and PSNR: Liu et al. \cite{LCW16}, Wang et al. \cite{WYW17} (SBDE), Kumar et al. \cite{KJ20} (TLE), and our proposed DTLE and DTLE-NoMED, as illustrated in Fig.~\ref{fig:PE}.
We execute two-layer embeddings of Liu et al.'s scheme for convenient comparison.
In terms of EC, the DTLE-NoMED method outperforms the other methods in all images except for the complex image $Peppers$. DTLE-NoMED achieves  higher PSNRs while preserving the same EC.
For the complex images $Peppers$ and $Barbara$, the initial PSNR of DTLE-NoMED is lower due to many overflow pixels resulting in huge auxiliary information after location map compression. The size of the location map for Liu et al.'s scheme is less affected by image variation, while Wang et al.'s approach doesn't create a location map and also is less affected. The location map of TLE only marks two possible overflow pixels while the size of the auxiliary information is lower than DTLE-NoMED. However, after MED pre-processing both test images turns to produce higher EC and PSNR than that of other schemes. This yields that DTLE has a better performance advantage for those (complex) images with many possibly overflow pixels and long compressed location maps. Meanwhile, the decreasing trend of PSNR for DTLE-NoMED is gentler and even eventually higher than the other schemes, which reflects the performance advantage of DTLE-NoMED in reducing image distortion. To sum up, the DTLE-NoMED is suitable for general images, while the DTLE works better for complex images.


\begin{figure*}[tb]
    \begin{minipage}[t]{0.3\linewidth}
		\centering
		\subfigure[][BOSSbase]{\includegraphics[width=1.0\linewidth]{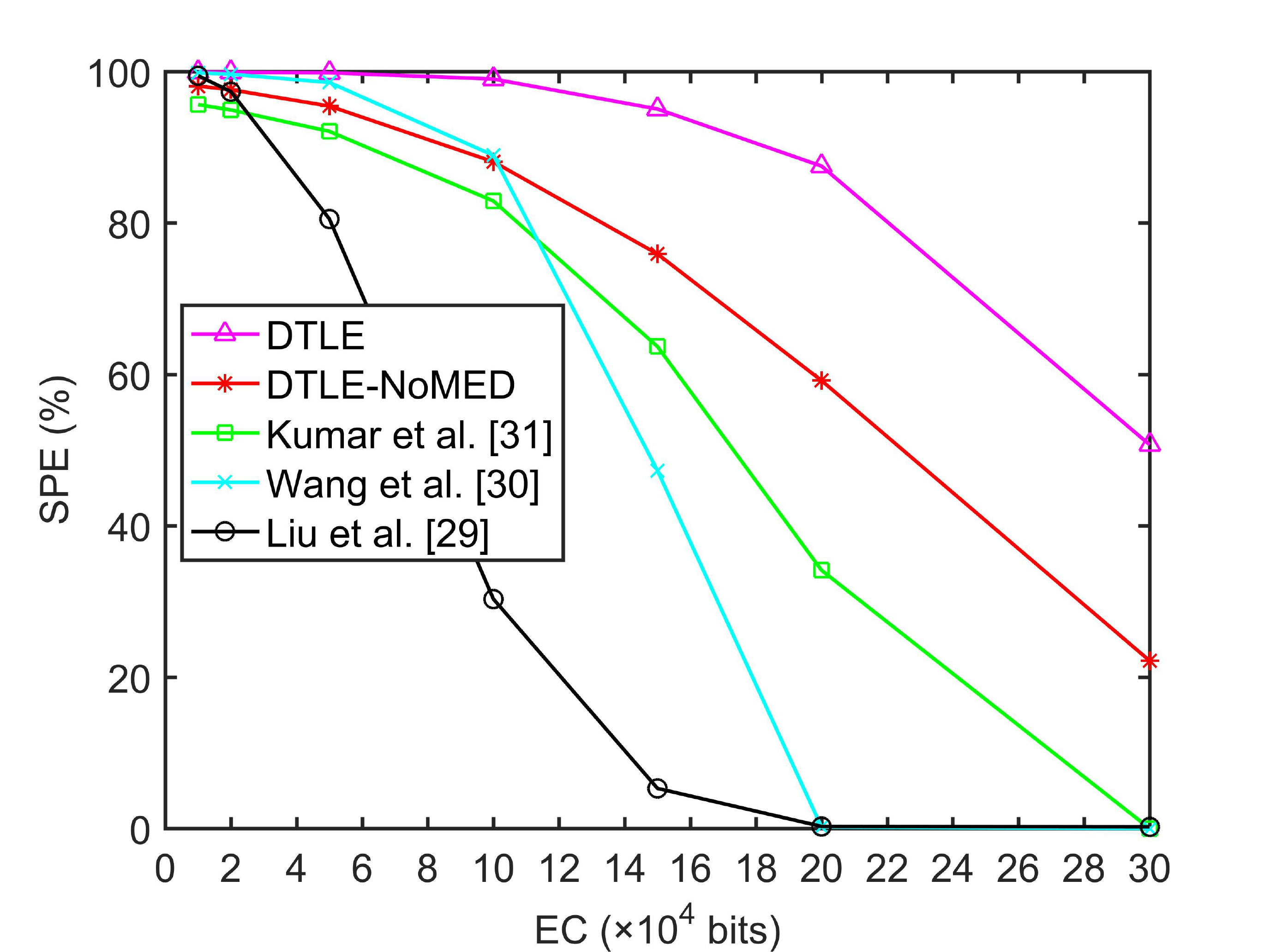}}
	\end{minipage}
\hfill
	\begin{minipage}[t]{0.3\linewidth}
		\centering
		\subfigure[][BOWS2]{\includegraphics[width=1.0\linewidth]{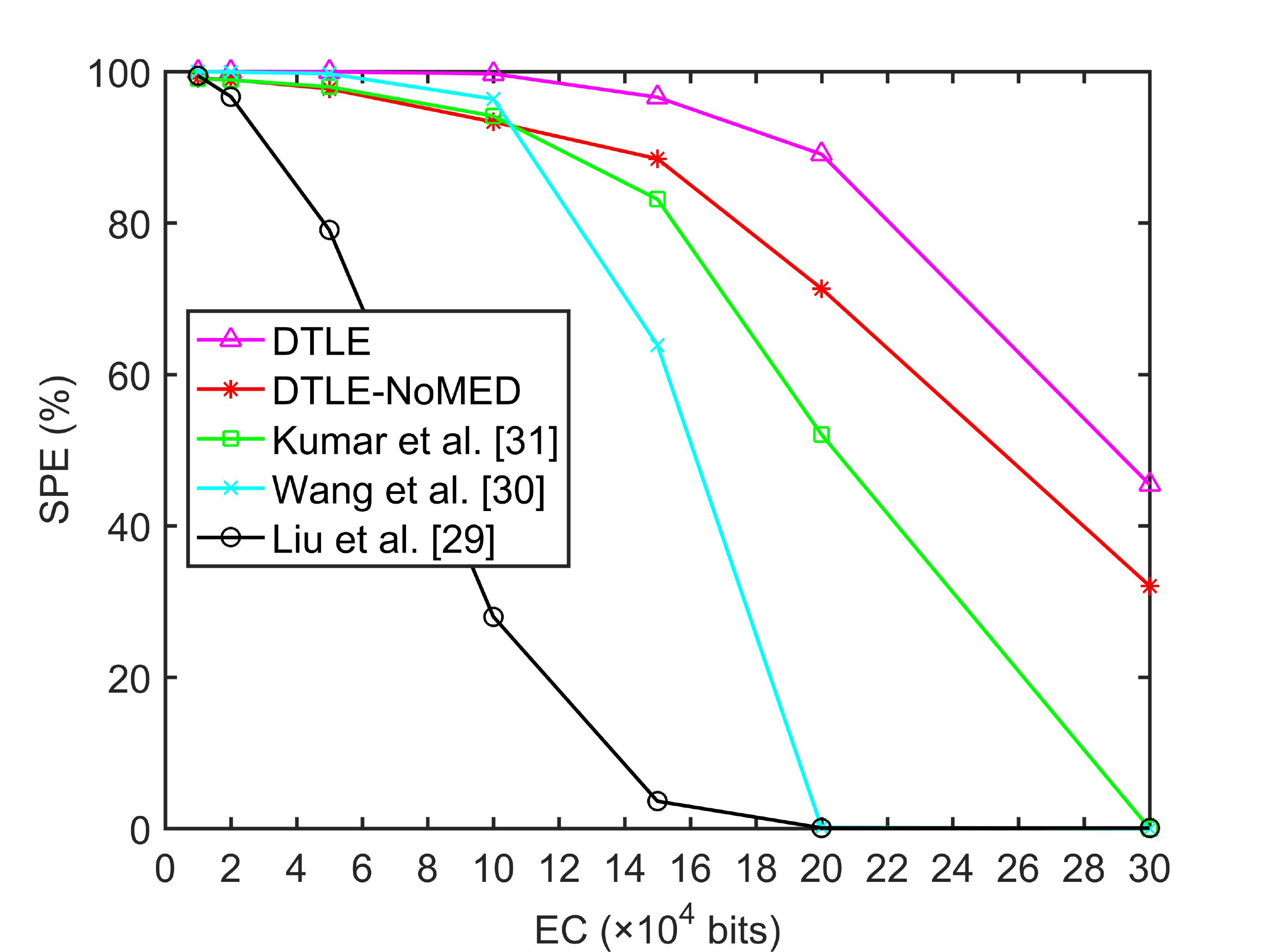}}
	\end{minipage}
\hfill
	\begin{minipage}[t]{0.3\linewidth}
		\centering
		\subfigure[][UCID]{\includegraphics[width=1.0\linewidth]{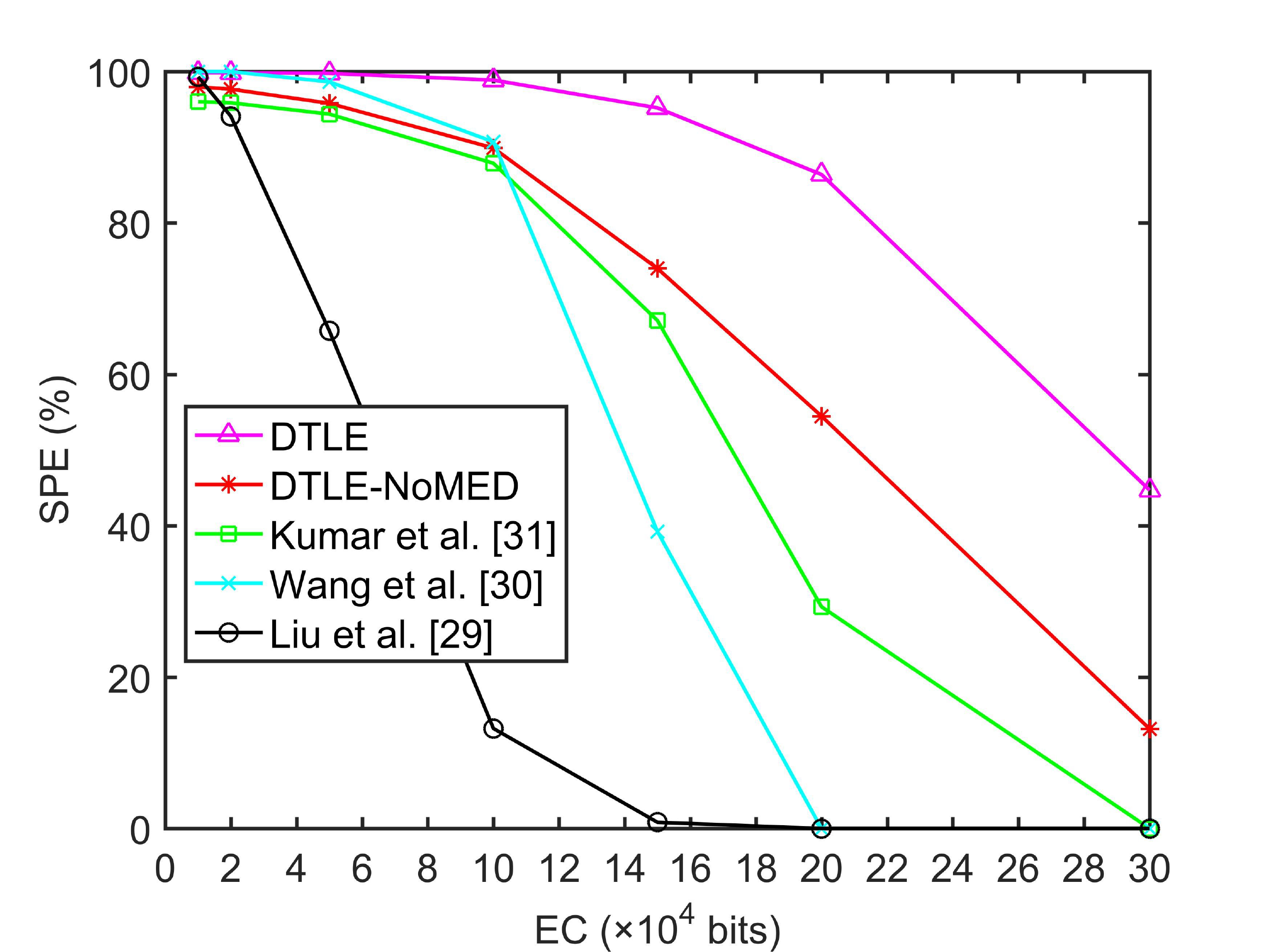}}
	\end{minipage}
    \caption{ SPE (success percentage of embeddings) comparisons of different schemes on datasets with varying EC.}
    \label{fig:CP}
\end{figure*}

\begin{figure*}[tb]
    \begin{minipage}[t]{0.3\linewidth}
		\centering
		\subfigure[][BOSSbase]{\label{fig:LMA}
        \includegraphics[width=1.0\linewidth]{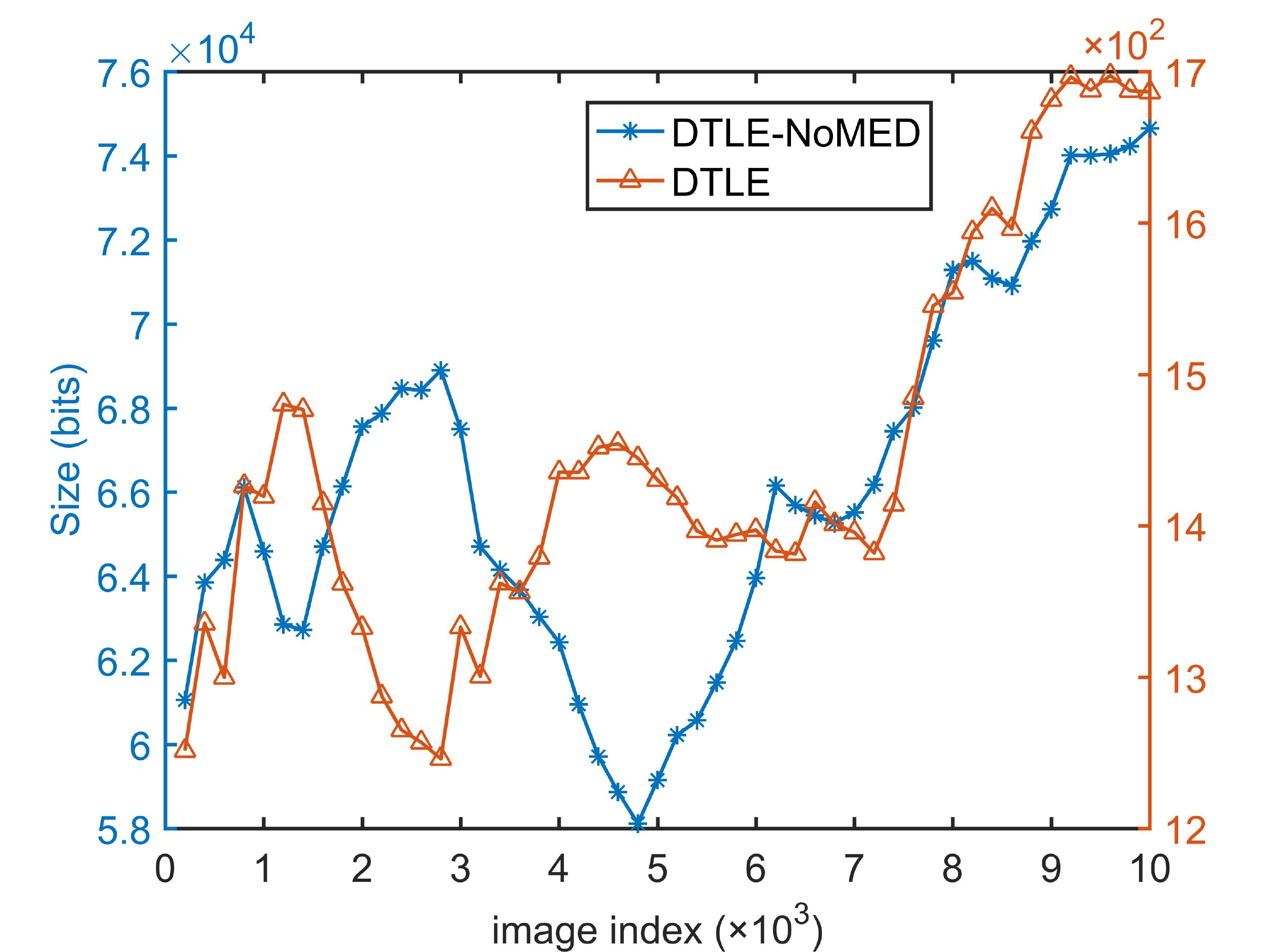}}
	\end{minipage}
\hfill
	\begin{minipage}[t]{0.3\linewidth}
		\centering
		\subfigure[][BOWS2]{\label{fig:LMB}
        \includegraphics[width=1.0\linewidth]{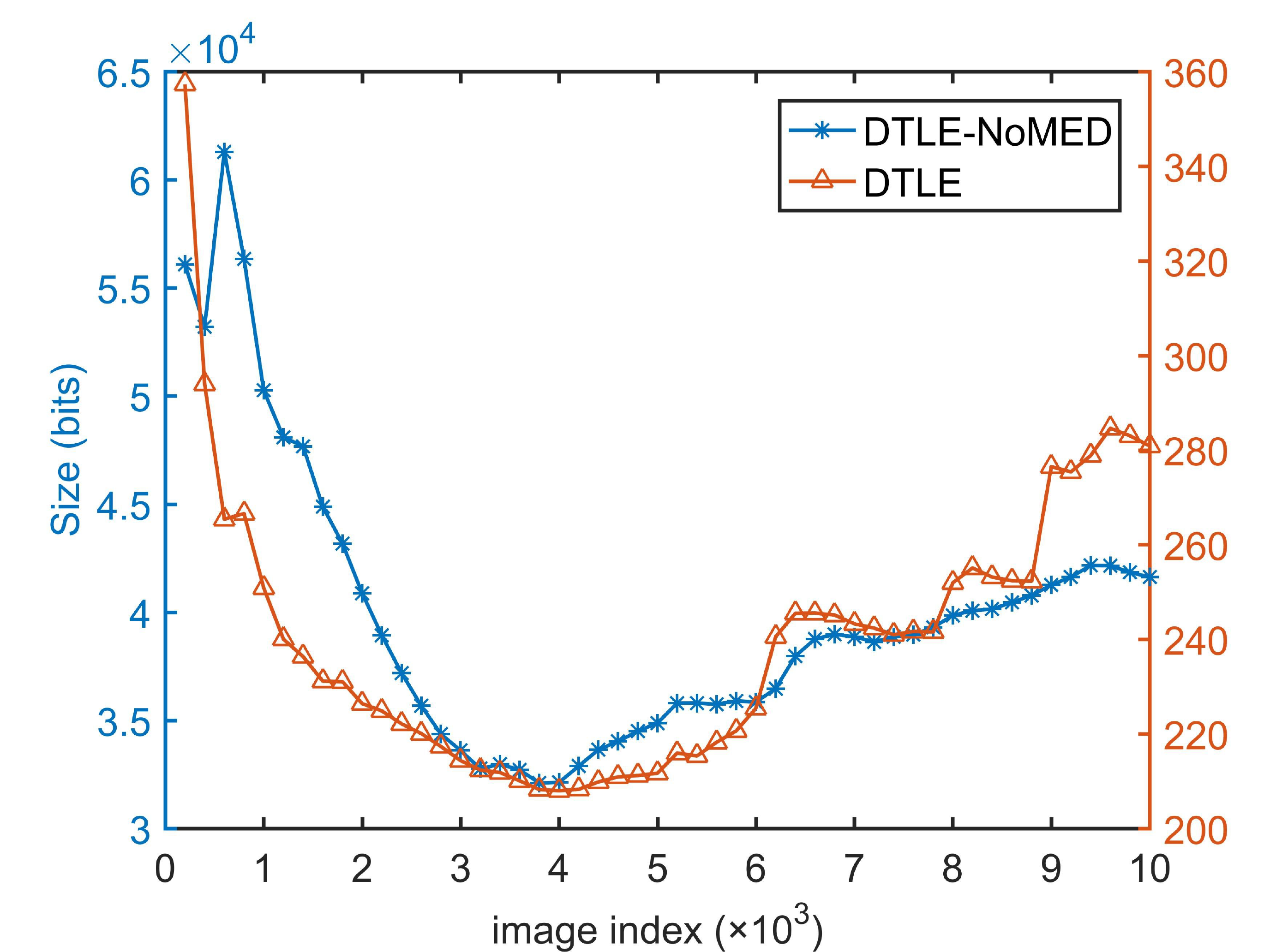}}
	\end{minipage}
\hfill
	\begin{minipage}[t]{0.3\linewidth}
		\centering
		\subfigure[][UCID]{\label{fig:LMC}
        \includegraphics[width=1.0\linewidth]{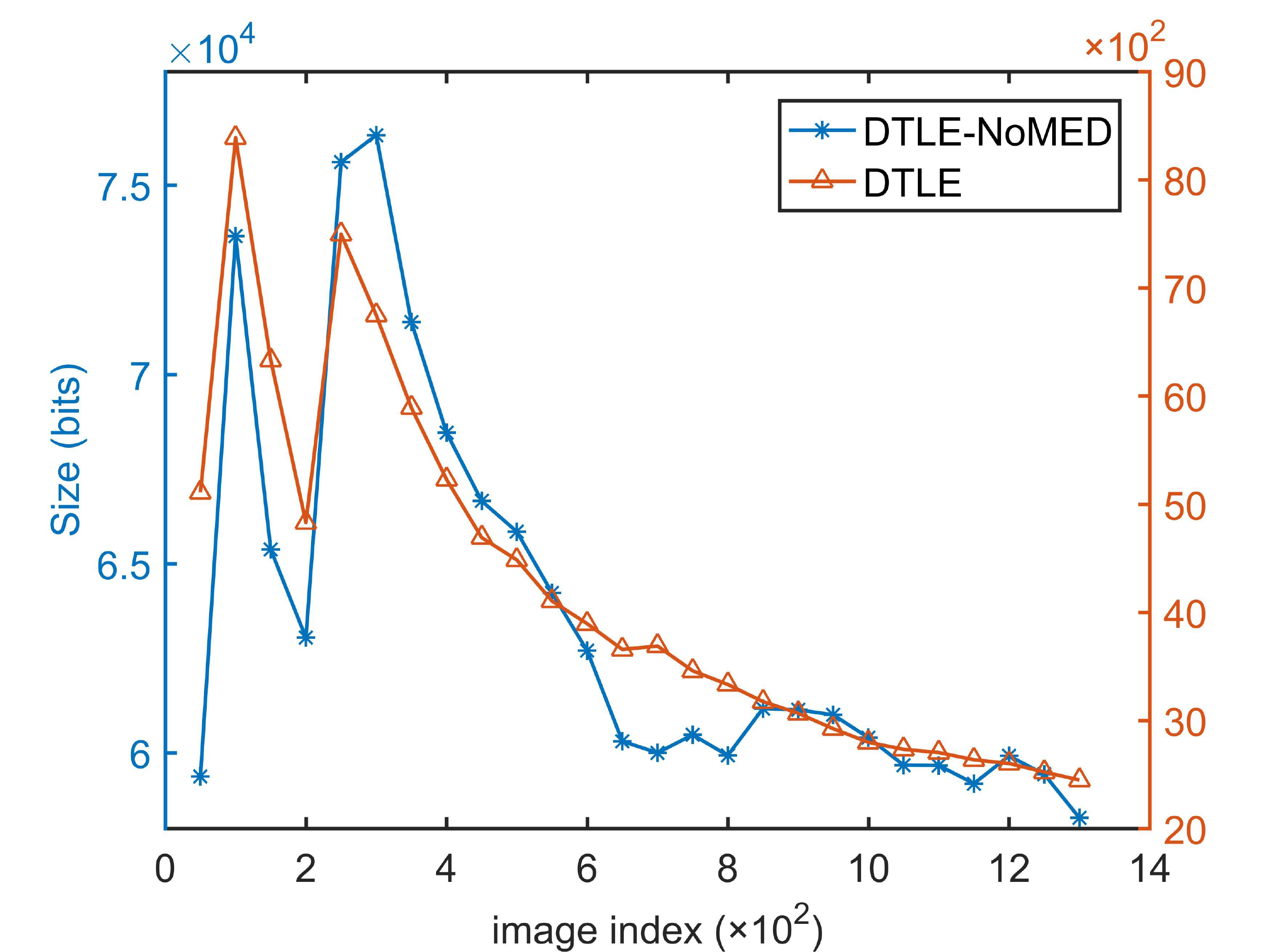}}
	\end{minipage}
    \caption{Comparisons of auxiliary information size over datasets.}
    \label{fig:LM}
\end{figure*}

\begin{figure*}[tb]
    \begin{minipage}[t]{0.3\linewidth}
		\centering
		\subfigure[][BOSSbase]{\includegraphics[width=1.0\linewidth]{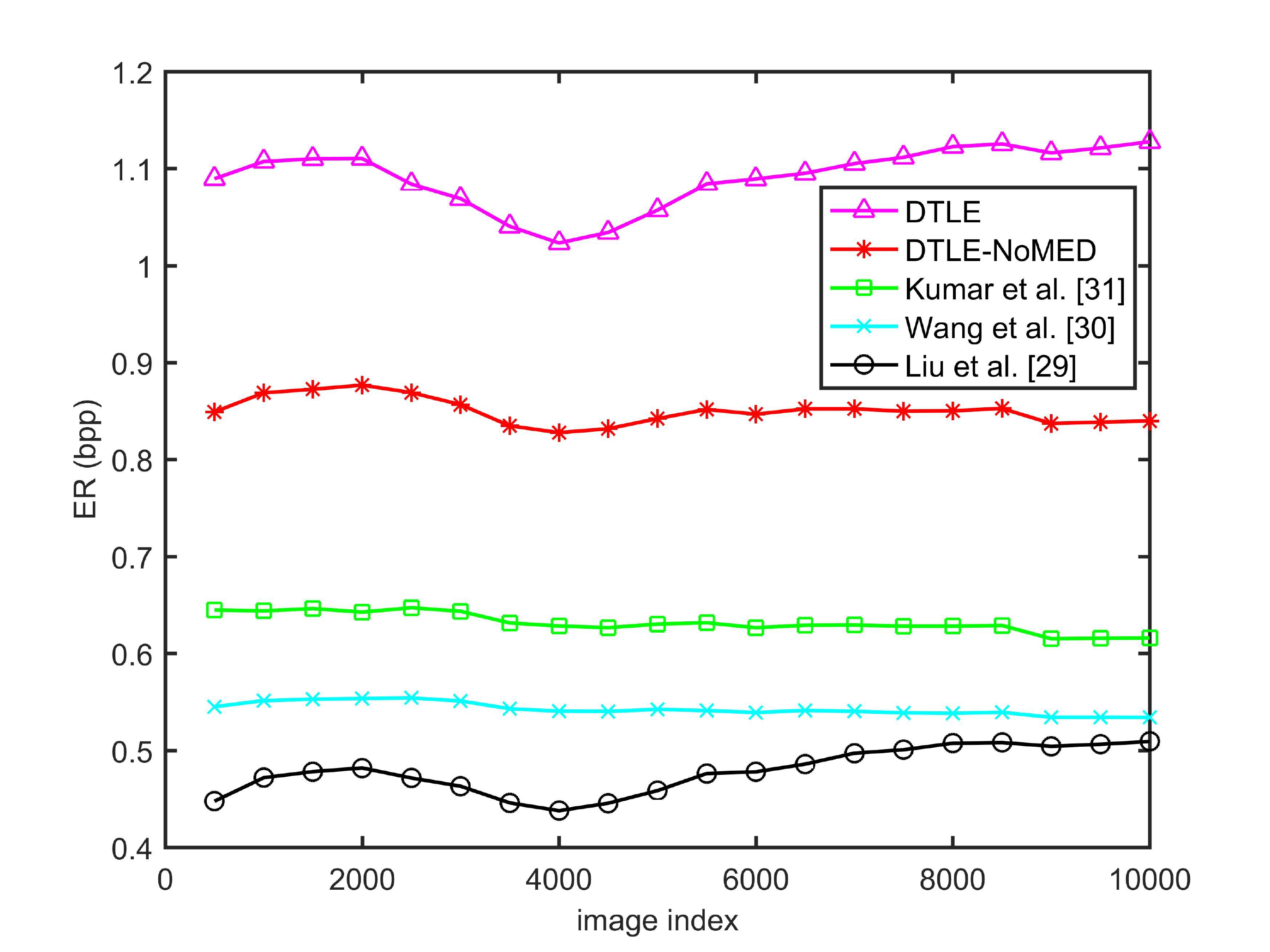}}
	\end{minipage}%
\hfill
	\begin{minipage}[t]{0.3\linewidth}
		\centering
		\subfigure[][BOWS2]{\includegraphics[width=1.0\linewidth]{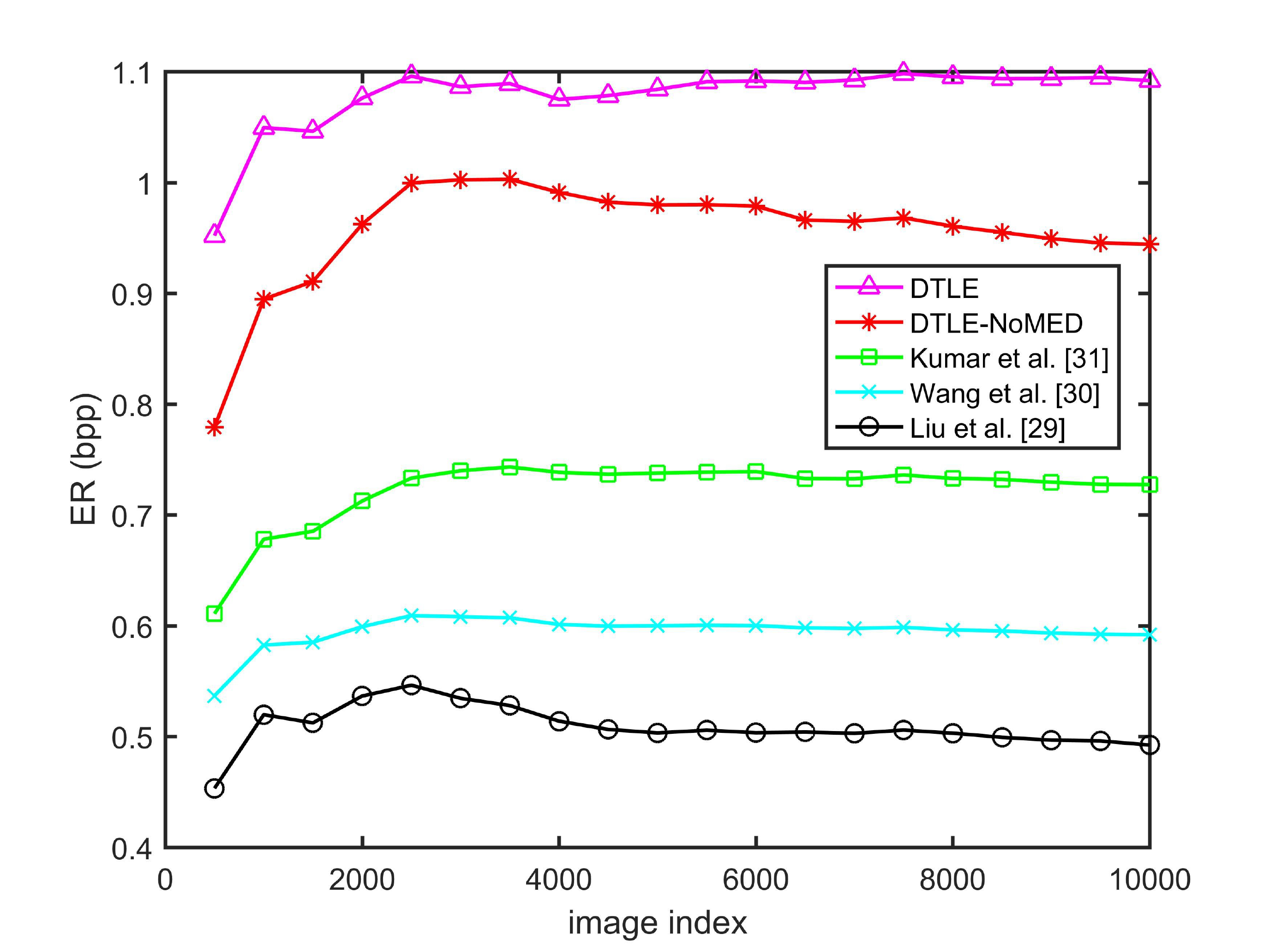}}
	\end{minipage}%
\hfill
	\begin{minipage}[t]{0.3\linewidth}
		\centering
		\subfigure[][UCID]{\includegraphics[width=1.0\linewidth]{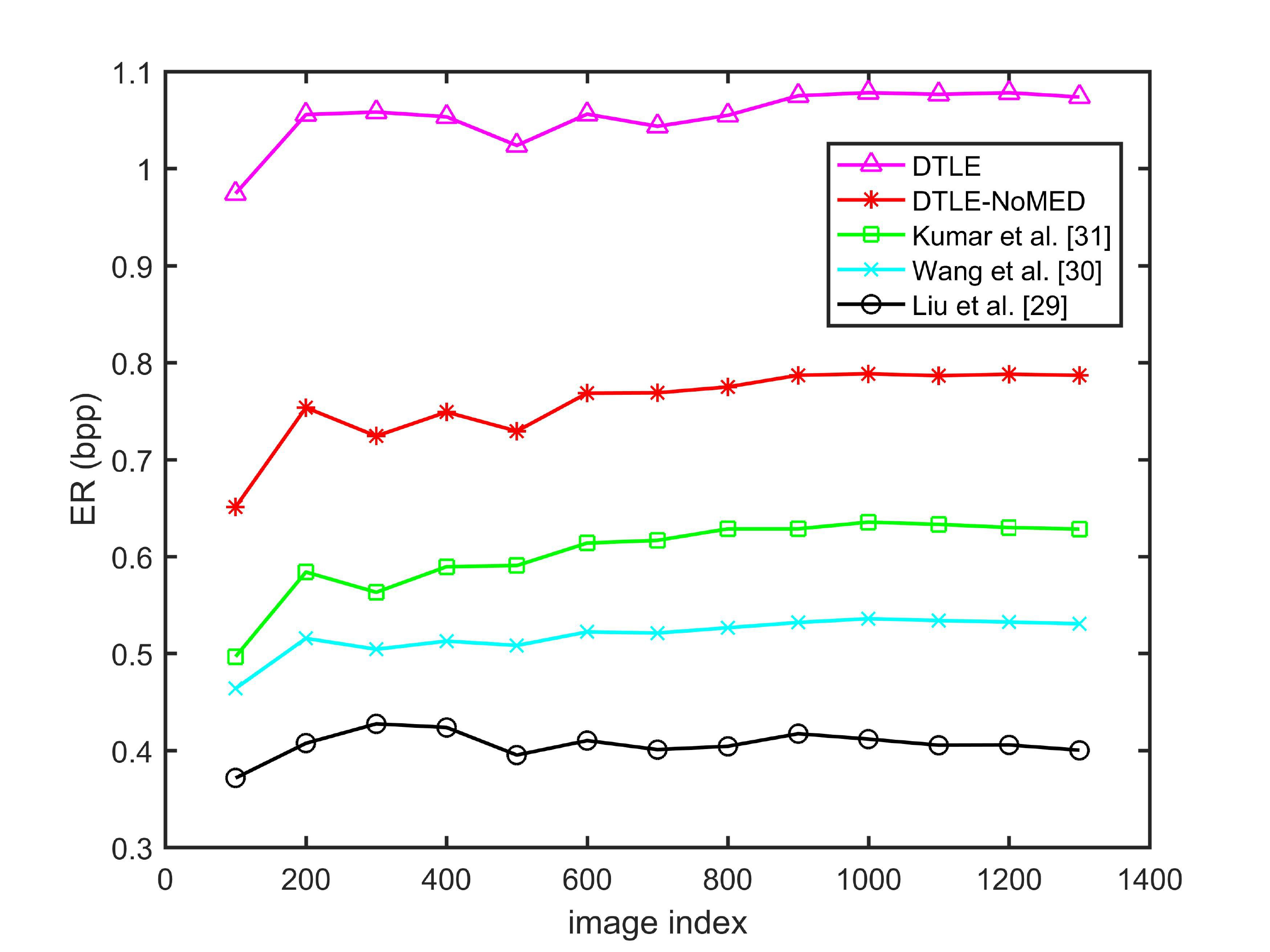}}
	\end{minipage}
    \caption{ER comparisons on the datasets with varying image index.}
    \label{fig:ER}
\end{figure*}

\subsection{Experimental results on datasets}

This section evaluates the performances of five schemes on the three datasets from three standpoints as follows.

\subsubsection{Universal applicability of algorithms}
We make statistics on the images embedded successfully with given bits of secret data (EC) in each dataset, and concentrate on the trend of success percentage of embeddings (SPE) with varying EC, as shown in Fig.~\ref{fig:CP}.
The SPE of Liu et al.'s scheme \cite{LCW16} drops down rapidly from the beginning for all datasets while
SPEs of the schemes from \cite{WYW17,KJ20} decrease significantly from EC $=10\times 10^4$ bits. At $20\times 10^4$ bits for EC, the SPE vanishes for the schemes of \cite{LCW16,WYW17} while staying $60\%$ and $80\%$ for DTLE-NoMED and DTLE, respectively. At an EC of $30\times 10^4$ bits, the DTLE  achieves almost a half on SPE ($50.7\%$, $45.4\%$ and $44.7\%$, respectively), while SPE of Kumar's scheme \cite{KJ20} vanishes. DTLE solves the issue of the large auxiliary information due to complex images in the datasets, which affects greatly the EC of DTLE-NoMED.
Hence, the DTLE attaches high EC for not only some general images but also some complex images, which confirms the universal applicability.

\subsubsection{Size of auxiliary information}
To compare the trend in the size of the auxiliary information conveniently, we deploy a coordinate system of two vertical axes for drawing curves. The left scale is used for DTLE-NoMED while the right one for DTLE.
As shown in Fig.~\ref{fig:LM}, the MED pre-processed images have greatly reduced size of auxiliary information. The size of auxiliary information for DTLE-NoMED is mainly influenced by the number of possible overflow pixels in the image, while that of DTLE depends on the smoothness of the image's texture. Images with complex textures have large $|{e_{\min }}|$ values through MED pre-processing, and more pixels may overflow after the histogram shift, producing more auxiliary information. At the image index of $1000$ in Fig.~\ref{fig:LMA}, the complexity of the image texture and the low similarity between pixels allows DTLE to maintain an upward trend on the size of auxiliary information. The number of possible overflow pixels in the original images gradually decreases, which leads to a decreasing trend for DTLE-NoMED. Then, the curves of the two approaches have opposite size trends for the index from $1000$ to $6000$.

\subsubsection{Embedding rate}
We also investigate the average ER of five approaches on the datasets, as shown in Fig.~\ref{fig:ER}. The DTLE-NoMED has a significant advantage of the average ER on the datasets, $0.8398$ bpp, $0.9443$ bpp and $0.7870$ bpp, respectively. The DTLE even reaches $1.1277$ bpp, $1.0918$ bpp and $1.0745$ bpp, with an improvement of $83.06\%$, $50.09\%$ and $70.85\%$, respectively, compared to TLE by Kumar et al. \cite{KJ20}. This indicates that our proposed DTLE achieves a superior performance over the existing schemes.
\section{Conclusion}
\label{sec:con}
In this paper, we design a reversible data hiding scheme DTLE with a novel two-layer embedding strategy.
The sixth-bit plane is taken into account for the HSB plane while two prediction error peaks are deployed
in either layer to increase the EC.
Moreover, MED pre-processing approach is utilized which effectively reduces the size of the auxiliary information particularly for complex images. Experimental results demonstrate that the DTLE has significant advantages over the existing schemes on both EC and PSNR.

\bibliography{sample-bibliography}{}
\bibliographystyle{IEEEtran}


\par\noindent
\parbox[t]{\linewidth}{
\noindent\parpic{\includegraphics[height=1.5in,width=1in,clip,keepaspectratio]{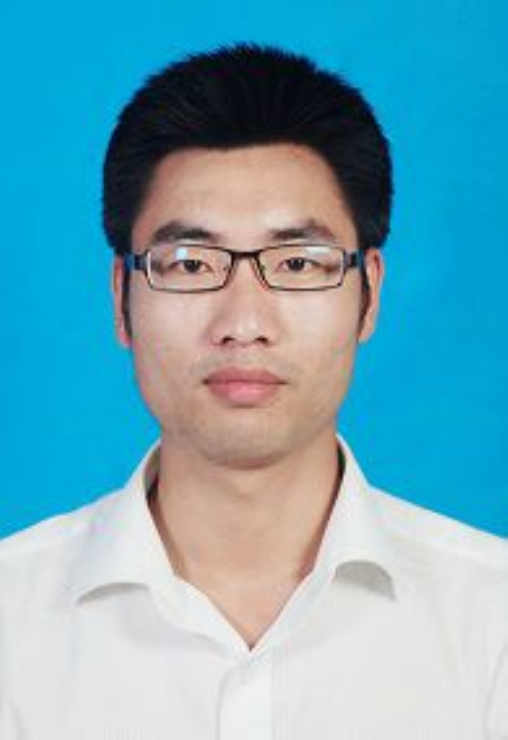}}
\noindent {\bf Fuhu Wu}\
received the B.Sc. and M.E. degrees from Anhui University, Hefei, China, in 2009 and 2012, respectively. He is currently
working toward the Ph.D. degree with the School of Computer Science and Technology, Anhui University, Hefei, China. His current research interests include image processing and reversible data hiding.}
\vspace{0.2\baselineskip}

\par\noindent
\parbox[t]{\linewidth}{
\noindent\parpic{\includegraphics[height=1.5in,width=1in,clip,keepaspectratio]{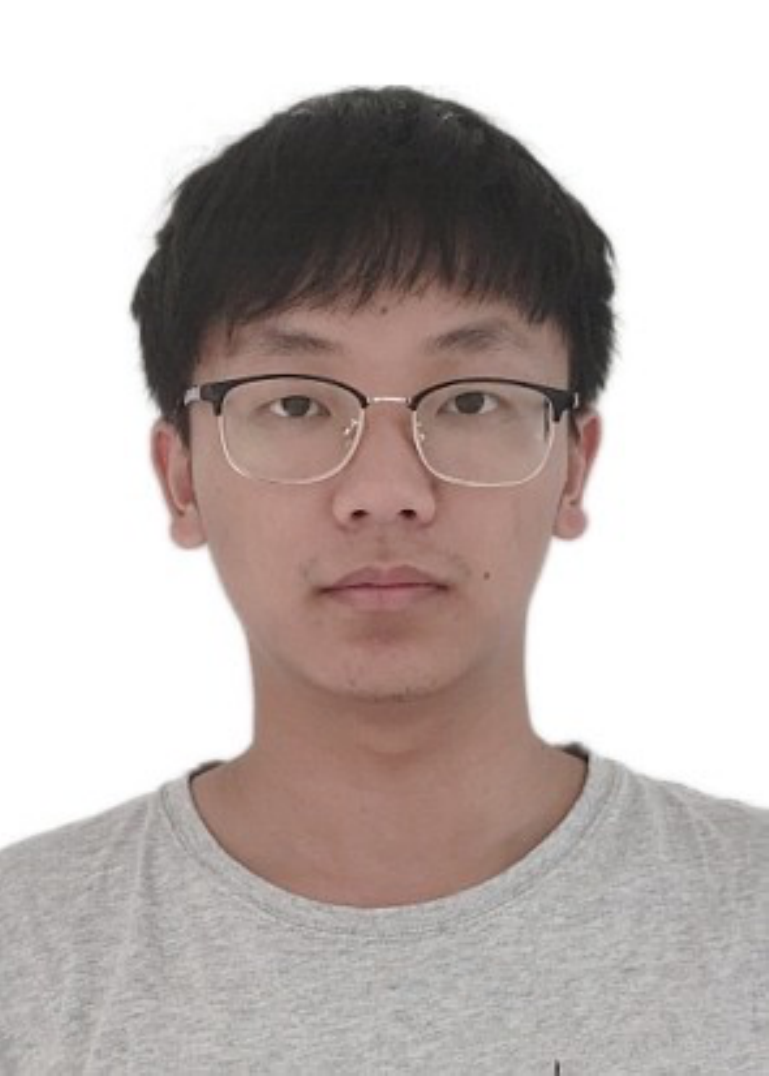}}
\noindent {\bf Jian Sun}\
received the bachelor's degree in Mechatronics in 2020 and currently, he is working toward the master's degree with the School of Computer Science and Technology, Anhui University, Hefei, China. His current research interests include reversible data hiding.}
\vspace{1\baselineskip}

\par\noindent
\parbox[t]{\linewidth}{
\noindent\parpic{\includegraphics[height=1.5in,width=1in,clip,keepaspectratio]{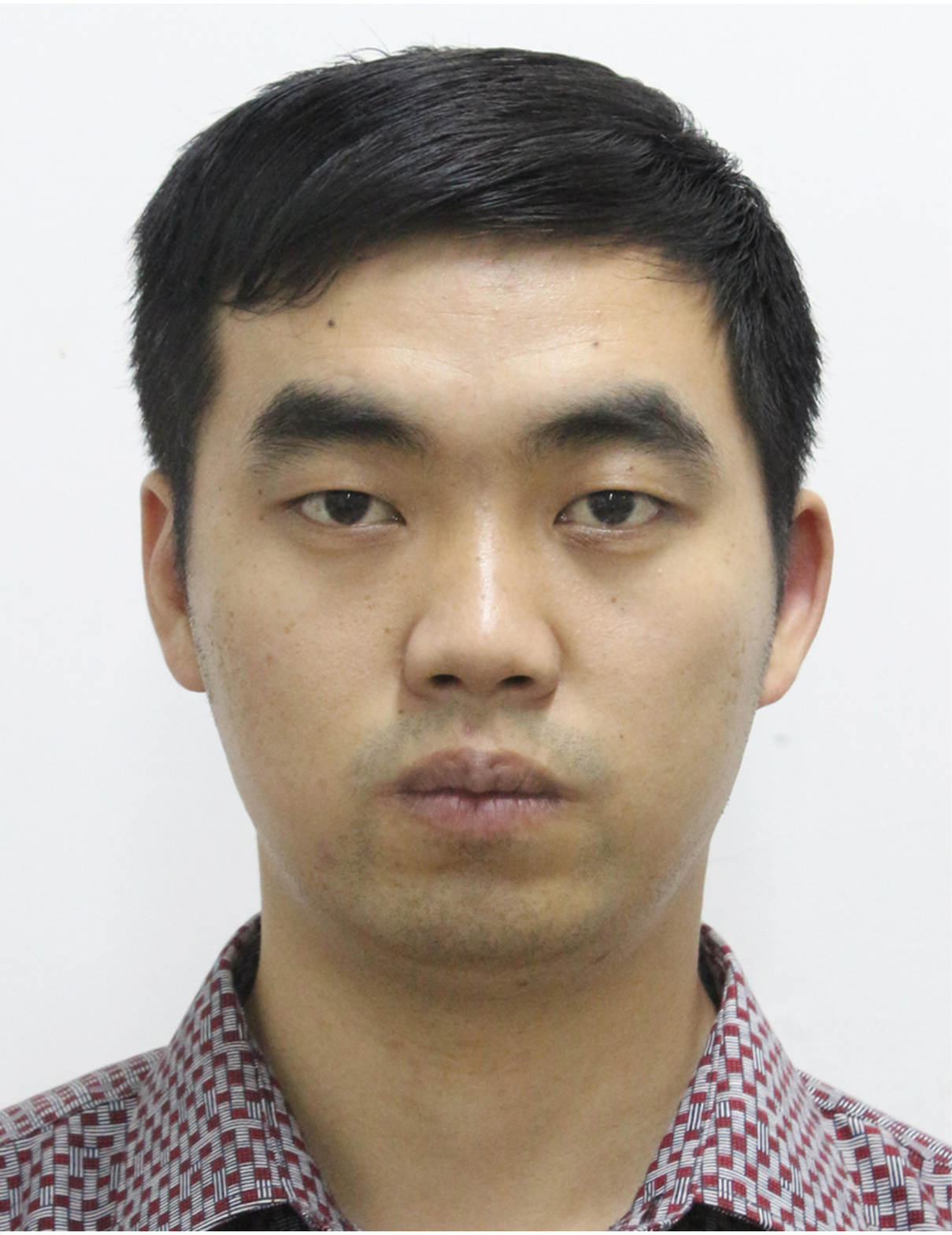}}
\noindent {\bf Shun Zhang}\
received his PhD degree in applied mathematics from Beijing Normal University in 2012. He was a visiting scholar at Friedrich-Schiller-Universitat Jena from 2014 to 2015. He is now an associate professor at Anhui University. His research interests include privacy preservation and computational complexity.}
\vspace{0.3\baselineskip}

\par\noindent
\parbox[t]{\linewidth}{
\noindent\parpic{\includegraphics[height=1.5in,width=1in,clip,keepaspectratio]{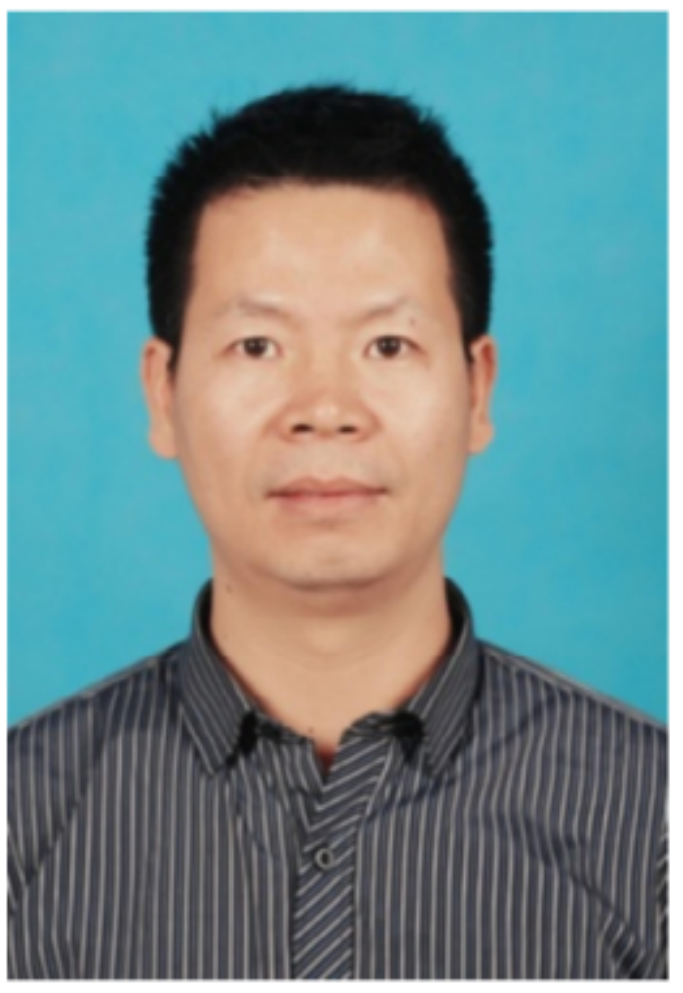}}
\noindent {\bf Zhili Chen}\
received his PhD degree in computer science from University of Science and Technology of China in 2009. He is currently a professor at East China Normal University. His main research interests include privacy preservation, secure multiparty computation, information hiding, spectrum auction and game theory in wireless communications.}
\vspace{0.2\baselineskip}

\par\noindent
\parbox[t]{\linewidth}{
\noindent\parpic{\includegraphics[height=1.5in,width=1in,clip,keepaspectratio]{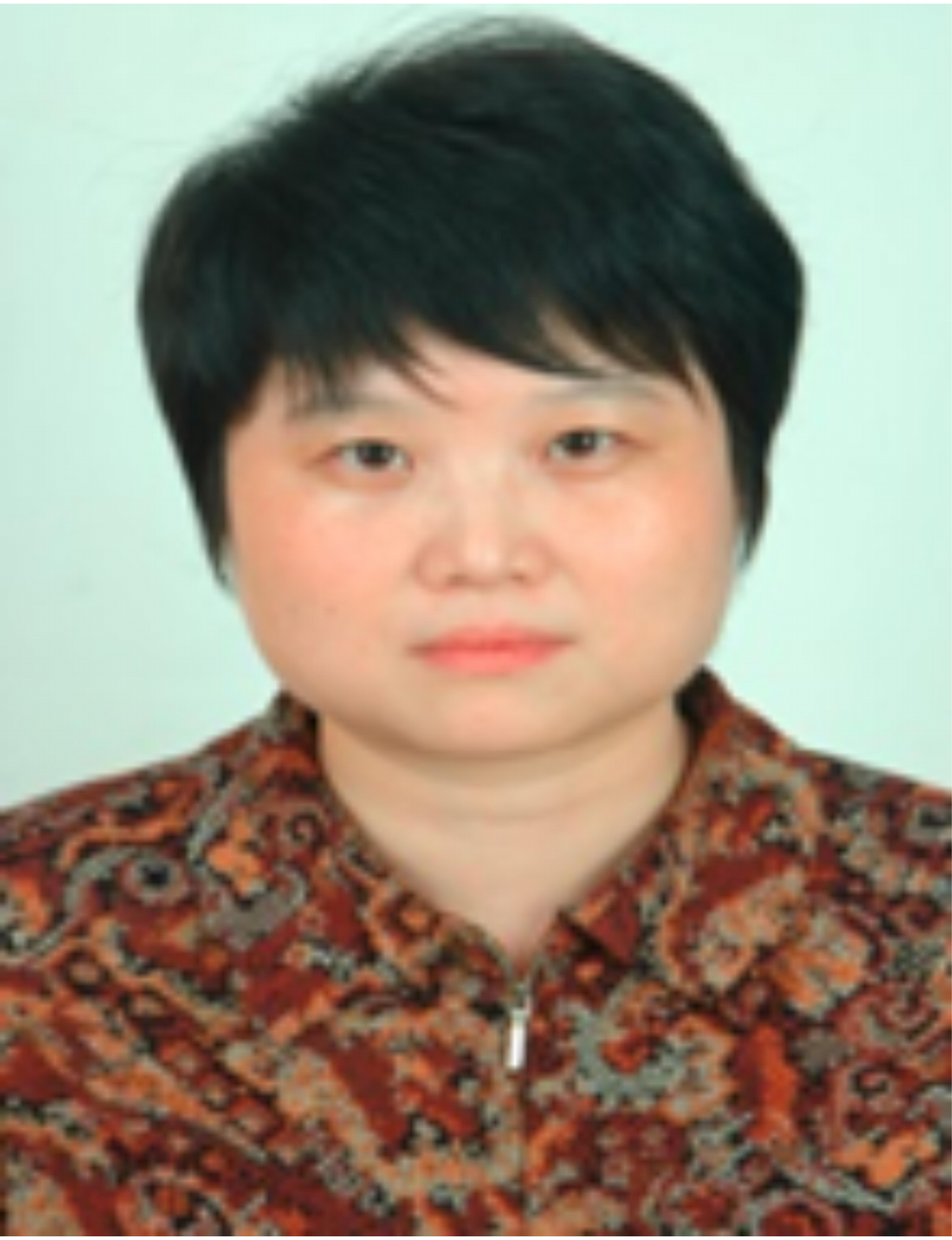}}
\noindent {\bf Hong Zhong}\
received her PhD degree in computer science from University of Science and Technology of China in 2005. She is currently a professor at School of Computer Science and Technology at Anhui University. Her research interests include applied cryptography, IoT security, vehicular ad hoc network, and software-defined networking (SDN).}



%


%
%

\end{document}